\documentclass[12pt]{article}%
\usepackage[nosort]{cite}
\usepackage{graphicx}
\usepackage{multicol}
\usepackage{amsfonts}
\usepackage{amssymb}
\usepackage{amsmath}
\usepackage{heck}
\usepackage{afterpage}
\usepackage{setspace}
\usepackage{verbatim}
\usepackage{color}
\usepackage{longtable}
\usepackage{subcaption}
\usepackage{epsfig}
\usepackage{epstopdf}
\usepackage{adjustbox}
\usepackage{tikz}
\usepackage[margin=1in]{geometry}
\usepackage{titletoc}
\usepackage{hyperref}%
\usepackage{mathrsfs}
\usepackage{dsfont}
\usepackage{algorithm}
\usepackage{algpseudocode}
\usepackage{pgfplots}
\usepackage{algorithmicx, algpseudocode,algorithm}
\usepackage{caption}
\usepackage{float}
\usepackage{makecell}
\usepackage{mathtools}
\usepackage{pdflscape}
\usepackage{array}
\usepackage{longtable}
\usepackage{braket}
\usepackage{multirow}
\newcolumntype{C}{>{\centering\arraybackslash}m{2cm}}
\newcommand{\Tr}{\, {\rm Tr}}
\DeclarePairedDelimiter\abs{\lvert}{\rvert}

\numberwithin{equation}{section}
\newsavebox{\mysavebox}

\hypersetup{colorlinks,citecolor=black,filecolor=black,linkcolor=black,urlcolor=black}

\usetikzlibrary{decorations.markings}
\usetikzlibrary{decorations.markings}
\usetikzlibrary{hobby}
\usepackage{tikz}
\usetikzlibrary{arrows}
\usetikzlibrary{arrows.meta}
\usetikzlibrary{arrows,decorations.pathmorphing}
\usetikzlibrary{shapes.geometric,calc,arrows, positioning,shapes.misc,decorations.markings}
\tikzset{
  big arrow/.style={
    decoration={markings,mark=at position 1 with {\arrow[scale=2,#1]{>}}},
    postaction={decorate},
    shorten >=0.4pt},
  big arrow/.default=black}
\usetikzlibrary{arrows.meta,positioning,calc}
\definecolor{Xcol}{HTML}{0E5403} 
\definecolor{Zcol}{HTML}{57FF48} 
\definecolor{Ycol}{HTML}{FF3232} 

\pgfdeclarelayer{edgelayer}
\pgfdeclarelayer{nodelayer}
\pgfsetlayers{edgelayer,nodelayer,main}
\pgfplotsset{compat=1.16}
\tikzstyle{none}=[inner sep=0pt]

\tikzstyle{NodeCross}=[draw, shape=circle, cross out, inner sep=0pt, minimum size=6pt,line width=0.25mm]
\tikzstyle{Circle}=[draw, shape=circle, black, fill=black, inner sep=0pt, minimum size=6pt]
\tikzstyle{circle}=[draw, shape=circle, black, fill=black, inner sep=0pt, minimum size=16pt]
\tikzstyle{Star}=[draw, shape=star, fill=black, star points=8, inner sep=0pt, minimum size=8pt]
\tikzstyle{CircleRed}=[draw, shape=circle, black, fill=red, inner sep=0pt, minimum size=6pt]
\tikzstyle{StarP}=[draw={rgb,255: red,128; green,0; blue,128}, shape=star, fill={rgb,256: red,128; green,0; blue,128}, star points=8, inner sep=0pt, minimum size=12pt]
\tikzstyle{ShadedCircRed}=[draw=red, shape=circle, fill={rgb, 255: red,255; green,114; blue, 118}, inner sep=0pt, minimum size=80pt, line width=0.5mm, fill opacity=0.2]
\tikzstyle{ShadedCircRed2}=[draw=red, shape=circle, fill={rgb, 255: red,255; green,114; blue, 118}, inner sep=0pt, minimum size=10pt]
\tikzstyle{ShadedCircRed3}=[draw=black, shape=rectangle, fill={rgb, 255: red,255; green,114; blue, 118}, inner sep=0pt, minimum size=113pt, line width=0.25mm]
\tikzstyle{ShadedCirc}=[draw=red, shape=circle, fill=white, inner sep=0pt, minimum size=45pt,  fill opacity=1.0,  line width=0.5mm]
\tikzstyle{CircleBlue}=[draw, shape=circle, fill=blue, inner sep=0pt, minimum size=6pt]
\tikzstyle{BigCirclePurple}=[draw, shape=circle, fill={rgb,255: red,191; green,0; blue,191}, inner sep=0pt, minimum size=12pt]
\tikzstyle{CirclePurple}=[draw, shape=circle, fill={rgb,255: red,191; green,0; blue,191}, inner sep=0pt, minimum size=5pt]
\tikzstyle{EmptyCircle}=[draw, shape=circle, inner sep=0pt, minimum size=4pt]
\tikzstyle{GreenCircle}=[draw, shape=circle,  fill={rgb,255: red,80; green,200; blue,120}, inner sep=0pt, minimum size=8pt]
\tikzstyle{BrownCircle}=[draw, shape=circle,  fill={rgb,255: red,210; green,105; blue,30}, inner sep=0pt, minimum size=8pt]
\tikzstyle{CirclePurpleSmall}=[draw, shape=circle, fill={rgb,255: red,191; green,0; blue,191}, inner sep=0pt, minimum size=4pt]
\tikzstyle{BigCircleGreen}=[draw, shape=circle, fill={rgb,255: red,0; green,191; blue,0}, inner sep=0pt, minimum size=12pt]
\tikzstyle{BigCircleBlue}=[draw, shape=circle, fill={rgb,255: red,0; green,0; blue,191}, inner sep=0pt, minimum size=12pt]
\tikzstyle{BigCircleRed}=[draw, shape=circle, fill={rgb,255: red,191; green,0; blue,0}, inner sep=0pt, minimum size=12pt]
\tikzstyle{BrownCircleSmall}=[draw, shape=circle,  fill={rgb,255: red,210; green,105; blue,30}, inner sep=0pt, minimum size=6pt]
\tikzstyle{SmallCircleBrown}=[draw, shape=circle,  fill={rgb,255: red,210; green,105; blue,30}, inner sep=0pt, minimum size=5pt]
\tikzstyle{SmallCircleRed}=[draw, shape=circle, fill={rgb,255: red,191; green,0; blue,0}, inner sep=0pt, minimum size=6pt]
\tikzstyle{DashedLine}=[-, densely dashed, line width=0.25mm]
\tikzstyle{DottedLine}=[-, dotted, line width=0.25mm]
\tikzstyle{ThickLine}=[-, line width=0.25mm]
\tikzstyle{ArrowLineRight}=[-, -{Stealth[scale=1.25]}, line width=0.25mm, scale=5]
\tikzstyle{ArrowLineRed}=[-, draw={rgb,255: red,191; green,0; blue,0}, -{Stealth[scale=1.75]}, line width=0.1mm, scale=5]
\tikzstyle{RedLine}=[-, draw={rgb,255: red,191; green,0; blue,0}, fill=none, line width=0.5mm]
\tikzstyle{DashedLineThin}=[-, densely dashed, line width=0.125mm, fill=none, draw=black]
\tikzstyle{DottedRed}=[-, dotted, draw={rgb,255: red,191; green,0; blue,0}, fill=none, line width=0.25mm]
\tikzstyle{DashedRed}=[-, densely dashed, draw={rgb,255: red,191; green,0; blue,0}, fill=none, line width=0.25mm]
\tikzstyle{BlueLine}=[-, draw={rgb,255: red,0; green,0; blue,191}, fill=none, line width=0.5mm]
\tikzstyle{ArrowLineBlue}=[-, draw={rgb,255: red,0; green,0; blue,191}, -{Stealth[scale=1.75]}, line width=0.1mm, scale=5]
\tikzstyle{GreenDoubleArrow}=[<->, draw={rgb,155: red,0; green,255; blue,0},  line width= 0.5mm, scale=5]
\tikzstyle{RedDoubleArrow}=[<->, draw={rgb,255: red,255; green,0; blue,0},  line width= 0.5mm, scale=5]
\tikzstyle{BlueDottedLight}=[-, dotted, draw={rgb,255: red,0; green,0; blue,191}, fill=none, line width=0.3mm]
\tikzstyle{BrownLine}=[-, draw={rgb,255: red,210; green,105; blue,30}, fill=none, line width=0.5mm]
\tikzstyle{DottedRed}=[-, dotted, draw={rgb,255: red,191; green,0; blue,0}, fill=none, dotted, line width=0.5mm]
\tikzstyle{DottedPurple}=[-, dotted, draw={rgb,255: red,191; green,0; blue,191}, fill=none, dotted, line width=0.5mm]
\tikzstyle{BlueDottedLight}=[-, dotted, draw={rgb,255: red,0; green,0; blue,191}, fill=none, line width=0.5mm]
\tikzstyle{ArrowLinePurple}=[-, draw={rgb,255: red,191; green,0; blue,191}, -{Stealth[scale=1.75]}, line width=0.5mm, scale=5]
\tikzstyle{DashedLineGreen}=[-, densely dashed, draw={rgb,255: red,74; green,103; blue,65}, line width=0.25mm]
\tikzstyle{LineGreen}=[-, draw={rgb,255: red, 74; green,200; blue,65}, line width=0.5mm]
\tikzstyle{ArrowLineGreen}=[-, draw={rgb,255: red,0; green,191; blue,0}, -{Stealth[scale=1.75]}, line width=0.5mm, scale=5]
\tikzstyle{GreenLine}=[-, draw={rgb,255: red,0; green,191; blue,0}, fill=none, line width=0.5mm]
\tikzstyle{PurpleLine}=[-, draw={rgb,255: red,191; green,0; blue,191}, fill=none, line width=0.5mm]
\tikzstyle{PPurpleLine}=[-, draw={rgb,255: red,191; green,0; blue,191}, fill=none, line width=2.5mm]
\tikzstyle{DPurpleLine}=[-, dotted, draw={rgb,255: red,191; green,0; blue,191}, fill=none, line width=0.5mm]
\tikzstyle{SBrownLine}=[-, draw={rgb,255: red,191; green,0; blue,191}, fill=none, opacity=0.35, line width=2.5mm]
\tikzstyle{DottedBlue}=[-, dotted, draw=blue, fill=none, dotted, line width=0.5mm]
\tikzstyle{DashedPurpleLine}=[-, densely dashed, draw={rgb,255: red,191; green,0; blue,191}, fill=none, line width=0.5mm]
\tikzstyle{SmallCircleBlue}=[draw, shape=circle, fill=blue, inner sep=0pt, minimum size=5pt]
\tikzstyle{SmallCirclePurple}=[draw, shape=circle, fill={rgb,255: red,191; green,0; blue,191}, inner sep=0pt, minimum size=5pt]
\tikzset{snake it/.style={decorate, decoration=snake}}

\tikzset{
  mid arrow/.style={
    postaction={
      decorate,
      decoration={
        markings,
        mark=at position 0.5 with {
          \arrow{Stealth[length=1.5mm,width=1.2mm]}
        }
      }
    }
  }
}
\tikzset{
dashstar/.style={
 dash pattern=on 5pt off 5pt,
 postaction={
  decorate,
  decoration={
   markings,
   mark=between positions 9pt and 1 step 10pt with {
     \node[color=red] {*};
   }
  }
 }
},
dashstarstar/.style={ 
 dash pattern=on 5pt off 10pt,
 postaction={
   decorate,
   decoration={
     markings,
     mark=between positions 10pt and 1
          step 15pt
           with {
            \node at (-2pt,0pt) {\pgfuseplotmark{star}};
            \node at (2pt,0pt) {\pgfuseplotmark{star}};
           }
   }
 }
}
}

\begin{document}

\begin{flushright}
   {\small \texttt{\hfill ZMP-HH/26-9}}
\end{flushright}

\date{June 2026}

\title{Controlled Chaos in 4D SCFTs}

\institution{HAMBURG}{\centerline{$^{1}$II. Institut für Theoretische Physik, Universität Hamburg, 22607 Hamburg, Germany}}
\institution{PENN}{\centerline{$^{2}$Department of Physics and Astronomy, University of Pennsylvania, Philadelphia, PA 19104, USA}}
\institution{PENNmath}{\centerline{$^{3}$Department of Mathematics, University of Pennsylvania, Philadelphia, PA 19104, USA}}

\authors{
Florent Baume\worksat{\HAMBURG}\footnote{e-mail: \texttt{florent.baume@desy.de}},
Atakan \c{C}avu\c{s}o\u{g}lu\worksat{\PENN}\footnote{e-mail: \texttt{atakanc@sas.upenn.edu}}, \\[4mm]
Vivek Chakrabhavi\worksat{\PENN}\footnote{e-mail: \texttt{vivekcm@sas.upenn.edu}}, and
Jonathan J. Heckman\worksat{\PENN,\PENNmath}\footnote{e-mail: \texttt{jheckman@sas.upenn.edu}}
}

\abstract{Chaotic dynamics play an important role in a number of physical systems.
One of the qualitative hallmarks of this behavior is the appearance of a sufficiently ``complex'' spectrum of energy levels. This also makes it challenging to directly verify the onset of chaos in interacting quantum field theories. We present a class of 4D superconformal field theories (SCFTs)
given by orbifolds of 4D $\mathcal{N} = 4$ Super Yang--Mills theory in which operator mixing in a controlled subsector is described by an effective spin chain in one spatial dimension with nearest neighbor interactions tuned by the marginal couplings of the SCFT. Tuning the marginal couplings results in a chaotic spectrum, while generically the spin chain exhibits Anderson localization. We diagnose the onset of chaos by analyzing the statistical distribution of eigenvalues of the dilatation operator, in particular properties such as eigenvalue level repulsion, spectral rigidity, and the spectral form factor. We also show that other diagnostics such as Krylov complexity sometimes do not faithfully capture this information. This structure defines a chaotic billiard in the target space of the stringy realization. We also comment on the large $N$ holographic dual description, where the controlled single spin chain approximation must be supplemented by multi-trace dynamics, i.e., the splitting and joining of multiple spin chains.}

\maketitle

\enlargethispage{\baselineskip}

\setcounter{tocdepth}{3}

\tableofcontents

\newpage

\section{Introduction} \label{sec:Intro}

Quantum chaos has emerged as a unifying framework for understanding the dynamics of complex quantum systems, with deep connections to statistical mechanics, quantum information, and gravity. In many-body quantum systems, it is believed that chaos is characterized by
level spacing correlations matching those of random matrix ensembles, while integrable systems have uncorrelated Poisson spectra \cite{Dyson1962I,Bohigas:1983er, Muller:2004nb, Guhr:1997ve,Haake:2010fgh, Berry85}. Other features of quantum chaos include operator growth and rapid delocalization in Hilbert space, which are quantified via out-of-time-order correlators  and Krylov complexity.\footnote{See for example \cite{1969JETP...28.1200L,Shenker:2013pqa, Maldacena:2015waa, Hashimoto:2017oit, Parker:2018yvk, Barbon:2019wsy, Avdoshkin:2019trj, Dymarsky:2021bjq, Caputa:2021sib, Balasubramanian:2022tpr, Baiguera:2025dkc, Nandy:2024evd, Rabinovici:2020ryf, Rabinovici:2021qqt, Rabinovici:2022beu,Balasubramanian:2022dnj, Avdoshkin:2022xuw, Balasubramanian:2023kwd, Balasubramanian:2024ghv, Balasubramanian:2025xkj,  Rabinovici:2025otw}.}

Despite this progress, for the most part, quantum chaos has only been studied in lower-dimensional models, and is best understood in quantum mechanical or effectively one-dimensional settings. 
It is natural to ask how these signatures of chaos arise from microscopic dynamics in quantum field theory, particularly in strongly coupled and higher-dimensional settings.

In general terms, a promising way to extend the many controlled lower-dimensional examples to the higher-dimensional setting is to seek out large charge subsectors where the effective Hamiltonian reduces to a better studied lower-dimensional system.
A celebrated example of this sort is operator mixing in large charge sectors of $\mathcal{N} = 4$ Super Yang-Mills theory.\footnote{For a
broader discussion of the utility of the large charge expansion, see in particular \cite{Hellerman:2015nra} and \cite{Gaume:2020bmp}
for a recent review.}
As discovered in \cite{Berenstein:2002jq} and heavily studied ever since,
there is a controlled subsector of operators where the anomalous dimensions of ``nearly'' $1/2$ BPS operators
are governed by an effective spin chain Hamiltonian. In the planar limit, this system exhibits integrability,
but deformations away from the original planar $\mathcal{N} = 4$ theory also exhibit quantum chaos \cite{McLoughlin:2022jyt}.

Motivated by these considerations, one of our aims in this note will be to establish the existence and properties of large charge
subsectors in 4D $\mathcal{N} = 1$ and $\mathcal{N} = 2$ SCFTs where the onset of quantum chaos is controlled by a spin chain system. Along these lines, we show that provided there is a suitable large charge single trace $1/2$ BPS operator in the spectrum, this can be interpreted as the ground state of a spin chain. Excitations of the spin chain correspond to non-BPS operators where operator mixing is governed by the spin chain Hamiltonian. In particular, the strength of this operator mixing is controlled by the marginal couplings of the SCFT. In a long spin chain (i.e. a large charge operator of the SCFT), all of these interactions amount to nearest neighbor interactions, and higher-loop effects specify longer range interactions. To keep things concrete, we focus on quiver gauge theories obtained from orbifolds of $\mathcal{N} = 4$ Super Yang--Mills theory. In string theory terms, these gauge theories arise from D3-branes probing the supersymmetric orbifold $\mathbb{C}^{3} / \Gamma$. The procedure for extracting the resulting quiver gauge theory follows from \cite{Douglas:1996sw, Kachru:1998ys, Lawrence:1998ja}. A single trace large charge operator of the parent $\mathcal{N} = 4$ theory which is invariant under $\Gamma$ naturally descends to a gauge invariant operator of the quiver gauge theory. Orbifold inheritance guarantees that at one loop order, operator mixing in the parent $\mathcal{N} = 4$ theory descends to operator mixing in the orbifold theory.\footnote{See reference \cite{Beisert:2005he} for a detailed analysis of the special case of abelian orbifolds of the form $\mathbb{C}^{3} / \mathbb{Z}_L$. The structure holds more generally, of course.} As such, there is a natural spin chain sector in all of these cases.

However, compared with the case of $\mathcal{N} = 4$ Super Yang-Mills theory, these 4D SCFTs can sometimes admit a much larger class of marginal deformations. In particular, for $\mathcal{N}=2$ theories, each gauge group factor of the quiver gauge theory yields a marginal gauge coupling. The counting is more involved for $\mathcal{N}=1$ theories, but in the case where $\mathbb{C}^3/\Gamma$ is a toric singularity, the number of marginal gauge couplings is (aside from a few special cases) set by $d_{\mathrm{ext}} -1$, where $d_{\mathrm{ext}}$ is the number of external edges in the toric diagram \cite{Imamura:2007dc}. As such, the corresponding class of spin chain Hamiltonians is quite rich, providing access to a much larger class of possible dynamics.

In this work we confine our analysis to one-loop mixing effects in the dilute gas regime, i.e., where the number of impurities is sufficiently small that scattering interactions amount to subleading perturbations of the energy spectrum. In the limit of a low number of impurities relative to the number of spin chain excitations (i.e., the dilute gas approximation), the hopping term for a single impurity takes the general form:
\begin{equation}
\widehat{H}_{\mathrm{hopping}} \sim
\begin{pmatrix}
	b_L + b_1 & -b_1 & 0 & \cdots & -b_L\\
	-b_1 & b_1+b_2 & -b_2 & \cdots & 0\\
	0 & -b_2 & b_2+b_3 & \ddots & \vdots\\
	\vdots & \vdots & \ddots & \ddots & -b_{L-1}\\
	-b_L & 0 & \cdots & -b_{L-1} & b_{L-1} + b_{L}
\end{pmatrix},
\end{equation}
where the $b_i$ are functions of the marginal gauge couplings. The well studied case of an integrable spin chain amounts to taking $b_{i} = b$ for all $i$ on a chain with periodic boundary conditions. Likewise, the case of an open spin chain
corresponds to $b_L = 0$, and the specialization to the integrable case requires setting $b_i = b$.

Provided we remain in the dilute gas regime, it thus suffices to determine possible values of the $b_i$ for which the spectrum of $\widehat{H}_{\mathrm{hopping}}$ exhibits chaotic dynamics. We study this problem from a number of complementary perspectives. One direct
way to access chaotic dynamics is to simply extract the full eigenvalue spectrum, and look for eigenvalue repulsion. We indeed find that an ensemble of random matrices with the $b_i$'s drawn from a power of the chi-distribution for particular choices of $\alpha$ and $p$:\footnote{We remark that this ensemble of matrices is different from the celebrated
tridiagonal ensemble found in \cite{Dumitriu:2002ntg} with entries of the form:
\begin{equation}\label{eq:DumitriuEnsemble}
\widehat{M} \sim\frac{1}{\sqrt{2}}
\begin{pmatrix}
	a_1 & b_1 & 0 & \cdots & 0\\
	b_1 & a_2 & b_2 & \cdots & 0\\
	0 & b_2 & a_3 & \ddots & \vdots\\
	\vdots & \vdots & \ddots & \ddots & b_{L-1}\\
	0 & 0 & \cdots & b_{L-1} & a_{L}
\end{pmatrix},
\end{equation}
where the entries are drawn from random distributions $b_{i} \sim \chi_{i \beta}$ and $a_{i} \sim \mathcal{N}(0,2)$. Here $\mathcal{N}(\mu,\sigma^2)$ refers to a normal distribution with mean $\mu$ and variance $\sigma^2$. For related discussions of chaotic dynamics, see also, for example, \cite{Parker:2018yvk, Balasubramanian:2022dnj}. In this case, the ensemble of matrices exhibits the same eigenvalue distribution as the $\beta$-ensemble of Gaussian matrices, where in our conventions $\beta = 1,2,4$, respectively refer to the Gaussian Orthogonal Ensemble (GOE), Gaussian Unitary Ensemble (GUE), and Gaussian Symplectic Ensemble (GSE). In our case, the eigenvalue spectrum is different, but the universality class for eigenvalue repulsion is the same, provided that the parameters $p$ and $\alpha$ are tuned appropriately.}
\begin{equation}
b_i \sim (\chi_{\alpha i})^p,
\end{equation}
exhibits chaotic spectral statistics. Here, $p$ and $\alpha$ are parameters that determine the properties of the eigenvalue spectrum.\footnote{Equivalently, this distribution corresponds to a generalized gamma distribution with $i$-dependent parameters,
\begin{equation}
b_i \sim \mathrm{GeneralizedGamma}\left(2^{p/2}, \frac{\alpha i}{p}, \frac{2}{p}\right).
\end{equation}
See Appendix \ref{app:Distributions} for more details on the distributions we use.} At more generic values of the $b_i$, we find an intermediate behavior that is neither integrable nor fully chaotic, namely behavior that exhibits Anderson localization \cite{Anderson:1958vr} and emergent integrability.

In addition to these local measures based on eigenvalue repulsion, we also study the ``spectral rigidity'' of the eigenvalue spectrum. This diagnostic probes more global properties of the spectrum and provides an additional way to diagnose the onset of chaos \cite{Dyson1962I, Dyson1962II, Dyson1962III, DysonMehta1963IV, Berry85}. We find that just as for our local measures of level statistics based on eigenvalue repulsion, our tuned system exhibits many of the expected hallmarks of a chaotic spectrum, including the signature logarithmic growth of the spectral rigidity as a function of energy window:
\begin{equation}
\Delta_{3}(\mathcal{E}) \propto \mathrm{log} (\mathcal{E} / \mathcal{E}_0).
\end{equation}
The precise constant of proportionality depends on the details of our tuned ensemble, and we find that as $p$ increases, it approaches the same universality class as the generalized Gaussian random matrices.

To further diagnose the onset of chaotic dynamics from the eigenvalue spectrum, we also study the spectral form factor \cite{Cotler:2016fpe}, i.e., we consider the ratio of generalized partition functions given by $\langle \vert Z(t,T^{-1}) \vert^2 \rangle / \langle Z(T^{-1})^2 \rangle$ as a function of time $t$ and temperature averaged over many draws from our ensemble of Hamiltonians (as denoted by the $\langle \bullet \rangle$ brackets). We find that for appropriately tuned Hamiltonians, the time dependence of the spectral form factors exhibits the expected dip, ramp, and plateau associated with a chaotic system. Combined with the local statistics and global spectral rigidity measures, this provides another strong piece of evidence that our tuned system is indeed chaotic.

Krylov complexity provides a complementary perspective which focuses on complexity spread.\footnote{See e.g. \cite{Parker:2018yvk, Barbon:2019wsy, Avdoshkin:2019trj, Dymarsky:2021bjq, Caputa:2021sib, Balasubramanian:2022tpr, Baiguera:2025dkc, Nandy:2024evd, Rabinovici:2020ryf, Rabinovici:2021qqt, Rabinovici:2022beu, Balasubramanian:2022dnj, Avdoshkin:2022xuw, Balasubramanian:2023kwd, Balasubramanian:2024ghv, Balasubramanian:2025xkj, Rabinovici:2025otw, Kar:2021nbm, Adhikari:2022whf, Liu:2022god, Bhattacharjee:2022ave, Du:2022ocp, Espanol:2022cqr, Camargo:2022rnt, Bhattacharya:2023zqt, Hashimoto:2023swv, Caputa:2024vrn, Chattopadhyay:2024pdj, Baggioli:2024wbz, Alishahiha:2024vbf, Craps:2024suj, Das:2024tnw, PG:2025ixk, Demulder:2025uda, Peacock:2025mwl, Grabarits:2026hjz, Camargo:2026szl, Alfinito:2026yex, Roychowdhury:2026vzq, Roychowdhury:2026igc, Nastase:2026lhz, Fatemiabhari:2025cyy, Pain:2026ssv, Fatemiabhari:2025usn, Fatemiabhari:2025poq, Fatemiabhari:2026rob, Fatemiabhari:2026goj, Caputa:2026hvh}.} This amounts to studying the exploration of operator and/or state growth as a function of time evolution of the underlying Hamiltonian. In our setup, the spin chain Hamiltonian defines a natural notion of time evolution, and we can study the growth of complexity starting from a ``generic'' initial state. Concretely, we consider generic linear combinations of a single localized impurity. To contrast this, we also study the case of a single localized excitation in the middle of the spin chain, comparing the growth of Krylov complexity in both situations. Curiously enough, we find that the qualitative form of Krylov complexity is rather insensitive to the choice of $b_i$'s, the exception being the integrable spin chain with all $b_i$ equal. We take this to be a general cautionary lesson that even in controlled situations eigenvalue statistics provide a more robust diagnostic for the onset of chaos.

This chaotic behavior of the spin chain also manifests itself geometrically. Indeed, in the stringy realization in terms of $N$ D3-branes probing a Calabi--Yau cone $X$, the excitations of the spin chain translate to a chaotic billiard bouncing in the background geometry. In the special case of 4D $\mathcal{N} = 2$ SCFTs which are also realized via M-theory realizations of compactifications of 6D SCFTs, this chaotic motion also manifests as fluctuations of an M2-brane in the extra-dimensional geometry.

At large $N$, the quiver gauge theories we consider have a semi-classical gravity dual. For example, in the case of a large number of D3-branes probing the singularity $\mathbb{C}^{3} / \Gamma$, this is type IIB string theory on AdS$_5 \times S^{5}/\Gamma$. Strictly speaking, this approximation is only valid provided the radius of curvature, as set by $\lambda^{1/4} / \vert \Gamma \vert \gg 1$ is sufficiently large. Otherwise, we are no longer in the semi-classical approximation. This turns out to be incompatible with the large 't Hooft coupling limit of the single spin chain sector approximation. Rather, the single spin chain approximation tells us about a highly curved gravity dual. To some extent, this is to be expected: in the gravity dual, chaotic dynamics is typically a hallmark of gravitational objects with a large number of microstates, as in a large black hole (see e.g., \cite{Shenker:2013pqa, Maldacena:2015waa, You_2017, Garcia-Garcia:2016mno, Cotler:2016fpe, Chen:2024oqv}). To access such sectors, we must instead entertain operators with even larger operator scaling dimensions, i.e., the multi-trace sector of the theory. In the quiver, we can still consider essentially the same class of spin chains, but where we now allow the operator of the quiver gauge theory to wind around the quiver multiple times. This leads to non-trivial dynamics where impurities can still propagate along a single spin chain, but also where the spin chains themselves can split and join. While the pure spin chain approximation is no longer entirely valid in this case, some of the operator mixing is still constrained by a large charge expansion. As such, we can still exhibit many similar qualitative features indicative of controlled chaotic dynamics.

The rest of this paper is organized as follows. In section \ref{sec:REVIEW}, we briefly review some general diagnostics of quantum chaos
which we shall use to study the dynamics of our 4D SCFTs. In section \ref{sec:QUIVER}, we turn to a class of 4D SCFTs which admit a large charge subsector with operator mixing captured by a spin chain. With this in place, we turn to an analysis of chaos for these spin chains in section \ref{sec:SpinChaos}. In section \ref{sec:BILLIARDS} we characterize these spin chain 
fluctuations in terms of a chaotic billiard which explores the extra-dimensional 
geometry of a string / M-theory background. In section \ref{sec:HOLO} we turn to the holographic interpretation of these systems. Section \ref{sec:CONC} presents our conclusions. We defer some additional review material on computing operator mixing / spin chain Hamiltonians in QFTs to Appendix \ref{app:SpinChainComputation}, on the chi / generalized gamma distributions to Appendix \ref{app:Distributions}, and on spectral rigidity to Appendix \ref{app:SpectralRigidity}.

\section{Signatures of Chaos} \label{sec:REVIEW}

To frame the analysis to follow, in this section we review the main signatures we use to diagnose the onset of quantum chaos. After this, we turn to the case of interest in this work, chaos in 4D SCFTs.

Chaos in classical mechanics is defined as the sensitivity of the trajectories to the initial conditions, which arises as a consequence of nonlinearities in the equations of motion of a system with multiple degrees of freedom. However, how classical chaos arises from quantum mechanics by the correspondence principle is not well understood, as the time evolution of states in quantum mechanics is inherently linear. There exist several different approaches to study this problem, the most common of which is the study of the statistical description of the spectrum of a quantum Hamiltonian \cite{Dyson1962I,Bohigas:1983er,Oganesyan:2007wpd,Atas:2013gvn}. Alternatively, another approach that has been proposed and has drawn much interest in recent years is the analysis of the Lanczos spectrum \cite{Lanczos:1950zz,Balasubramanian:2022dnj,Balasubramanian:2023kwd} and various notions of complexity for chaotic systems, including Krylov complexity \cite{Parker:2018yvk,Rabinovici:2020ryf,Balasubramanian:2022tpr,Rabinovici:2022beu,Rabinovici:2025otw,Avdoshkin:2022xuw,Pain:2026ssv}. We will compare and contrast both approaches.

\subsection{Level Statistics and Gaussian Ensembles}\label{sec:Gaussian-Ensembles}

The most common statistical approach to identify chaotic quantum systems is to relate the local features of their spectra to random matrix theory (RMT),\footnote{See for example \cite{Guhr:1997ve,Haake:2010fgh, Akemann:2011csh} for some reviews of Random Matrix Theory.} as it is conjectured that local features of the spectra of chaotic systems are well-described by RMT \cite{Dyson1962I,Bohigas:1983er}. This is sometimes even taken to be the definition of chaotic quantum systems. Qualitatively, correlations between the eigenvalues of random matrices result in ``repulsions'' between the eigenvalues in the spectra. The natural quantity to investigate this repulsion is the spacing between consecutive eigenvalues $E_i$,
\begin{equation}
    s_i = E_{i+1} - E_i \, ,
\end{equation}
which is called the level spacing. Chaotic systems are conjectured to exhibit the same type of correlation and repulsion between the eigenvalues in the spectra as random matrix theories. In contrast, integrable systems have an infinite number of conserved quantities. For this reason, different sectors of the Hilbert space with different values for those conserved quantities do not mix under Hamiltonian time evolution, and the energy levels of integrable systems are uncorrelated with no repulsion between them.

The correlations between the energy eigenvalues are expected to be only local, which means that the global shape or structure of the density of states, $\rho(E)$, is not an appropriate quantifier for quantum chaos. Therefore, while studying the statistical properties of the spectra of chaotic systems, it is customary to locally scale the spectrum to set the density of states $\rho(E)$ uniform, which is sometimes called ``unfolding'' the spectrum.\footnote{To be precise, if we denote by $\bar{N}(E)$ the smoothed-out cumulative density of states denoting the (expected) number of eigenvalues $E_i$ with values less than $E$, then the unfolded spectrum $\{x_i\}$ can be obtained as $x_i = \bar{N}(E_i)$. This way, the mean spacing between consecutive unfolded eigenvalues $x_i$ is one, and spacings $s_i = x_{i+1} - x_i$ encode information about level repulsion in the system. See appendix \ref{app:SpectralRigidity} for more details.} Another approach to get rid of the dependence on the density of states is to investigate the ratios between consecutive level spacings instead of the level spacings themselves:
\begin{equation}
    r_i = \frac{s_{i+1}}{s_i} \, ,
\end{equation}
which is a quantity independent of the density of states $\rho(E)$. It would be equally valid to define this quantity as $s_i / s_{i+1}$, and the probability density function for $r_i$ for random matrices is invariant under $r_i \rightarrow 1/r_i$. Therefore, it is customary to restrict the domain of $r$ to $[0,1]$ by defining
\begin{equation} \label{eq:DefinitionRtilde}
    \Tilde{r}_i = \min\left(r_i, \frac{1}{r_i}\right) \, ,
\end{equation}
which is a quantity commonly used for the statistical analysis of chaotic systems \cite{Oganesyan:2007wpd}.

For integrable systems, the energy levels are uncorrelated and can be treated as independent variables; they become uniformly distributed after the unfolding of the density of states. Therefore, the density of the unfolded level spacings $s$ for integrable systems follows the Poisson distribution
\begin{equation}
    \rho_{\text{Poisson}}(s) = e^{-s} \, ,
\end{equation}
from which the probability distribution of the spacing ratios $r$ can be calculated as
\begin{equation} \label{eq:IntegrableRStatistics}
    P_{\text{Poisson}}(r) = \frac{1}{\left(1+r\right)^2} \, .
\end{equation}

A canonical example of chaotic dynamics is the level spacing statistics of Gaussian ensembles. Gaussian ensembles are probability distributions over self-adjoint matrices whose entries are independently sampled from the Gaussian distribution. The three main types are Gaussian Orthogonal Ensemble (GOE), Gaussian Unitary Ensemble (GUE), and Gaussian Symplectic Ensemble (GSE), whose entries are respectively real, complex, and quaternionic, with each independent real component drawn from the standard normal distribution $\mathcal{N}(0,1)$. These are part of a larger class of distributions called Gaussian $\beta$ ensemble labelled by the Dyson index $\beta$, for which GOE, GUE and GSE correspond to $\beta = 1,2,4$, counting the number of real components of each entry. For an $L \times L$ self-adjoint matrix $W_L$ drawn from these distributions, the probability density function is given by
\begin{equation}
    P(W_L) = \frac{1}{Z} e^{-\frac{\beta}{4} \Tr(W_L^2)} \quad \text{with} \quad Z = 2^{\frac{1}{2}L}\left(\frac{2\pi}{\beta}\right)^{\frac{1}{2}L + \frac{1}{4}\beta L (L-1)} \, .
\end{equation}

For large matrix sizes $L$ and after normalization, the distribution of the eigenvalues of these ensembles converges to the Wigner semicircle
\begin{equation}
    \rho(\lambda) = \frac{1}{2 \pi}\sqrt{4 - \lambda^2} \, ,
\end{equation}
which is called the Wigner semicircle law.

\begin{figure}[!t]
	\centering
	\includegraphics[width=.8\linewidth]{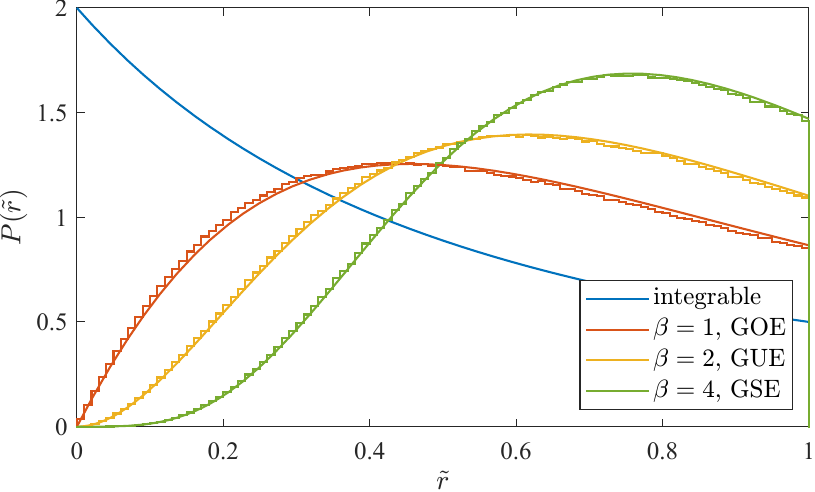}
	\caption{Distributions of the level spacing ratios $\tilde{r}$ for the integrable case and Gaussian beta ensembles. The solid lines are the approximate formulas given by equations \eqref{eq:IntegrableRStatistics} and \eqref{GE_rStatistics}, and the histograms are the numerical distributions, which were obtained by diagonalizing the random tridiagonal matrix given in equation \eqref{Gaussian_ensemble_tridiagonalization}, for $L=1000$ with $10^4$ runs. As can be seen, the analytical formula \eqref{GE_rStatistics} is a very good approximation for the $r$-statistics of Gaussian ensembles.}
	\label{fig:distrib-theory}
\end{figure}

Rather than dealing directly with large $L$ matrices, many aspects of the level spacing statistics are remarkably well captured by random $2 \times 2$ matrices. In this case, the level spacings obey the Wigner--Dyson distribution, which is also called the Wigner surmise:
\begin{equation}
    \rho_\beta(s) = \begin{cases}
        \frac{\pi}{2} s e^{-\frac{\pi}{4} s^2} & \beta = 1 \\
        \frac{32}{\pi^2} s^2 e^{-\frac{4}{\pi} s^2} & \beta =2 \\
        \frac{2^{18}}{3^6 \pi^3} s^4 e^{-\frac{64}{9\pi} s^2} & \beta = 4
    \end{cases}
\end{equation}
A similar formula for the distribution of the spacing ratios can also be derived \cite{Atas:2013gvn} for $3\times3$ matrices
\begin{equation} \label{GE_rStatistics}
    P_\beta (r) \propto \frac{(1+r)^\beta}{(1+r+r^2)^{1+\frac{3}{2}\beta}} \, ,
\end{equation}
which also gives a good approximation for large matrices, as can be seen in figure \ref{fig:distrib-theory}. For any chaotic system, the expectation is that the distribution of the level spacing ratios matches that of a Gaussian beta ensemble, as expressed above.

An alternative way to study Gaussian ensembles is by tridiagonalization. Given a matrix $W_L$ sampled from a Gaussian beta ensemble, there exists a transformation into a tridiagonal form with the same distribution \cite{Dumitriu:2002ntg}:
\begin{equation} \label{Gaussian_ensemble_tridiagonalization}
    W_L = \frac{1}{\sqrt{2}}\begin{pmatrix}
        a_0 & b_1 & 0 & \cdots & 0 \\
        b_1 & a_1 & b_2 & \cdots & 0 \\
        0 & b_2 & a_2 & \ddots & \vdots \\
        \vdots & \vdots & \ddots &  \ddots & b_{L-1} \\
        0 & 0 & \cdots & b_{L-1} & a_{L-1}
    \end{pmatrix} \, ,
\end{equation}
where the diagonal components $a_i$ are sampled from a normal distribution and the off-diagonal components $b_i$ from the chi distribution $\chi_k$ with growing degrees of freedom $k$. Specifically, if the components of this tridiagonal matrix are independently sampled from
\begin{equation}
    a_i \sim \mathcal{N}(0,2) \, , \quad b_i \sim \chi_{i\,\beta} \, ,
\end{equation}
its eigenvalue statistics is equivalent to that of the corresponding Gaussian beta ensemble.

Eigenvalue level repulsion provides a local diagnostic for the onset of chaotic dynamics. It is also helpful to directly probe more global properties of the spectrum. This includes spectral rigidity, in the sense of \cite{Dyson1962I, Dyson1962II, Dyson1962III, DysonMehta1963IV, Berry85}, as well as the spectral form factor associated with the time / temperature dependence of the partition function \cite{Cotler:2016fpe}. We discuss these diagnostics in more detail in section \ref{sec:SpinChaos} and Appendix \ref{app:SpectralRigidity}.

\subsection{Krylov Complexity and the Lanczos Approach} \label{sec:Krylov}
Another quantifier of chaos that has been widely studied in the literature in recent years is the notion of complexity growth.
There have been attempts to define various notions of complexity for quantum processes in the literature \cite{Baiguera:2025dkc, nielsen2005,Dowling:2006tnk}, most of which deal with the action of the time evolution operator $U(t) = e^{-i H t}$.

A common way to quantify the complexity of a given quantum state is to compute how much it is spread over a given basis. Given a state $\ket{\psi}$ and an ordered basis $\mathcal{B} = \{\ket{B_n}: n = 0,1,2,\dots\}$ for the Hilbert space, one can expand $\ket{\psi}$ in this basis as $\ket{\psi} = \sum_n \psi_n \ket{B_n}$, where $\psi_n = \braket{B_n | \psi}$. Then, one can define the complexity of $\ket{\psi}$ as
\begin{equation}
    C_\mathcal{B} = \sum_n n \abs{\psi_n}^2 = \sum_n n \abs{\braket{B_n |\psi}}^2 \, .
\end{equation}

However, this definition of complexity is basis-dependent, and the complexity in any basis $\mathcal{B}$ with $\ket{B_0} = \ket{\psi}$ trivially is zero. Instead, one is usually interested in how much the state $\ket{\psi}$ gets spread over the basis $\mathcal{B}$ with time. After evolving the state $\ket{\psi}$ with time as $\ket{\psi(t)} = U(t) \ket{\psi} = e^{-i H t} \ket{\psi}$ and writing it in the basis $\mathcal{B}$ as $\ket{\psi(t)} = \sum_n \psi_n(t) \ket{B_n}$ with $\psi_n(t) = \braket{B_n | \psi(t)}$, one can define the complexity of the initial state $\ket{\psi}$ as a function of time as
\begin{equation} \label{state_complexity}
    C_{\mathcal{B}}(t) = \sum_n n \abs{\psi_n(t)}^2 = \sum_n n \abs{\bra{B_n} e^{-i H t} \ket{\psi}}^2 \,.
\end{equation}

Alternatively, one can also compute the Shannon entropy of the state $\ket{\psi(t)}$ as
\begin{equation}
    S_{\mathcal{B}}(t) = - \sum_n \abs{\psi_n(t)}^2 \log\left(\abs{\psi_n(t)}^2\right) \, ,
\end{equation}
and define the complexity as the exponential of the Shannon entropy \cite{Balasubramanian:2022tpr}:
\begin{equation} \label{Shannon_complexity}
    C_{\mathcal{B}}^H(t) = e^{S_{\mathcal{B}}(t)} \, .
\end{equation}

In chaotic systems, one expects a sufficiently generic state to explore most of the Hilbert space under time evolution,\footnote{This is not necessarily true for any state. For instance, eigenstates of the Hamiltonian only pick up a complex phase and do not evolve into any other state. This statement can therefore only be made for sufficiently generic states with minimal symmetry.} as there are no symmetries or conserved quantities that restrict an initial state to a region of the Hilbert space. Therefore, one expects those states to spread over any given basis, and the complexity of states in chaotic systems is expected to be high. On the other hand, in integrable systems, the large number of conserved quantities restricts the states to stay in a smaller region of the Hilbert space under time evolution, and the complexity is expected to be smaller.

The state complexity, as defined in \eqref{state_complexity} or \eqref{Shannon_complexity}, is still a quantity that depends on the basis $\mathcal{B}$. However, it was proven in \cite{Balasubramanian:2022tpr} that there is an ordered basis $\mathcal{K}$ that minimizes the complexity in the vicinity of $t=0$, which is called the Krylov basis, and the state complexity $C_{\mathcal{K}}(t)$ in this basis was named ``spread complexity.'' From now on, we will refer to this notion of complexity as ``Krylov state complexity.''

The definition of the Krylov basis follows from the time evolution of an initial state. The state $\ket{\psi}$ evolves with time as
\begin{equation} \label{state_time_evolution_taylor_expansion}
    \ket{\psi(t)} = e^{-i H t} \ket{\psi} = \sum_{n=0}^{\infty} \frac{(-i t)^n}{n!} H^n \ket{\psi} \, .
\end{equation}
Then, the Krylov basis $\mathcal{K} = \{ \ket{K_n}: n=0,1,2,\dots\}$ is defined as the basis obtained from the application of the Gram--Schmidt procedure to the ordered set of states $\{ H^n \ket{\psi} \}$. The first state of this basis is then $\ket{K_0} = \ket{\psi}$.

The iterative application of the Gram--Schmidt procedure to derive the Krylov basis $\mathcal{K}$ is also called the Lanczos algorithm \cite{Lanczos:1950zz}. The Krylov basis is obtained by the recursion relation
\begin{equation}
    \ket{A_{n+1}} = (H - a_n) \ket{K_n} - b_n \ket{K_{n-1}} \, , \quad \ket{K_n} = \frac{1}{b_n} \ket{A_n} \, ,
\end{equation}
where $a_n = \bra{K_n} H \ket{K_n}$ and $b_n = \sqrt{\braket{A_n|A_n}}$ are called the Lanczos coefficients. By definition, one sets $b_0 = 0$, and the first state of the basis is $\ket{K_0} = \ket{\psi}$. The algorithm terminates if any $b_n = 0$.

The Krylov basis is not necessarily complete, as the algorithm can terminate before generating a basis for the full Hilbert space, which means that the initial state $\ket{\psi}$ does not explore the full Hilbert space under time evolution. However, the Krylov basis can be extended to obtain a complete basis $\mathcal{K}_c = \mathcal{K} \cup \{\ket{B_n}: n = \abs{\mathcal{K}}, \abs{\mathcal{K}}+1,\dots\}$, and all possible extensions give the same result for the Krylov state complexity $C_{\mathcal{K}_c}(t)$.

In the Krylov basis, the Hamiltonian is also tridiagonal, as can be seen from the relation
\begin{equation}
    H \ket{K_n} = a_n \ket{K_n} + b_{n+1} \ket{K_{n+1}} + b_n \ket{K_{n-1}} \, ,
\end{equation}
which puts the Hamiltonian in the form
\begin{equation}
    H = \begin{pmatrix}
        a_0 & b_1 & 0 & 0 &\cdots \\
        b_1 & a_1 & b_2 & 0 & \ddots \\
        0 & b_2 & a_2 & b_3 & \ddots \\
        0 & 0 & b_3 & a_3 & \ddots \\
        \vdots & \ddots & \ddots & \ddots & \ddots
    \end{pmatrix} \, .
\end{equation}

\begin{figure}[!t]
	\centering
	\includegraphics[width=.8\linewidth]{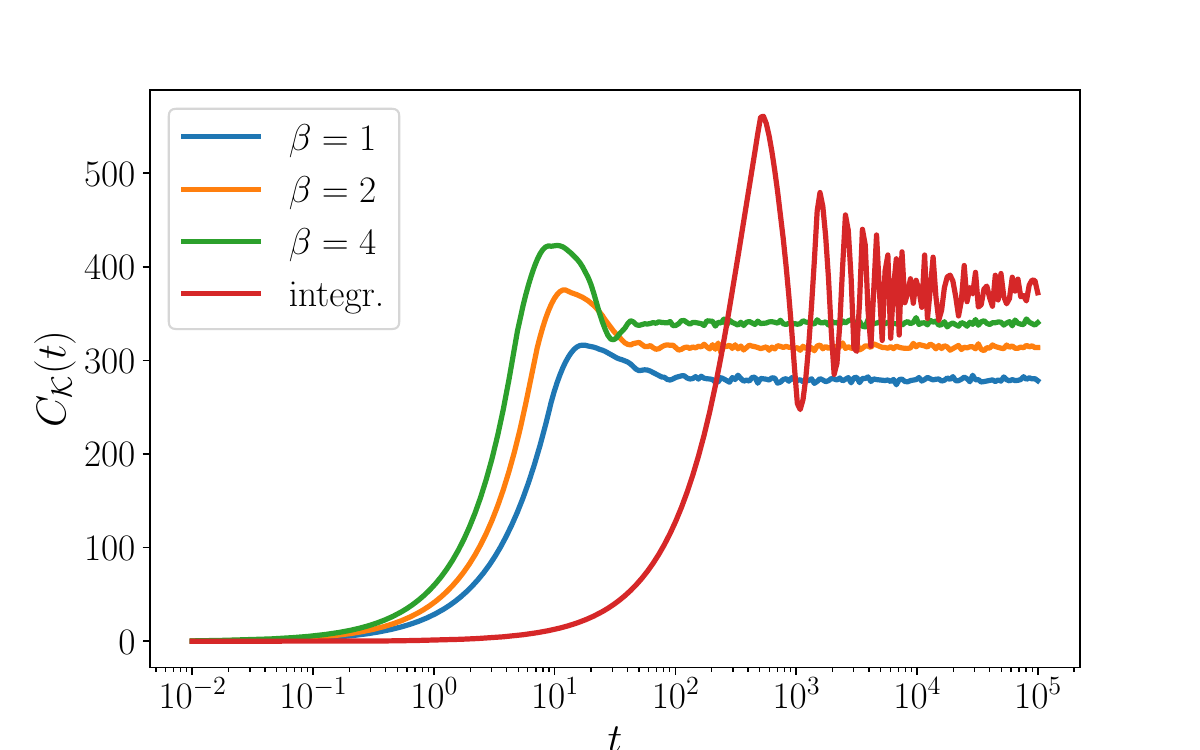}
	\caption{Evolution of Krylov state complexity in integrable and chaotic cases. In all cases, the Hamiltonian is tridiagonal and of length $L=1000$, and the seed state is the one with a single impurity in the middle of the spin chain $|\phi(t=0)\rangle = |500\rangle$.}
	\label{fig:krylov-theoric}
\end{figure}

In figure \ref{fig:krylov-theoric}, we show the typical behavior of Krylov
state complexity for the tridiagonal Hamiltonians associated with an integrable XXX
spin chain\footnote{In the 1-impurity sector, the integrable XXX spin chain Hamiltonian is tridiagonal with $b_i = b$ and $a_i = a$ for all $i$. See section \ref{sec:SpinChaos} for more details on spin chains.} and Gaussian beta ensembles (equation \eqref{Gaussian_ensemble_tridiagonalization}). We can see that in the integrable
case, $C_{\mathcal{K}}(t)$ grows slowly and reaches a long-lived oscillatory
plateau. On the other hand, in the chaotic regime, the rapid growth starts very
early and quickly saturates to an almost-constant plateau.

Instead of the complexity of states, one can also consider the complexity of operators, which quantifies the spread of operators with time in the Heisenberg picture. Given an operator $\mathcal{O}$, it evolves with time in the Heisenberg picture as $\mathcal{O}(t) = U^\dagger(t) \mathcal{O} U(t) = e^{i H t} \mathcal{O} e^{-i H t}$. Its Taylor expansion is
\begin{equation}
    \mathcal{O}(t) = \sum_{n=0}^\infty \frac{(i t)^n}{n!} \mathcal{L}^n\{\mathcal{O}\} \,
\end{equation}
where $\mathcal{L}\{\mathcal{O}\} = [H,\mathcal{O}]$ is the action of the Liouvillian on the operator $\mathcal{O}$, which is given by the commutator of the Hamiltonian $H$ with $\mathcal{O}$. The $n$'th power of $\mathcal{L}$ is understood as $n$ successive applications of $\mathcal{L}$ to $\mathcal{O}$.

Comparison with equation \eqref{state_time_evolution_taylor_expansion} readily shows that this expression can also be understood as the time evolution of a state in the Hilbert space of operators, on which the Hamiltonian, the generator of time evolution, is given by the Liouvillian. Concretely, starting from a Hilbert space $\mathcal{H}$ and operators $\mathcal{O}$ acting on the states in this Hilbert space, one can define a new auxiliary Hilbert space $\mathcal{H}_{\mathcal{O}}$, where a state $|\mathcal{O})$ in this Hilbert space corresponds to the operator $\mathcal{O}$ acting on the states in the original Hilbert space $\mathcal{H}$.\footnote{We denote the states in this auxiliary Hilbert space $\mathcal{H}_\mathcal{O}$ with round brackets $|\cdot)$ to distinguish them from the states in the original Hilbert space $\mathcal{H}$, which are denoted by $\ket{\cdot}$.} Then, the Liouvillian $\mathcal{L}$ is the generator of the time evolution, i.e. the Hamiltonian, acting on the states in the Hilbert space $\mathcal{H}_{\mathcal{O}}$ as $\mathcal{L} | \mathcal{O}) = | [H,\mathcal{O}])$.

Having defined the auxiliary Hilbert space $\mathcal{H}_{\mathcal{O}}$, it is possible to investigate the complexity of states following the same procedure outlined in this section, which would give a measure for the complexity of operators acting on the original Hilbert space. However, to define an orthonormal basis in $\mathcal{H}_{\mathcal{O}}$, it is necessary to specify an inner product, which introduces an ambiguity to the notion of operator complexity. A common choice is to define the inner product of two states $|\mathcal{O}_1)$ and $|\mathcal{O}_2)$ as $(\mathcal{O}_1 | \mathcal{O}_2) = \frac{1}{D} \Tr(\mathcal{O}_1^\dagger \mathcal{O}_2)$, where $D$ is the dimension of the original Hilbert space $\mathcal{H}$.\footnote{It would be interesting to investigate more general choices of inner products / distance measures geared towards gate complexity. See e.g., \cite{Heckman:2026beg} for a recent discussion.}

With this definition of inner product, one can apply the Lanczos algorithm starting with any operator $\mathcal{O}$, and compute the Krylov basis $\mathcal{K} = \{|K_n): n = 0,1,\dots; |K_0) = |\mathcal{O}) \}$ and the Lanczos coefficients $a_n$ and $b_n$ in the Hilbert space $\mathcal{H}_{\mathcal{O}}$. Then, the Krylov operator complexity can be defined in the same manner as the Krylov state complexity as
\begin{equation} \label{Krylov_operator_complexity}
    K_{\mathcal{O}}(t) = \sum_n n \abs{(K_n|\mathcal{O}(t))}^2 = \sum_n n \abs{(K_n| e^{i H t} \mathcal{O} e^{- i H t})}^2 \, ,
\end{equation}
which can be seen to be a special case of state complexity.

It should be noted that if the starting operator $\mathcal{O}_0$ is Hermitian, then the states $(i\mathcal{L})^n|\mathcal{O}_0)$ correspond to Hermitian operators, and as a consequence, the states $|K_n)$ in the Krylov basis all correspond to (anti-)Hermitian operators. Therefore, the Lanczos coefficients $a_n = (K_n| \mathcal{L} | K_n)$ all vanish if the operator $\mathcal{O}_n$ is Hermitian, and in the Krylov basis the Liouvillian takes the tridiagonal form
\begin{equation}
    \mathcal{L} = \begin{pmatrix}
        0 & b_1 & 0 & 0 &\cdots \\
        b_1 & 0 & b_2 & 0 & \ddots \\
        0 & b_2 & 0 & b_3 & \ddots \\
        0 & 0 & b_3 & 0 & \ddots \\
        \vdots & \ddots & \ddots & \ddots & \ddots
    \end{pmatrix} \, .
\end{equation}

The notion of operator complexity defined in \eqref{Krylov_operator_complexity} was named ``Krylov complexity'' or ``K-complexity'' in \cite{Parker:2018yvk}, which is usually what is understood as Krylov complexity in the literature. However, we will make a distinction between ``Krylov state complexity'' $C_{\mathcal{K}}(t)$, which is the state complexity in \eqref{state_complexity} evaluated in the Krylov basis, and the ``Krylov operator complexity'' $K_{\mathcal{O}}(t)$ given in \eqref{Krylov_operator_complexity}, and we will mostly be interested in the behavior of Krylov state complexity for generic initial states in chaotic systems, as we expect the behavior of Krylov state complexity and the Lanczos coefficients for generic states to encode the same information as Krylov operator complexity for Hermitian operators. Indeed, in the situations we consider in this paper, the states of the spin chain will be interpreted as 4D SCFT operators, and in radial quantization these can also be interpreted as states of the 4D SCFT.

\section{Spin Chain Sectors of 4D SCFTs} \label{sec:QUIVER}

In the previous section we discussed some generic diagnostics for determining the onset of chaotic dynamics in a quantum system. Now, quantum field theories are sufficiently complicated that it is rather difficult to extract all of the necessary spectral data to reliably find signatures of chaotic behavior. Our aim in this section will be to identify an isolated subsector where the structure reduces to a far simpler lower-dimensional system.

Indeed, a celebrated way to make progress is the controlled setting of $1+1$-dimensional spin chains, i.e., spins arranged along a spatial one-dimensional lattice. The main idea we shall pursue here is that we can identify subsectors of particular 4D SCFTs in which the dynamics is essentially controlled by the dynamics of this lower-dimensional system. The class of 4D SCFTs we focus on are quiver gauge theory SCFTs, where the arrangement of links between gauge groups provides a natural way to identify a long spin chain ground state, with excitations of the spin chain corresponding to ``similar'' gauge invariant operators. In this setting, the Hamiltonian of the spin chain will be identified with the dilatation operator of the conformal field theory, i.e., extracting the anomalous dimension matrix associated with operator mixing:
\begin{equation}
\langle \mathcal{O}_i^\dagger(x)\, \mathcal{O}_{j}(0)\rangle
\sim \frac{1}{|x|^{2\Delta_0}} \left(\delta_{ij} - \gamma_{ij} \log(|x|^2 \Lambda^2) + \cdots \right),
\label{eq:2PointCorrelator}
\end{equation}
specifies a Hamiltonian for excitations in a spin chain:
\begin{equation}
\widehat{\gamma} = \widehat{H}_{\mathrm{spin \, chain}}.
\end{equation}
Since the anomalous dimension matrix depends on the tunable marginal couplings of the 4D SCFT, there are a correspondingly
large class of possible spin chain Hamiltonians. Our aim will be to identify integrable and chaotic dynamics within
this parametric family, a task we turn to in section \ref{sec:SpinChaos}.

Perhaps the most celebrated case of a spin chain subsector of a 4D QFT comes from $\mathcal{N} = 4$ Super Yang--Mills theory with gauge group $SU(N)$. To illustrate, consider the $1/2$ BPS operator $\mathrm{Tr}(X^{J})$, where $X$ denotes a complex adjoint-valued scalar of the theory. This specifies the ground state of an effective spin chain. Indeed, excitations of the spin chain amount to introducing ``impurities,'' i.e., considering operators where $X$ is replaced by another field. A canonical example is the $SU(2)$ spin chain where we introduce impurities associated with the $SU(2)$ subgroup of the full $SO(6)$ R-symmetry which rotates $X$ and another complex scalar $Y$. Introducing impurities amounts to inserting $Y$'s at different locations instead of the $X$'s: $\mathrm{Tr}(YX^{j_1}....YX^{j_I})$, with $I$ impurities such that $I + j_1 + ... + j_I = J$. The $X$'s and $Y$'s fill out a doublet representation of $SU(2)$, so we can think of $X$ as specifying a $\ket{\downarrow}$ state, and $Y$ as specifying a $\ket{\uparrow}$ state. The ground state is then $\mathrm{Tr}(X^{J}) \sim \ket{\downarrow \cdots \downarrow}$, and impurities amount to spin-up states which propagate due to a local interaction Hamiltonian. Working to one loop order in the gauge coupling, the resulting anomalous dimension matrix is controlled by the spin chain Hamiltonian:
\begin{equation}\label{eq:SpinHop}
    \widehat{\gamma} = \widehat{H}_{\mathrm{spin}} = \sum_{j=1}^{J-1} H_j \qquad \qquad H_j = -\kappa_j \left( \Vec{S}_j \cdot \Vec{S}_{j+1} - \frac{1}{4} \right) \,,
\end{equation}
where in the case of planar $\mathcal{N} = 4 $ SYM the $\kappa_{i}$'s are all equal. Higher-loop contributions simply correspond to adding longer range interactions to the spin chain.

More generally, we can start with the ground state and consider excitations of the entire $\mathfrak{psu}(2,2 \vert 4) \simeq \mathfrak{sl}(4\vert 4)$ superconformal algebra. For example, instead of inserting the complex scalar $Y$, we could consider any other member of the same superconformal multiplet for a complex scalar, e.g., a fermionic insertion, or $D_{\mu} X$, a covariant derivative of $X$. All of this is controlled by the same sort of operator mixing structure.  In the planar limit, this provides a controlled example of integrability which has been extensively studied. See e.g., \cite{Berenstein:2002jq, Minahan:2002ve, Beisert:2002bb, Beisert:2003jj, Beisert:2003tq, Beisert:2003yb, Beisert:2004hm, Beisert:2004ry, Beisert:2005he, Beisert:2005if, Gadde:2010zi, Pomoni:2013poa, Pomoni:2019oib} for a partial list of references. It is worth noting that while the detailed properties of operator mixing clearly depend on the specific type of impurity being inserted, the leading order hopping term behavior controlled by the Hamiltonian of equation $(\ref{eq:SpinHop})$ is universal. As such, in the dilute gas approximation where impurities rarely scatter, the leading order behavior of the energy spectrum is detected by the spectrum of spin waves propagating independently on the spin chain.

Deformations of this integrable structure also exhibit chaotic behavior. This was specifically studied for certain $\mathcal{N} = 4 \rightarrow \mathcal{N} = 1$ marginal deformations in reference \cite{McLoughlin:2020zew}. However, the number of such deformations is rather limited \cite{Leigh:1995ep}.

\begin{figure}[t!]
\begin{center}
	$$\begin{matrix}\begin{tikzpicture}[
			node/.style={circle, thick, draw=black!100,fill=white!100,  minimum size=2mm, inner sep=0pt}
	]
	\node[node, fill=black, label=above:{\footnotesize $SU(N_1)$}] (A1)  {};
    
	\node[] (A2) [right=6mm of A1] {$\cdots$};
	\node[node, fill=black, label=above:{\footnotesize $SU(N_i)$}] (A3) [right=6mm of A2] {};
	\node[] (A4) [right=6mm of A3] {$\cdots$};
	\node[node, fill=black, label=above:{\footnotesize $SU(N_L)$}] (A5) [right=6mm of A4] {};
    
    \node[rectangle, fill=black, label=above:{\footnotesize $SU(F_L)\phantom{M}$}] (F1) [left=10mm of A1] {};
	\node[rectangle, fill=black, label=above:{\footnotesize $\phantom{M}SU(F_R)$}] (F5) [right=10mm of A5] {};
    
    \draw (A1.east) -- (A2.west);
	\draw (A2.east) -- (A3.west);
    \draw (A3.east) -- (A4.west);
    \draw (A4.east) -- (A5.west);
    
    \draw (A1.west) -- (F1.east);
    \draw (A5.east) -- (F5.west);
    
	\end{tikzpicture}\end{matrix}
	\qquad
	\begin{matrix}\begin{tikzpicture}[
    node/.style={
        circle,
        thick,
        draw=black,
        fill=black,
        minimum size=2mm,
        inner sep=0pt
    }
]
    \def\n{6}
    \def\r{1.1}
    \def\rlab{1.6}

    \foreach \j in {1,...,\n} {
        \pgfmathsetmacro{\ang}{90 - 360/\n*(\j-1)}
        \coordinate (P\j) at (\ang:\r);
        \coordinate (L\j) at (\ang:\rlab);
    }

    \draw (90:\r) arc[start angle=90, end angle=30, radius=\r];
    \draw (30:\r) arc[start angle=30, end angle=-30, radius=\r];
    \draw (-30:\r) arc[start angle=-30, end angle=-90, radius=\r];
    \draw (-90:\r) arc[start angle=-90, end angle=-150, radius=\r];
    \draw (-150:\r) arc[start angle=-150, end angle=-210, radius=\r];
    \draw (-210:\r) arc[start angle=-210, end angle=-270, radius=\r];

    \node[node] at (P1) {};
    \node[node] at (P2) {};
    \node[node] at (P4) {};
    \node[node] at (P6) {};

    \node[fill=white, inner sep=1pt, font=\footnotesize] at (P3) {$\cdots$};
    \node[fill=white, inner sep=1pt, font=\footnotesize] at (P5) {$\cdots$};

    \node[font=\footnotesize] at (L1) {$SU(N_1)$};
    \node[font=\footnotesize] at (L2) {$\phantom{MM}SU(N_2)$};
    \node[font=\footnotesize] at (L4) {$SU(N_i)$};
    \node[font=\footnotesize] at (L6) {$SU(N_L)\phantom{MM}$};
\end{tikzpicture}\end{matrix}$$
\end{center}
\caption{Linear or closed 4D $\mathcal{N}=2$ quivers leading to a sector described by a spin chain. Each edge denotes a $\mathcal{N}=2$ hypermultiplet, with square nodes depicting a flavor symmetry. The value of $F_L, F_R$ is set by demanding cancellation of $\beta$-functions.}
		\label{fig:quivers}
\end{figure}
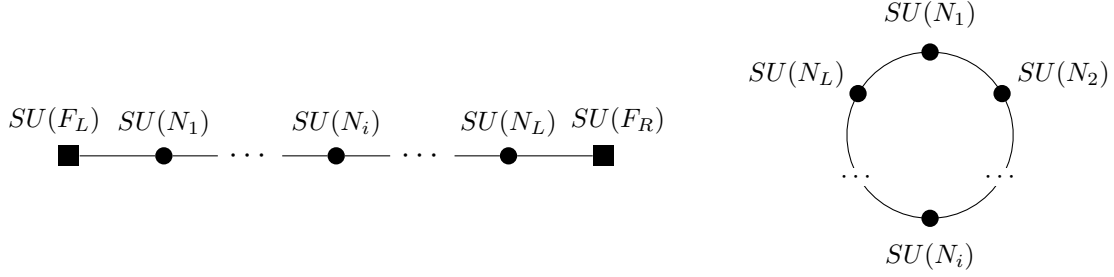

With this in mind, we now seek out a similar class of spin chain subsectors in a broader class of SCFTs. Let us illustrate the general idea with a representative example. Consider a 4D $\mathcal{N} = 2$ quiver gauge theory with all gauge groups\footnote{Since we are interested in local operator statements, we shall not concern ourselves with the global form of the gauge group as opposed to the gauge algebra.} $SU(N)$, which we index as $G_{i} = SU(N_i)$ for $i = 1,...,L$, with $i = L+1$ identified with $i = 1$, in the obvious notation. Each such gauge group comes with a vector multiplet. We also introduce a collection of bifundamental hypermultiplets which connect $SU(N_i)$ with $SU(N_{i+1})$. In terms of $\mathcal{N} = 1$ multiplets, we have a bifundamental $X_{i}$ in the $(N_{i}, \overline{N}_{i+1})$ and a bifundamental $Y_{i}$ in the $(\overline{N}_{i}, N_{i+1})$. As found in \cite{Douglas:1996sw}, this system naturally arises from $N$ D3-branes probing the transverse geometry $\mathbb{C} \times \mathbb{C}^{2} / \mathbb{Z}_{L}^{(1,-1)}$. Related quiver gauge theories which do not form a closed loop also naturally arise from the dimensional reduction of 6D $\mathcal{N} = (1,0)$ SCFTs.\footnote{See e.g., \cite{Heckman:2013pva, DelZotto:2014hpa, Heckman:2015bfa, Baume:2020ure, Baume:2022cot} and \cite{Heckman:2018jxk, Argyres:2022mnu} for recent reviews.} In this case, conformal invariance requires that we have flavor symmetries on the left and right of the linear quiver. See figure \ref{fig:quivers} for a depiction of these cases. The important point for us is that this quiver gauge theory has a much larger number of marginal couplings:
\begin{equation}
\tau_{j} = \frac{4 \pi i }{g_{j}^{2}} + \frac{\theta_{j}}{2 \pi},
\end{equation}
in the obvious notation. In the case of $N$ D3-branes probing $\mathbb{C}^{2} / \mathbb{Z}_L$, the IIB axio-dilaton is related to these parameters via:
\begin{equation}
\underset{j}{\sum} \tau_{j} = \tau_{\mathrm{IIB}}.
\end{equation}

The marginal couplings directly determine the strength of nearest neighbor interactions in an effective spin chain system, a fact we now explain.
First of all, we note that in this 4D $\mathcal{N} = 2$ gauge theory, we can indeed form a spin chain ground state by taking the product of all the $X_{i}$. For a closed loop in the quiver, this is the operator $\mathrm{Tr}(X_1 ... X_L)$, and in the linear quiver it is the operator $X_1 ... X_L$. Excitations above the ground state proceed much as in the $\mathcal{N} = 4$ case; we can swap out the $X_{i}$ for $Y_{i}^{\dag}$, as well as inserting covariant derivative insertions and / or adjoint valued chiral superfields $Z_{i}$ associated with each gauge group factor. The local operator mixing is again controlled by a spin chain Hamiltonian \cite{Baume:2020ure}, and this extends to impurity excitations of the corresponding $\mathfrak{su}(2,2\vert 2)$ superconformal algebra.

Marginal deformations amount to altering the strength of nearest neighbor interactions of the one-loop approximation to the spin chain Hamiltonian. Indeed, working in terms of canonically normalized superfields, observe that in $\mathcal{N} = 1$ terms, the superpotential is fixed by $\mathcal N=2$ supersymmetry to have the form:
\begin{equation}
W_{\mathcal N=2}=\sqrt{2}\sum_{i=1}^{L}g_i\Tr\Big(\Phi_i Y_i X_i-\Phi_i X_{i-1}Y_{i-1}\Big),
\label{eq:N2Wmain}
\end{equation}
where the $g_{i}$ denote the marginal gauge couplings of the theory. The important point is that once we move away from the special value $g_{i} = g$ inherited from the $\mathcal{N} = 4$ theory, the corresponding strength of nearest neighbor interactions in the spin chain Hamiltonian of line (\ref{eq:SpinHop}) will also be non-uniform: the $\kappa_{i}$ are generically distinct, and we do not have integrable dynamics. Rather, one should expect more general operator mixing to emerge, albeit in a controlled fashion.

The considerations presented here clearly extend to a far broader class of 4D $\mathcal{N} = 1$ superconformal quiver gauge theories. For example, we can construct many similar systems by considering $N$ D3-branes probing supersymmetric orbifolds $\mathbb{C}^{3} / \Gamma$ with $\Gamma$ of sufficiently high order. All that we require is that the spin chain ground state $\mathrm{Tr}(X^{L})$ of the parent $\mathcal{N} = 4$ theory is invariant under this orbifold group action, and at the very least we can always take $L = \vert \Gamma \vert$. Excitations of the parent theory then descend via orbifold inheritance to the orbifold theory. The generic form of excitations assemble in representations of $\mathfrak{su}(2,2\vert 1)$, the 4D $\mathcal{N} = 1$ superconformal algebra. Generically, this involves covariant derivatives and gauginos of the vector multiplet acting on the $X$ bifundamentals, as well as swapping out the $X$ for their fermionic superpartners. In all of these cases, the leading order hopping terms of the spin chain Hamiltonian are the same, as dictated by orbifold inheritance from the $\mathcal{N} = 4$ SYM theory.

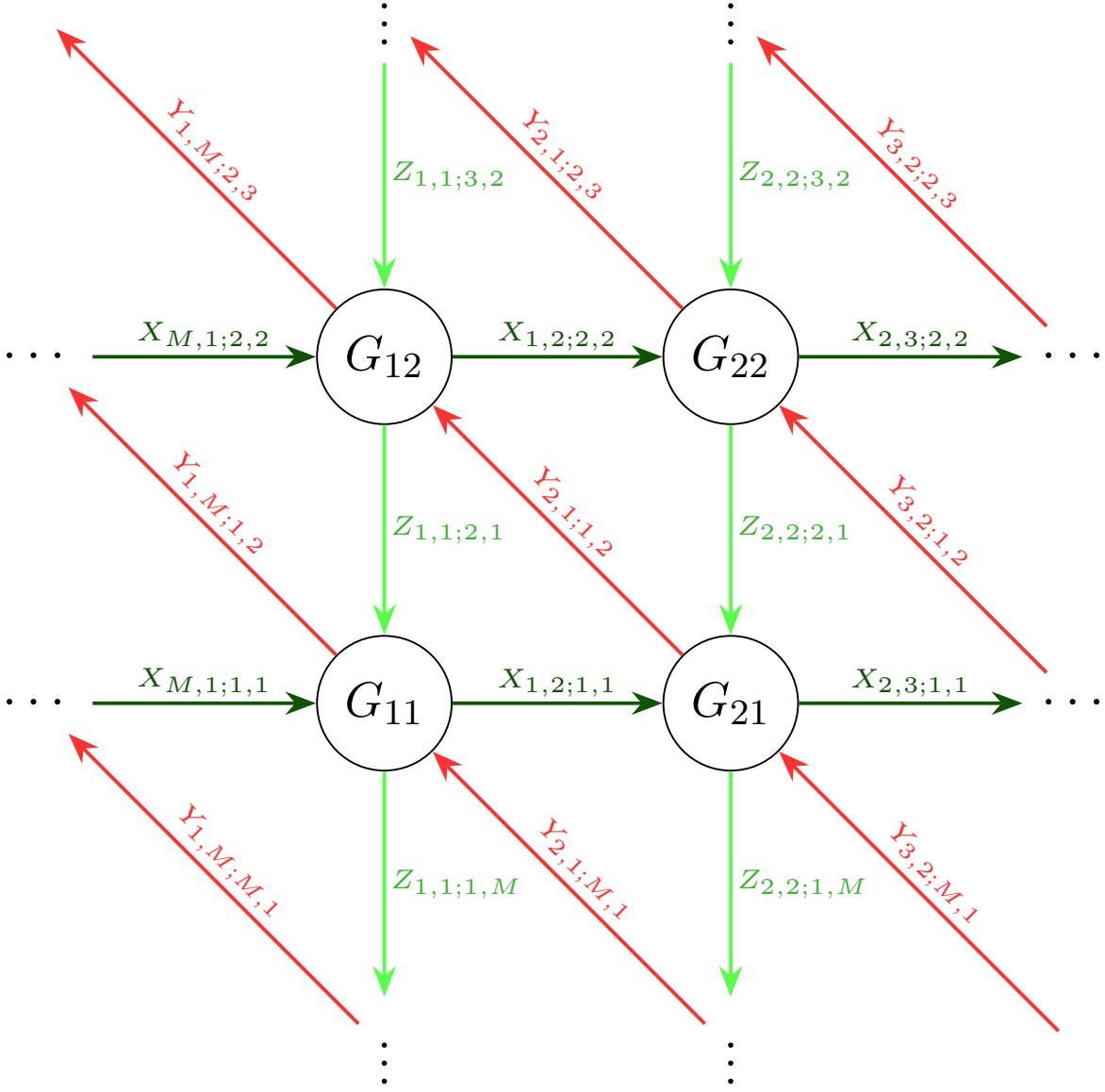
\begin{figure}[t!]
    \centering
    \scalebox{1.8}{
\begin{tikzpicture}[
    >={Stealth[length=2.4mm]},
    x=2.7cm, y=2.7cm,
    Gnode/.style={circle,draw,fill=white,minimum size=1.05cm,inner sep=1pt,font=\small},
    Xlink/.style={->,thick,Xcol},
    Zlink/.style={->,thick,Zcol},
    Ylink/.style={->,thick,Ycol},
    elab/.style={font=\tiny,inner sep=1pt},
  ]

  \foreach \j in {1,2} {
    \node (G0\j) at (0,\j) {$\cdots$};
    \foreach \i in {1,2} {
      \node[Gnode] (G\i\j) at (\i,\j) {$G_{\i\j}$};
    }
    \node (G3\j) at (3,\j) {$\cdots$};
  }
  \foreach \i in {1,2} {
    \node (G\i0) at (\i,0) {$\vdots$};
    \node (G\i3) at (\i,3) {$\vdots$};
  }

  \node (G03) at (0,3) {};
  \node (G30) at (3,0) {};

  \foreach \j in {1,2} {
    \draw[Xlink] (G0\j) -- (G1\j) node[elab,Xcol,pos=0.5,above]{$X_{M,1;\j,\j}$};
    \draw[Xlink] (G1\j) -- (G2\j) node[elab,Xcol,pos=0.5,above]{$X_{1,2;\j,\j}$};
    \draw[Xlink] (G2\j) -- (G3\j) node[elab,Xcol,pos=0.5,above]{$X_{2,3;\j,\j}$};
  }

  \foreach \i in {1,2} {
    \draw[Zlink] (G\i3) -- (G\i2) node[elab,Zcol!70!black,pos=0.5,right]{$Z_{\i,\i;3,2}$};
    \draw[Zlink] (G\i2) -- (G\i1) node[elab,Zcol!70!black,pos=0.5,right]{$Z_{\i,\i;2,1}$};
    \draw[Zlink] (G\i1) -- (G\i0) node[elab,Zcol!70!black,pos=0.5,right]{$Z_{\i,\i;1,M}$};
  }

  \draw[Ylink] (G21) -- (G12)
     node[elab,Ycol,pos=0.5,sloped,above,inner sep=1.5pt]{$Y_{2,1;1,2}$};
  \draw[Ylink] (G31) -- (G22)
     node[elab,Ycol,pos=0.5,sloped,above,inner sep=1.5pt]{$Y_{3,2;1,2}$};
  \draw[Ylink] (G32) -- (G23)
     node[elab,Ycol,pos=0.5,sloped,above,inner sep=1.5pt]{$Y_{3,2;2,3}$};
  \draw[Ylink] (G22) -- (G13)
     node[elab,Ycol,pos=0.5,sloped,above,inner sep=1.5pt]{$Y_{2,1;2,3}$};
  \draw[Ylink] (G11) -- (G02)
     node[elab,Ycol,pos=0.5,sloped,above,inner sep=1.5pt]{$Y_{1,M;1,2}$};
  \draw[Ylink] (G12) -- (G03)
     node[elab,Ycol,pos=0.5,sloped,above,inner sep=1.5pt]{$Y_{1,M;2,3}$};
  \draw[Ylink] (G20) -- (G11)
     node[elab,Ycol,pos=0.5,sloped,above,inner sep=1.5pt]{$Y_{2,1;M,1}$};
  \draw[Ylink] (G30) -- (G21)
     node[elab,Ycol,pos=0.5,sloped,above,inner sep=1.5pt]{$Y_{3,2;M,1}$};
  \draw[Ylink] (G10) -- (G01)
     node[elab,Ycol,pos=0.5,sloped,above,inner sep=1.5pt]{$Y_{1,M;M,1}$};
\end{tikzpicture}}
    \caption{Quiver for $\mathbb{C}^3/\mathbb{Z}_M \times \mathbb{Z}_M$. Nodes $G_{ij}$ form a periodic square grid, and horizontal, vertical, and diagonal links are the fields $X_{i,i+1;j,j}$ (dark green), $Z_{i,i;j,j-1}$ (light green), and $Y_{i+1,i;j+1,j}$ (red) respectively. Each row of horizontal $X$ links closes into a loop through $M$ gauge groups, giving the spin-chain ground states $\mathrm{Tr}(X^M) \rightarrow \sum_j \mathrm{Tr}\left(\prod_i X_{i,i+1;j,j} \right)$. See \cite{DewPalette} to implement this color palette choice.}
    \label{fig:ZMZMQuiver}
\end{figure}

There is an additional subtlety in the $\mathcal{N}=1$ case since the number of marginal deformations of the SCFT is typically far smaller than the number of gauge groups. Indeed, for $N$ D3-branes probing a sufficiently generic\footnote{i.e., no additional isometries beyond the $\mathbb{C}^{\ast}$ actions.} toric singularity $X$, the number of marginal deformations is $d_{\mathrm{ext}} - 1$, where $d_{\mathrm{ext}}$ is the number of external legs in the toric diagram.\footnote{For a systematic discussion of the orbifold case see reference \cite{Razamat:2002tm}, and \cite{Imamura:2007dc} for the more general toric case. This counting follows from the procedure of \cite{Leigh:1995ep, Green:2010da}.} This number is smaller than the total number of gauge groups in a quiver. That being said, there are clearly many examples which still accommodate a suitable spin chain sector. For example, the orbifold $\mathbb{C}^3 / \mathbb{Z}_M \times \mathbb{Z}_M$ with weights $(1,-1,0)$ and $(0,1,-1)$ has a total of $\vert \mathbb{Z}_M \times \mathbb{Z}_M \vert = M^2$ gauge group factors, but only $3M - 1$ marginal deformations. In the resulting quiver, we can label the gauge group factors as $G_{ij}$, arranged along a square grid, with $i$ a horizontal coordinate and $j$ a vertical coordinate. In this terminology, the descendants of $X$ define horizontal links $X_{i,i+1;j,j}$, the descendants of $Z$ define vertical links $Z_{i,i;j,j-1}$, and the $Y$ define diagonal links $Y_{i+1,i;j,j+1}$ (see figure \ref{fig:ZMZMQuiver}). By inspection, a ground state such as $\mathrm{Tr}(X^M)$ of the parent $\mathcal{N} = 4$ SYM theory descends to the protected operators:
\begin{equation}
\mathrm{Tr}(X^M) \rightarrow \sum_j \mathrm{Tr}\left(  \prod_i X_{i,i+1;j,j} \right), 
\end{equation}
in the obvious notation. Each one of these summands defines an independent spin chain ground state in the orbifold theory. Observe that for a fixed $j$, the operator $\mathrm{Tr}\left( \prod_i  X_{i,i+1;j,j} \right)$ is constructed from bifundamentals which pass through only $M$ gauge groups. 
As such, we can tune the nearest neighbor hopping terms in this case to produce any desired spin chain Hamiltonian. At the other extreme, an orbifold such as $\mathbb{C}^3 / \mathbb{Z}_K$ with $K$ prime results in a two-dimensional conformal manifold \cite{Razamat:2002tm}. For more general $K$, and appropriately chosen orbifold weights we can increase the number of marginal couplings. Similar considerations hold for related quivers such as those with discrete torsion.

One can also consider non-orbifold examples with D3-branes probing canonical singularities of a Calabi--Yau threefold with a large number of collapsed divisors. Likewise, we can construct many 4D $\mathcal{N} = 1$ SCFTs from compactification of 6D SCFTs on a (punctured) Riemann surface, where the directions of the quiver lift geometrically to that of the internal geometry. While we typically cannot appeal to orbifold inheritance in such situations, we broadly expect the same sort of spin chain structure to emerge in these cases as well. In what follows we shall assume that we have a sufficient number of marginal parameters to tune the interactions of the spin chain.

In all of these cases, there will be quiver specific scattering / interaction terms amongst the spin chain excitations, but these contributions to the energy spectrum are all suppressed in the dilute gas / low impurity regime. As such, the basic form of local hopping for a quiver gauge theory with tuned gauge couplings dictates the structure of operator mixing.

\subsection{Dilute Gas Approximation}

In the dilute gas regime, the number of impurities $I \ll L$, the length of the spin chain. To leading order, then, the operator mixing
problem amounts to a nearest neighbor hopping problem. As such, we can work in any subsector to extract the generic dependence for operator
mixing. It thus suffices to consider the local hopping term for a single impurity. Since we have already argued that the structure of hopping is essentially generic across all such quivers, it suffices to consider local impurity hopping in the $SU(2)_R$ subsector of the 4D $\mathcal{N} = 2$ quiver gauge theory obtained from $N$ D3-branes probing $\mathbb{C} \times \mathbb{C}^{2} / \mathbb{Z}_{L}$.

This hopping term analysis was carried out in \cite{Baume:2020ure}. The derivation for this sector as well as the derivative impurity / $SL(2)$ sector is reviewed in Appendix \ref{app:SpinChainComputation}; here we summarize the main points. Consider, then, a basis of one-impurity operators $|i\rangle$, i.e., we insert the impurity at the lattice site $i$. The one-loop dilatation operator acts as
\begin{equation}\label{su2-hopping-hamiltonian}
		\gamma |i\rangle = H_{i,i}|i\rangle + H_{i,i-1}|i-1\rangle + H_{i,i+1}|i+1\rangle,
\end{equation}
so that the anomalous-dimension matrix is tridiagonal. In particular, the $\mathcal N=2$ anomalous-dimension matrix takes the constrained form
\begin{equation}
\gamma_{ij}=\frac{1}{8\pi^2}\left(- \widetilde{C}_{i} g_i^2 \delta_{i,j-1}+ (\widetilde{C}_{i} g_i^2+ \widetilde{C}_{i-1} g_{i-1}^2)\delta_{ij}-\widetilde{C}_{i-1} g_{i-1}^2 \delta_{i,j+1}\right),
\label{eq:N2gammaMainText}
\end{equation}
where the $\widetilde{C}_{j} = (N_j^2 - 1) / 2N_j $ are group theory dependent factors, which in the case of the $\mathcal{N} = 2 $ quiver are all equal. We remark that in quiver gauge theories where the ranks are not all equal (as happens in non-abelian orbifolds and more general Calabi--Yau singularities) the same general structure persists.

By inspection, we therefore see that the Hamiltonian forms a tridiagonal matrix in the single impurity basis:
\begin{equation} \label{eq:AnomalousDimensionMatrix}
\widehat{H}_{\mathrm{spin}} = \gamma_{\mathrm{eff}} \times
\begin{pmatrix}
b_{L} + b_1 & -b_1 & 0 & \cdots & -b_L \\
-b_1 & b_1 + b_2 & -b_2 & \cdots & 0\\
0 & -b_2 & b_2 + b_3 & \ddots & \vdots\\
\vdots & \vdots & \ddots & \ddots & -b_{L-1} \\
-b_L & 0 & \cdots & -b_{L-1} & b_{L-1} + b_{L}
\end{pmatrix}.
\end{equation}
Here, we have factored out an overall $\gamma_{\mathrm{eff}}$ associated with the ``typical'' size of the anomalous dimensions in the operator mixing problem: we will suppress this in our discussion of the pure spin chain energy spectrum problem, but of course it is important to
reintroduce it in the 4D SCFT analysis as well as the match to a candidate gravity dual.
We have presented the form of operator mixing for a circular quiver, i.e., a spin chain on a closed loop.
The case with open boundary conditions follows as a special case by setting $g_{L}^{2} = 0$, i.e., switching off $b_{L}$.
This difference is highly localized, and is a subleading contribution to the eigenvalues in the large $L$ limit. The important
point for us is that all of the $b_{j}$'s are tunable parameters, corresponding to marginal couplings of the SCFT.

Summarizing, we have found a tridiagonal matrix structure which is suggestively similar to what one expects in the Krylov basis.
That being said, we caution that the impurity basis and Krylov basis are generically distinct. We now turn to an explicit analysis of chaotic dynamics for operator mixing.

\section{Chaotic Spin Chains} \label{sec:SpinChaos}

In the previous section we showed that a broad class of 4D SCFT quiver gauge theories admit a spin chain subsector, with excitations corresponding to directions in the superconformal algebra $\mathfrak{su}(2,2\vert \mathcal{N})$ for $\mathcal{N} = 1,2$. These gauge theories have marginal couplings which we can use to tune the strength of nearest neighbor interactions, as generated by one-loop contributions to the anomalous dimension matrix. Our aim in this section will be to establish different qualitative properties of the energy spectrum of this class of spin chains as we move through the space of marginal couplings of the 4D SCFT.

In general, this is a challenging task, so to simplify our analysis we shall assume from the outset that the number of impurity excitations $I$ above the ground state is small, i.e., we always assume $I \ll L$, with $L$ the length of the spin chain. In this approximation, impurities will still scatter off one another, but such effects will affect the energy spectrum at subleading order. For this reason, it will suffice to study the energy spectrum in the single impurity approximation. In this case, the single impurity spin chain Hamiltonian assumes the form advertised in section \ref{sec:QUIVER}:
\begin{equation}
\widehat{H}_{\mathrm{spin}} =
\begin{pmatrix}
b_{L} + b_1 & -b_1 & 0 & \cdots & -b_L \\
-b_1 & b_1 + b_2 & -b_2 & \cdots & 0\\
0 & -b_2 & b_2 + b_3 & \ddots & \vdots\\
\vdots & \vdots & \ddots & \ddots & -b_{L-1} \\
-b_L & 0 & \cdots & -b_{L-1} & b_{L-1} + b_{L}
\end{pmatrix},
\label{eq:VariableCouplingHamiltonian}
\end{equation}
where in comparing with equation (\ref{eq:AnomalousDimensionMatrix}) to focus the analysis on the essential points
we have dropped $\gamma_{\mathrm{eff}}$. We restore this contribution when we return to the
4D SCFT / holographic interpretation in section \ref{sec:HOLO}. The integrable point corresponds
to taking all $b_{j}$ equal for a closed spin chain, and to taking $b_L = 0$ and all other $b_j$ equal for an open spin chain.
These two special choices are also the ones one naturally gets from the orbifold procedure;
deformations of the marginal couplings move us to more generic non-integrable values.

By inspection, this is superficially similar to the structure of the matrices encountered in equation (\ref{eq:DumitriuEnsemble}), since we can effectively tune / randomly draw elements from a preferred distribution for the off-diagonal elements $b_{j}$. Note, however, that in contrast to equation (\ref{eq:DumitriuEnsemble}), the diagonal elements are not independent. As a consequence, we will identify a different ensemble of such tridiagonal matrices which exhibits chaotic level repulsion. Our general plan will be to produce a tuned set of $b_{j}$'s such that the eigenvalue spectrum exhibits the universal energy level repulsion expected in chaotic dynamics. In general, we find that the scaling of the $b_j$ must be fairly strong as a function of $j$, i.e., it grows as a power law $b_j \sim j^{p}$ for $p > 1$, and in practice we find that a numerical match to a Gaussian ensemble typically requires $ p > 10$. This is to be compared with reference \cite{Parker:2018yvk} where a universal operator growth hypothesis was proposed. The general expectation presented there is that there is a bound on the off-diagonal entries of a many-body Hamiltonian written in the tridiagonal Lanczos basis, i.e., $b^{\mathrm{Lanczos}}_{j} \sim O(j)$, i.e., at most a first order power law. The two situations are compatible, since in \cite{Parker:2018yvk} one makes the (natural) assumption that nearest neighbor interactions of the Hamiltonian are all $O(1)$, whereas we have ``by hand'' tuned the couplings to increase as we move through the spin chain.\footnote{It is also worth noting that even in the celebrated ensemble of line (\ref{eq:DumitriuEnsemble}), the typical off-diagonal entries grow to be quite large.} Additionally, it is worth emphasizing that the tridiagonal structure of our single impurity / dilute gas approximation for the spin chain Hamiltonian is not directly the many-body Hamiltonian in the Lanczos basis.

Our plan in the remainder of this section will be to analyze possible chaotic dynamics by directly analyzing the energy level spectrum, as well as more computationally convenient measures such as Krylov complexity. We determine choices of the $b_{j}$'s which result in an a chaotic energy level spectrum. Additionally, we find that Krylov complexity diagnostics can produce false positives and false negatives for chaotic behavior.

\subsection{Spectral Diagnostics}

\begin{figure}[htpb]
    \centering

    \begin{subfigure}{0.45\textwidth}
        \centering
        \includegraphics[width=\linewidth]{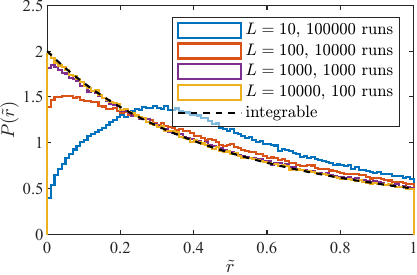}
        \caption{$b_i \sim \chi_1$}
    \end{subfigure}
    \hfill
    \begin{subfigure}{0.45\textwidth}
        \centering
        \includegraphics[width=\linewidth]{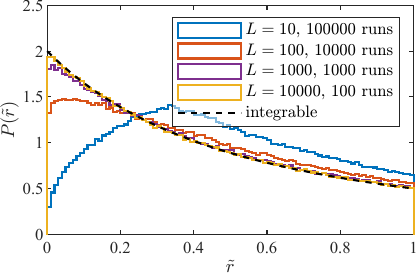}
        \caption{$b_i \sim \text{Uniform}(0,1)$}
    \end{subfigure}
    \hfill

    \vspace{0.5cm}
    \begin{subfigure}{0.45\textwidth}
        \centering
        \includegraphics[width=\linewidth]{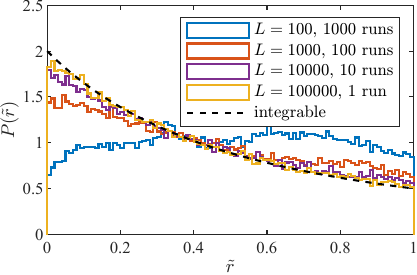}
        \caption{$b_i \sim \chi_5$}
    \end{subfigure}
    \hfill
    \begin{subfigure}{0.45\textwidth}
        \centering
        \includegraphics[width=\linewidth]{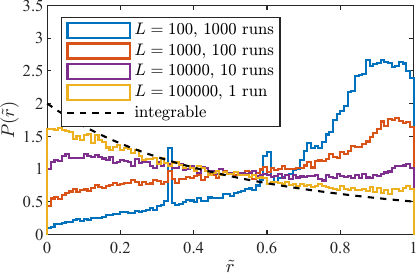}
        \caption{$b_i \sim \text{Uniform}(4,5)$}
    \end{subfigure}
    \hfill

    \vspace{0.5cm}
    \begin{subfigure}{0.45\textwidth}
        \centering
        \includegraphics[width=\linewidth]{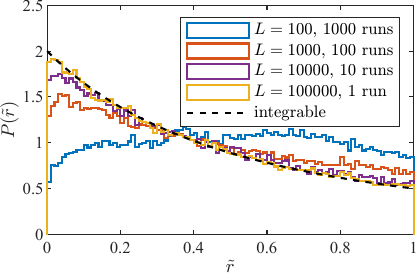}
        \caption{$b_i \sim \chi_{1+10x}$}
    \end{subfigure}
    \hfill
    \begin{subfigure}{0.45\textwidth}
        \centering
        \includegraphics[width=\linewidth]{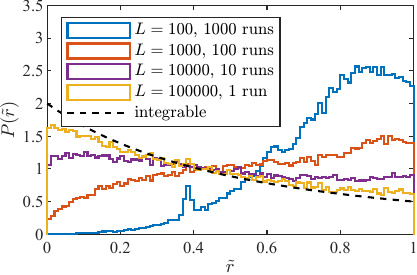}
        \caption{$b_i \sim \text{Uniform}\left(10x,1+10x\right)$}
    \end{subfigure}

	\caption{Distribution of the level spacing ratios $\tilde{r}$ for the tridiagonal Hamiltonian \eqref{eq:VariableCouplingHamiltonian} when the off-diagonal elements $b_i$ are independently and identically drawn from the same distribution, or if the distribution varies slowly in $x=i/L$. In the large spin chain limit $L \rightarrow \infty$, the distribution for $r$ converges to the integrable case \eqref{eq:IntegrableRStatistics} independent of the probability distribution for the couplings $b_i$, exhibiting Anderson localization, but the convergence rate depends on the chosen distribution.}
    \label{fig:rStatisticsWithIIDCouplings}
\end{figure}

As a warmup, it is natural to ask what happens if all couplings $b_i$ are independently drawn from the same distribution, and whether the resulting level statistics for a typical draw depend on the chosen distribution. As can be seen in figure \ref{fig:rStatisticsWithIIDCouplings}, independent of the distribution from which the couplings $b_i$ are drawn, the $r$-statistics for a typical draw always converges to the integrable statistics in the large chain limit, $L \rightarrow \infty$. However, the convergence rate depends on the chosen probability distribution, in particular, on the ratio of the mean to the standard deviation. Sharply peaked distributions for $b_i$ lead to slower convergence for the $r$-statistics. The same phenomena is also observed if the distribution for $b_i$ depends on $i$, but varies slowly with $i$ in the large $L$ limit, as can be seen from the last two subfigures in figure \ref{fig:rStatisticsWithIIDCouplings}. Concretely, defining $x = i / L$, if there exists a large $L$ limit for the distribution $\rho_{b_{xL}}$ continuous in $x$, the $r$-statistics converges to the integrable statistics. This can be understood as an example of emergent integrability \cite{Serbyn:2013kka,Huse:2014tqa,Imbrie:2016fbg,Imbrie:2016eae} in Anderson or many-body localized systems first proposed by Anderson \cite{Anderson:1958vr}.\footnote{See e.g. \cite{Nandkishore:2014kca,Abanin_2019} for reviews.} Examples of many-body localization have been observed in disordered spin chains \cite{Oganesyan:2007wpd,Znidaric:2008vkt,Pal_2010}, and was proven in \cite{Imbrie:2016fbg,Imbrie:2016eae}.

\begin{figure}[!t]
	\centering
	\includegraphics[width=.8\linewidth]{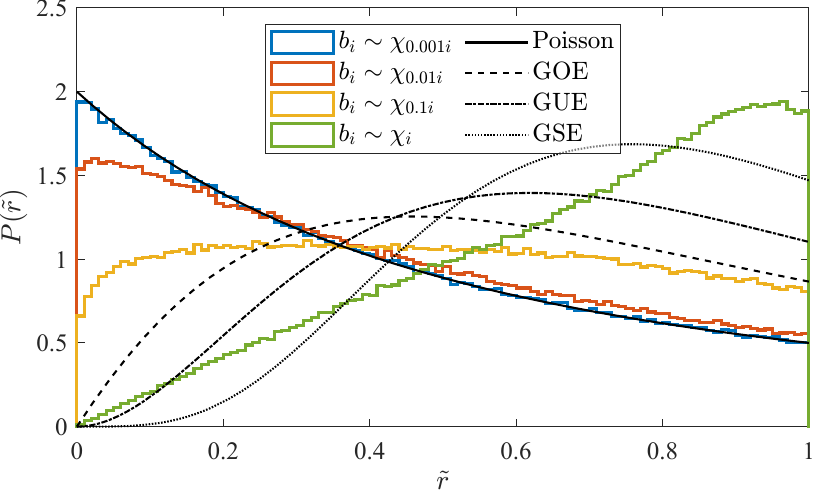}
	\caption{Distribution of the level spacing ratios $\tilde{r}$ for the tridiagonal Hamiltonian \eqref{eq:VariableCouplingHamiltonian} when the off-diagonal elements $b_i$ are drawn from the chi distribution $\chi_{\alpha i}$. The simulations were performed for $L=1000$ with 1000 runs, and the edge modes (the smallest and largest 10\% of eigenvalues) were discarded. It can be seen that the distribution does not match the Gaussian beta ensembles for any value of $\alpha$.}
	\label{fig:p-1-figure}
\end{figure}

As an alternative, inspired by the tridiagonalization of Gaussian ensembles in equation \eqref{Gaussian_ensemble_tridiagonalization}, one can try to draw the off-diagonal elements $b_i$ from the distribution $\chi_{\beta i}$, and investigate the statistics of the level spacing ratios $\tilde{r}$ for a generic draw for long spin chains. However, unlike the tridiagonal form of the Gaussian beta ensembles, the diagonal elements of the spin chain Hamiltonian are not independent, which leads to different $r$-statistics. In figure \ref{fig:p-1-figure}, we plot the $r$-statistics of the spin chain Hamiltonian when the off-diagonal elements are drawn from the chi distribution as $b_i \sim \chi_{\alpha i}$, for various values of $\alpha$. Already for $\alpha=1$, the distribution is highly skewed to $r = 1$. Decreasing $\alpha$ also decreases the mean $\tilde{r}$, but the curve does not fit the Gaussian beta ensembles for any value of $\alpha$. For very small values of $\alpha$, the distribution converges to the integrable case, as the change between the distributions that two neighboring couplings are drawn from get very small, and looks continuous if $\alpha \sim 1/L$, similar to the cases in figure \ref{fig:rStatisticsWithIIDCouplings}.

This observation motivates us to search for an alternative distribution from which the couplings $b_i$ are drawn. We propose that if the coupling parameters $b_i$ are drawn from the powers of the chi distribution\footnote{We note that powers of the chi distribution are special cases of the generalized gamma distribution, so the distribution we draw the couplings from can also be written as $b_i \sim \text{GeneralizedGamma}\left(2^{\frac{p}{2}}, \frac{\alpha  \beta i}{p}, \frac{2}{p} \right)$. We refer the reader to Appendix \ref{app:Distributions} for more details on these distributions.}
\begin{equation}\label{eq:gamma-dist}
    b_i \sim (\chi_{\alpha  \beta i})^p
\end{equation}
with a large enough exponent $p$, the distribution of the level spacing statistics of the system empirically matches that of the corresponding Gaussian beta ensemble, approximated by equation \eqref{GE_rStatistics}. In this expression, $\alpha$ and $p$ are two parameters that respectively control the mean $\langle \tilde{r} \rangle$ and the shape of the distribution of $\tilde{r}$.

In the large $L$ limit, drawing the couplings $b_i$ from the powers of the chi distribution as in \eqref{eq:gamma-dist} leads to a probability distribution for the level spacing ratios $\tilde{r}$, which we will denote as $P^{\alpha,p}_\beta(\tilde{r})$. To conclude that the dynamics of the Hamiltonian with these coupling constants is chaotic, it is necessary to compare this probability distribution to the statistics of Gaussian ensembles, $P^{\text{GE}}_\beta(\tilde{r})$ as given in Eq.~\eqref{GE_rStatistics}.

To quantify the proximity of our constructed probability distribution to that of $P^{\text{GE}}_{\beta}(\tilde{r})$, we consider two natural measures, namely the $L^2$ norm and the relative entropy, i.e., Kullback--Leibler divergence. The functional $L^2$ norm of the difference between the two distributions is:
\begin{equation} \label{eq:functional_l2_norm}
    \| \Delta P^{\alpha,p}_\beta \| = \left\| P^{\alpha,p}_\beta(\tilde{r}) - P^{\text{GE}}_\beta(\tilde{r}) \right\| = \left( \int_0^1 \left[ P^{\alpha,p}_\beta(\tilde{r}) - P^{\text{GE}}_\beta(\tilde{r}) \right]^2 \mathrm{d}\tilde{r} \right)^{1/2} \, .
\end{equation}
It is natural to ask how this quantity changes with $\alpha$ and $p$. Numerically, we observe that for a fixed exponent $p$, there exists a value for $\alpha$ that minimizes this norm, which we will denote as $\alpha_0$. The value of $\alpha_0$ depends on $p$,\footnote{In principle, $\alpha_0$ also depends on $\beta$, but we expect this dependence to be very small and irrelevant. Therefore, when numerically computing $\alpha_0$, we take $\beta=1$.} therefore we can consider $\alpha_0$ to be a function of $p$ and write $\alpha_0(p)$.

\begin{figure}[!t]
	\centering
	\includegraphics[width=.8\linewidth]{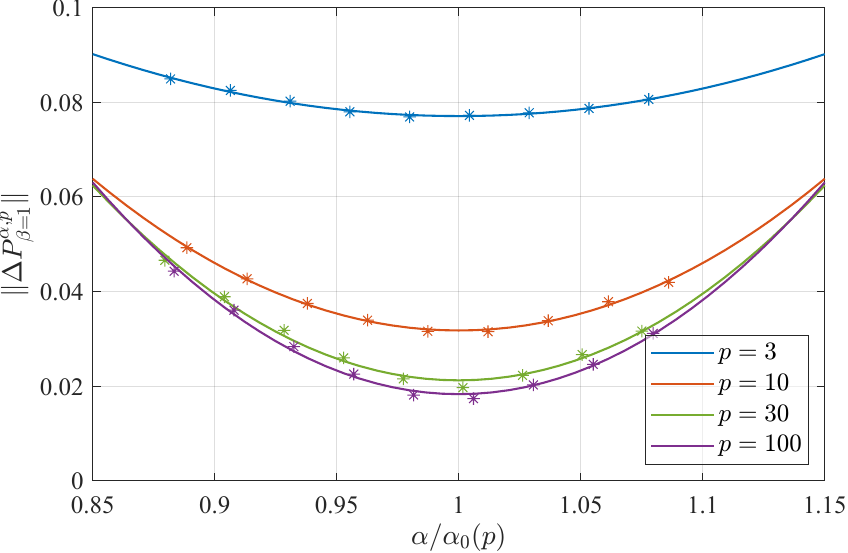}
	\caption{Numerical simulation of the $L^2$ norm difference $\| \Delta P^{\alpha,p}_\beta \|$ for four different values of $p$. The simulations were run for nine different equally spaced values of $\alpha$, and then the value of $\alpha$ that minimizes the norm, $\alpha_0(p)$, was estimated by a least-mean-squares fit of a parabola to the nine data points, and is given in table \ref{tab:p_versus_chaosQuantifiers}. 1000 draws were performed for the coupling constants $b_i$ of a Hamiltonian of size $L=10000$, and the smallest and largest 10\% of eigenvalues were discarded. As can be seen, the level statistics approximate chaotic statistics more closely as $p$ increases.}
	\label{fig:l2_norm}
\end{figure}

In figure \ref{fig:l2_norm}, we plot the simulation results for the $L^2$ norm for various values of $p$ and $\alpha$. To estimate $P^{\alpha,p}_{\beta}(\tilde{r})$, we perform 1000 draws for the couplings $b_i$ of a spin chain of length $L=10,000$ and compute the eigenvalues. To compute the level spacing ratios $\tilde{r}$, we discard the largest and smallest 10\% of eigenvalues to get rid of the effect of the edge modes. Then, we obtain a discrete estimate for $P^{\alpha,p}_{\beta}(\tilde{r})$ by counting the number of spacing ratios $\tilde{r}$ in intervals of width 0.01, and perform a Riemann sum to estimate the integral in \eqref{eq:functional_l2_norm}. For each $p$, we compute the norm for nine equally spaced values of $\alpha$, and estimate the minimum $\alpha_0$ by finding the best-fit parabola for these data points.

As an alternative measure of proximity, we also compute the relative entropy:
\begin{equation} \label{eq:KL_divergence}
    D_{\text{KL}}(P \, \| \,  Q) = \int_{-\infty}^\infty P(\tilde{r}) \log \frac{P(\tilde{r})}{Q(\tilde{r})} \mathrm{d}\tilde{r},
\end{equation}
where we allow both the target distribution as well as our tuned distribution to both play the roles of $P$ and $Q$ (since the relative entropy is asymmetric).

\begin{table}[!t]
\centering
\begin{tabular}{|c|c|C|C|C|C|}

\hline
\multicolumn{2}{|c|}{$p$} & 3 & 10 & 30 & 100 \\ \hline
\multicolumn{2}{|c|}{$\alpha_0(p)$} & 1.531 & 7.090 & 23.53 & 81.51 \\ \hline

\multicolumn{1}{|c|}{\multirow{3}{*}{$\big\| \Delta P^{\alpha_0(p),p}_\beta \big\|$}} & $\beta=1$ & 0.0769 & 0.0318 & 0.0200 & 0.0168 \\ \cline{2-6} \multicolumn{1}{|c|}{} & $\beta=2$ & 0.0873 & 0.0335 & 0.0230 & 0.0202 \\ \cline{2-6}
\multicolumn{1}{|c|}{} & $\beta = 4$ & 0.0891 & 0.0302 & 0.0221 & 0.0197 \\ \hline

\multicolumn{1}{|c|}{\multirow{3}{*}{$D_{\text{KL}}\big(P^{\alpha_0(p), p}_\beta \, \big\| \,  P^{\text{GE}}_\beta\big)$}} & $\beta=1$ & $5.85 \times 10^{-3}$ & $1.14 \times 10^{-3}$ & $4.91 \times 10^{-4}$ & $3.60 \times 10^{-4}$ \\ \cline{2-6} \multicolumn{1}{|c|}{} & $\beta=2$ & 0.0104 & $1.69 \times 10^{-3}$ & $7.17 \times 10^{-4}$ & $5.21 \times 10^{-4}$ \\ \cline{2-6}
\multicolumn{1}{|c|}{} & $\beta = 4$ & 0.0125 & $1.47 \times 10^{-3}$ & $5.78 \times 10^{-4}$ & $3.84 \times 10^{-4}$ \\ \hline

\multicolumn{1}{|c|}{\multirow{3}{*}{$D_{\text{KL}}\big(P^{\text{GE}}_\beta \, \big\| \,  P^{\alpha_0(p), p}_\beta\big)$}} & $\beta=1$ & $5.00 \times 10^{-3}$ & $1.04 \times 10^{-3}$ & $4.59 \times 10^{-4}$ & $3.38 \times 10^{-4}$  \\ \cline{2-6} \multicolumn{1}{|c|}{} & $\beta=2$ & $8.08 \times 10^{-3}$ & $1.46 \times 10^{-3}$ & $6.55 \times 10^{-4}$ & $4.82 \times 10^{-4}$ \\ \cline{2-6}
\multicolumn{1}{|c|}{} & $\beta = 4$ & $9.10 \times 10^{-3}$ & $1.25 \times 10^{-3}$ & $5.25 \times 10^{-4}$ & $3.58 \times 10^{-4}$ \\ \hline
\end{tabular}
\caption{Comparison of the different quantifiers (the $L^2$ norm difference defined in equation \eqref{eq:functional_l2_norm} and the Kullback--Leibler divergence defined in equation \eqref{eq:KL_divergence}) for the difference between the true Gaussian ensemble level spacing statistics given by equation \eqref{GE_rStatistics}, and the level spacing statistics of a tridiagonal spin chain Hamiltonian with the couplings drawn from the distribution $b_i \sim (\chi_{\alpha_0(p)\beta\,i})^p$. As $p$ increases, the level spacing statistics approximate the universality class of random Gaussian matrices.}
\label{tab:p_versus_chaosQuantifiers}
\end{table}

\begin{figure}[htpb]
    \centering
    \begin{tabular}{c@{\hspace{1.25em}}|@{\hspace{1.25em}}c}
    \begin{minipage}[t]{0.45\textwidth}
    \centering
    \begin{subfigure}[t]{\textwidth}
        \centering
        \includegraphics[width=\linewidth]{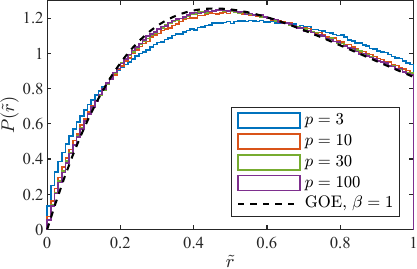}
        \caption{$\beta=1$}
    \end{subfigure}
    \hfill

    \vspace{0.5cm}
    \begin{subfigure}[t]{\textwidth}
        \centering
        \includegraphics[width=\linewidth]{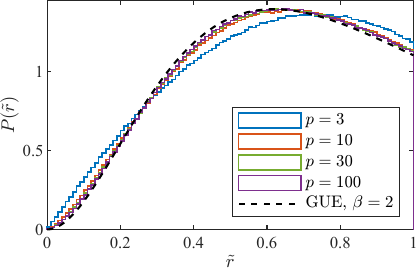}
        \caption{$\beta=2$}
    \end{subfigure}
    \hfill

    \vspace{0.5cm}
    \begin{subfigure}[t]{\textwidth}
        \centering
        \includegraphics[width=\linewidth]{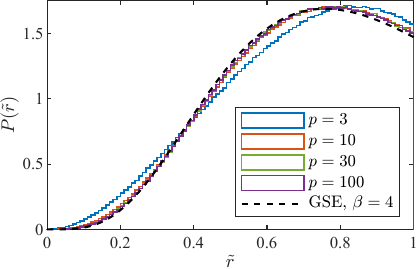}
        \caption{$\beta=4$}
    \end{subfigure}

	\caption{Distribution of the level spacing ratios $\tilde{r}$ for tridiagonal matrices of the form given in equation \eqref{eq:VariableCouplingHamiltonian} with the couplings drawn from the distribution $b_i \sim (\chi_{\alpha_0(p)\beta\,i})^p$, with $\alpha_0(p)$ given in table \ref{tab:p_versus_chaosQuantifiers}. A total of $1000$ draws were performed for $L=10^4$, with the smallest and largest $10\%$ of eigenvalues discarded.}
    \label{fig:r-without-edge-modes}
\end{minipage}
&
\begin{minipage}[t]{0.45\textwidth}
    \centering
    \begin{subfigure}[t]{\textwidth}
        \centering
        \includegraphics[width=\linewidth]{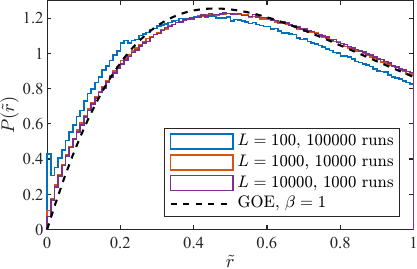}
        \caption{$p=10$}
    \end{subfigure}
    \hfill

    \vspace{0.5cm}
    \begin{subfigure}[t]{\textwidth}
        \centering
        \includegraphics[width=\linewidth]{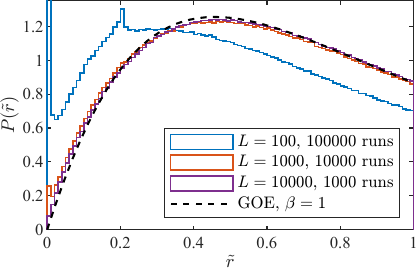}
        \caption{$p=30$}
    \end{subfigure}
    \hfill

    \vspace{0.5cm}
    \begin{subfigure}[t]{\textwidth}
        \centering
        \includegraphics[width=\linewidth]{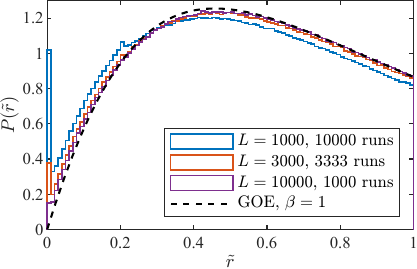}
        \caption{$p=100$}
    \end{subfigure}
	\caption{Distribution of the level spacing ratios $\tilde{r}$ for tridiagonal matrices of the form given in equation \eqref{eq:VariableCouplingHamiltonian} with the couplings drawn from the distribution $b_i \sim (\chi_{\alpha_0(p)\beta\,i})^p$, with $\alpha_0(p)$ given in table \ref{tab:p_versus_chaosQuantifiers} and $\beta = 1$. We see that the effects of the largest and smallest eigenvalues are suppressed for larger spin chains, but their impact increases with $p$.}
    \label{fig:r-with-edge-modes}
\end{minipage}
\end{tabular}
\end{figure}

In table \ref{tab:p_versus_chaosQuantifiers}, we compare the optimum $\alpha_0$, the $L^2$ norm $\| \Delta P^{\alpha,p}_\beta \|$, as well as the Kullback--Leibler divergence for various values of $p$. In figure \ref{fig:r-without-edge-modes}, we compare the probability distributions for various values of $p$. As can be seen, we obtain a very good match to Gaussian statistics. This provides strong evidence that the dynamics of the single-impurity / dilute gas approximation to the tuned spin chain Hamiltonian exhibits a chaotic energy spectrum.

It should be noted that figure \ref{fig:r-without-edge-modes} was obtained by running the simulations for a fixed spin chain length $L$, and excluding the smallest and largest 10\% of the eigenvalues from the computations of the $r$-statistics. It is necessary to exclude these ``outliers'' to make a conclusion about the infinite spin chain limit, as they can skew the $r$-statistics of finite spin chains, which become more impactful for large $p$ or small $L$. Figure \ref{fig:r-with-edge-modes} shows the decreasing effect of the largest and smallest eigenvalues as the spin chain length $L$ increases for a fixed exponent $p$, and it can also be seen that as $p$ increases, it is also necessary to take $L$ to be larger to make a conclusion about the $L\rightarrow \infty$ limit.

An important numerical observation we have is that the fit to Gaussian ensembles gets better as the exponent $p$ increases, which can be seen both from figure \ref{fig:r-without-edge-modes} and table \ref{tab:p_versus_chaosQuantifiers}. However, as mentioned before, for very large values of $p$, we run into numerical challenges, as the effect of edge modes becomes more and more prominent, and it is necessary to take the spin chain length $L$ to be larger to make a conclusion about the $L \rightarrow \infty$ limit, which becomes numerically challenging to compute. However, already for $p = 10$, we have a very good fit to Gaussian ensembles.

Furthermore, from table \ref{tab:p_versus_chaosQuantifiers}, we observe that the relation $\alpha_0(p)$ is almost linear in $p$ for the four data points we present, given by the least-squares line $\alpha_0(p) = 0.8257 p - 1.1023$. However, we emphasize that the precise relation between $\alpha_0$ and $p$ is likely not exactly linear in the large $p$ limit. It would be interesting to determine a numerical estimate for $\alpha_0(p)$ by finding a best-fit function over many data points. In general, $\alpha_0(p)$ will be some function of $p$ determined by this fit.\footnote{We pause here to note that while we have chosen a particular probability distribution for the couplings $b_i$, as in equation \eqref{eq:gamma-dist}, to exhibit the onset of chaotic dynamics, we do not expect this distribution to be the only choice that leads to chaotic dynamics. One reason to suspect this is that the $\chi$ and powers-of-$\chi$ distributions only explore a particular slicing of the generalized gamma distribution (see Appendix \ref{app:Distributions}), and we suspect that there are other slices of this distribution or even other entirely different distributions where the onset of chaos is less tuned.}

To make these arguments more precise, and to connect back to the QFTs discussed in section \ref{sec:QUIVER}, we need to determine the proper scaling for the couplings in the large $L$ limit. The expectation value and variance for $b_i$ drawn from the $(\chi_{\alpha \beta i})^p$ distribution in the large $L$ limit is given by
\begin{equation}
    \mu(b_i) =  (\alpha \beta xL)^{p/2}\,,\qquad
    \sigma^2(b_i) = \frac{1}{2}p^2(\alpha \beta xL)^{p-1} \, ,
\end{equation}
with $x = i/L$. By inspection, the $b_i$ for large $i \sim L$ can grow to be rather large and scales as $i^{p}$.

Of course, we are free to also consider a rescaling of the $b_i$'s so that the largest values still remain well-behaved in the large $i \rightarrow L$ limit. Introducing:
\begin{equation}
    \widetilde{b}_i = \frac{b_i}{(\alpha \beta L)^{p/2}},
\end{equation}
we can then proceed as before, with a Hamiltonian constructed from the $\widetilde{b}_i$'s rather than the $b_{i}$'s. Observe that
the mean and standard deviation are:
\begin{equation}
    \mu(\widetilde{b}_i) =  x^{p/2}\,,\qquad
    \sigma^2(\widetilde{b}_i) = \frac{p^2x^{p-1}}{2\alpha \beta L}\,.
\end{equation}
In particular, note that the variance is very small in the large $L$ limit. See Appendix \ref{app:Distributions} for more details.

It is also of interest to extract the average value of the $\tilde{b}_{j}$'s since this tells us the ``typical'' strength of the 't Hooft coupling amongst the gauge groups participating in the spin chain:
\begin{equation}
\tilde{b}_{\mathrm{chain\,avg}} = \frac{1}{L}\sum_{j = 1}^{L} b_{j} = \int_{0}^{1} x^{p/2} dx = \frac{2}{2 + p} + O(1/L).
\end{equation}

Summarizing, we have identified the onset of chaotic dynamics, provided we determine $\alpha_0(p)$ appropriately. This function
is something we have empirically tuned as a function of $p$, though it would of course be interesting to extract a closed form
expression for it.

\paragraph{Further Tests: Spectral Rigidity}

\begin{figure}[t!]
    \centering
    \begin{tabular}{|c|c|c|c|}
    \hline
         Asymptotes & $\beta = 1$ & $\beta = 2$ & $\beta = 4$ \\ \hline
         $p = 3$ & $0.187 \log \mathcal{E} - 0.202$ & $0.096 \log \mathcal{E} - 0.015$ & $ 0.050 \log \mathcal{E} + 0.043 $ \\ \hline
         $p = 10$ & $0.139 \log \mathcal{E} - 0.111$ & $0.071 \log \mathcal{E} + 0.010$ & $0.037 \log \mathcal{E} + 0.052$ \\ \hline
         $p = 30$ & $0.123 \log \mathcal{E} - 0.079$ & $0.063 \log \mathcal{E} + 0.022$ & $0.032 \log \mathcal{E} + 0.059$ \\ \hline
         theoretical & $0.101 \log \mathcal{E} - 0.007$ & $0.051 \log \mathcal{E} + 0.059$ & $0.025 \log \mathcal{E} + 0.078$ \\ \hline
    \end{tabular} \\ [1cm]
    \begin{subfigure}{0.49\textwidth}
        \centering
        \includegraphics[width=\linewidth]{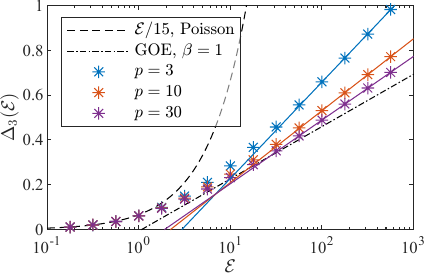}
        \caption{$\beta=1$}
        \label{subfig:Beta=1}
    \end{subfigure}
    \hfill
    \begin{subfigure}{0.49\textwidth}
        \centering
        \includegraphics[width=\linewidth]{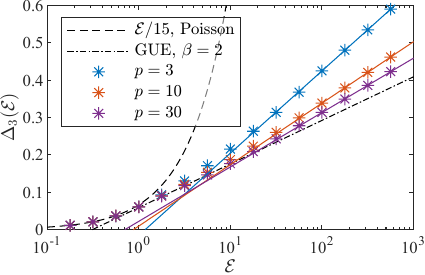}
        \caption{$\beta=2$}
    \end{subfigure}
    \\[0.5cm]
    \begin{subfigure}{0.49\textwidth}
        \centering
        \includegraphics[width=\linewidth]{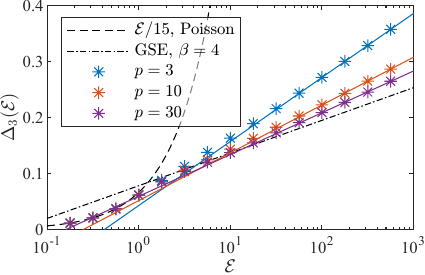}
        \caption{$\beta=4$}
    \end{subfigure}

	\caption{Spectral rigidity $\Delta_3(\mathcal{E})$ for (a) $\beta = 1$ (GOE), (b) $\beta = 2$ (GUE), (c) $\beta = 4$ (GSE), shown for $p = 3, 10, 30$. The solid lines are the approximations for the asymptotic behavior of $\Delta_3(\mathcal{E})$ obtained by a best-fit line to the data for $\mathcal{E} \geq 100$, given in the table above. The dashed curve is the Poisson result $\Delta_3(\mathcal{E}) = \mathcal{E}/15$, characteristic of uncorrelated levels. The dashed-dotted lines show the Wigner--Dyson asymptotics of equations \eqref{eq:Delta3_GOE}--\eqref{eq:Delta3_GSE}. For window sizes below the mean level spacing, $\mathcal{E} \lesssim 1$, all spectra follow the Poisson curve, as rigidity is insensitive to level correlations at these scales. For $\mathcal{E} \gtrsim 1$, the data cross over to the logarithmic growth characteristic of a chaotic system. As $p$ increases, this approaches the universality class for random matrices drawn from a Gaussian ensemble: the $p = 30$ data lie closest to the random matrix prediction across the asymptotic regime in all three symmetry classes.}
    \label{fig:SpectralRigidities}
\end{figure}

Thus far we have studied the $r$-statistics for our chaotic spin chain Hamiltonians and found excellent agreement with RMT. Since the $r$-statistics probe correlations between adjacent levels, this establishes that the short-range spectral properties of our models are of Wigner--Dyson type. However, random matrix universality is a statement about correlations on all scales, and we can also test the long-range behavior of the spectrum. The standard diagnostic for this is the Dyson--Mehta spectral rigidity $\Delta_3(\mathcal{E})$, where $\mathcal{E}$ is the length of an energy window in units of the mean level spacing, see e.g. \cite{DysonMehta1963IV, Berry85, MehtaBook}.\footnote{See also \cite{Balasubramanian:2022tpr, Balasubramanian:2023kwd} for recent discussions on this point. We thank V. Balasubramanian for comments on this point.} We review the details of spectral rigidity in Appendix \ref{app:SpectralRigidity}. 

For uncorrelated (Poisson) levels the rigidity grows linearly, $\Delta_3(\mathcal{E}) = \mathcal{E}/15$, whereas for Wigner--Dyson ensembles it grows only logarithmically, $\Delta_3(\mathcal{E}) \simeq \log(\mathcal{E})/\beta\pi^2$ up to a known ($\beta$-dependent) constant. We report our results for the spectral rigidity in figure \ref{fig:SpectralRigidities}, where we compute $\Delta_3(\mathcal{E})$ for $p = 3, 10, 30$.\footnote{We omit $p=100$ from spectral rigidity computations as we run into numerical issues with the procedure for unfolding the spectrum.} Our procedure for numerically extracting $\Delta_3(\mathcal{E})$ is presented in Appendix \ref{app:SpectralRigidity}. With this data in place, we perform a linear regression:
\begin{equation}
\Delta_{3}(\mathcal{E}) = a \log \mathcal{E} + b \equiv a \log (\mathcal{E} / \mathcal{E}_0),
\end{equation}
i.e., we can absorb the constant offset $b$ into a rescaling of $\mathcal{E}$. By inspection, we observe that the spectrum is far away from Poisson. Additionally, we find that the match to Wigner--Dyson ensembles is stronger as $p$ increases. For $p = 30$ both the universal logarithmic slope and the ensemble-dependent constant are reproduced with high accuracy, establishing that the long-range rigidity of the spectrum of the spin chain Hamiltonians in this case is also of Wigner--Dyson type.

\paragraph{Even Further Tests: Spectral Form Factor}

\begin{figure}[!t]
    \centering
    \begin{subfigure}{0.49\textwidth}
        \centering
        \includegraphics[width=\linewidth]{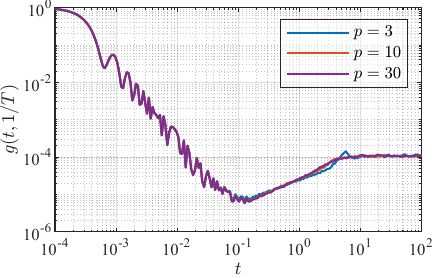}
        \caption{$1/T=0.0001$}
    \end{subfigure}
    \hfill
    \begin{subfigure}{0.49\textwidth}
        \centering
        \includegraphics[width=\linewidth]{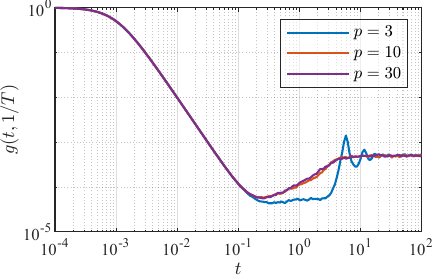}
        \caption{$1/T = 0.001$}
    \end{subfigure}
    \\[0.5cm]
    \begin{subfigure}{0.49\textwidth}
        \centering
        \includegraphics[width=\linewidth]{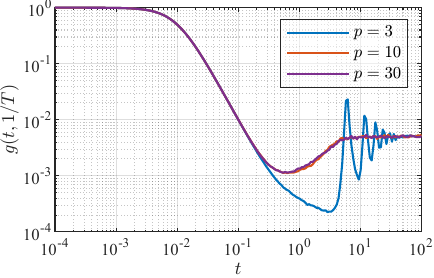}
        \caption{$1/T = 0.01$}
    \end{subfigure}
    \hfill
    \begin{subfigure}
        {0.49\textwidth}
        \centering
        \includegraphics[width=\linewidth]{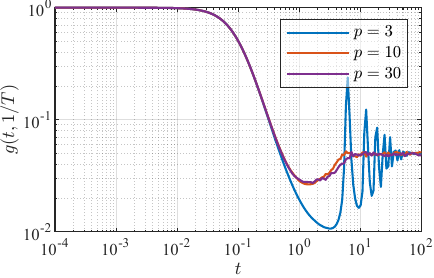}
        \caption{$1/T = 0.1$}
    \end{subfigure}
    \caption{Spectral form factor $g(t,1/T)$ shown for $p=3,10,30$ and various values of the inverse temperature $1/T$. The spectral form factor was calculated using the unfolded spectrum, the procedure for which is explained in Appendix \ref{app:SpectralRigidity}. After unfolding, the mean eigenvalue spacing becomes 1, and the largest (unfolded) eigenvalue is of the order $L = 10^4$. We see that the spectral form factor exhibits the ramp, dip, and the plateau characteristic of chaotic systems and Gaussian ensembles. All plots are for ensembles with $\beta = 1$ (i.e., the GOE case). Similar results hold for other values of $\beta$.}
    \label{fig:Spectral_Form_Factor}
\end{figure}

Another diagnostic of chaotic dynamics from the eigenvalue spectrum is the spectral form factor $g(t, 1/T)$, which can be defined in terms of the analytic continuation of the partition function
\begin{equation}
    Z(t,1/T) \equiv \Tr\left(e^{- (1/T + i t)H}\right) = \sum_n e^{- (1/T + i t)E_n} \, ,
\end{equation}
as
\begin{equation}
    g(t,1/T) \equiv \frac{\left\langle \abs{Z(t,1/T)}^2\right\rangle}{\left\langle \abs{Z(0,1/T)}^2\right\rangle} \, ,
\end{equation}
where $\langle \bullet \rangle$ denotes ensemble averaging.\footnote{Namely, we are computing an annealed average. Compared with \cite{Cotler:2016fpe} we have normalized our expressions by dividing by $\langle Z(T^{-1})^2 \rangle$ rather than $\langle Z(T^{-1}) \rangle^2$ so that we can compare different ensembles more easily and have $g(t=0) = 1$. In practice, the numerical distinction between the two averaged quantities is rather small.}

A characteristic of Gaussian ensembles is that the spectral form factor exhibits a dip, followed by a ramp, followed by a plateau \cite{Cotler:2016fpe}. We aim to show that a similar behavior is also observed for the class of matrices we give in equation \eqref{eq:VariableCouplingHamiltonian}. However, to make the comparison more precise, it is also necessary to unfold the spectrum similarly to the computation of spectral rigidity, as explained in Appendix \ref{app:SpectralRigidity}. After unfolding, $Z(t,1/T)$ is defined as
\begin{equation}
    Z(t,1/T) = \sum_n e^{-(1/T + i t)x_n} \, ,
\end{equation}
where $\{x_n\}$ are the unfolded eigenvalues. After unfolding, the mean spacing between the unfolded eigenvalues $\{x_n\}$ becomes 1, and the density of states is uniform. The unfolding allows us to directly compare systems with different densities of states, as it eliminates the impact of the density of states on the spectral form factor and sets a common scale.

In figure \ref{fig:Spectral_Form_Factor}, we show the spectral form factor when the couplings are drawn from the powers of the chi distribution. As expected for chaotic systems, the spectral form factor exhibits a dip, followed by a ramp, followed by a plateau, characteristic of chaotic dynamics and Gaussian ensembles. Moreover, as $p$ increases, fluctuations / ringing in this profile are also suppressed.

\subsection{Krylov Complexity}

\begin{figure}[t!]
    \centering
    \begin{subfigure}{0.49\textwidth}
        \centering
        \includegraphics[width=\linewidth]{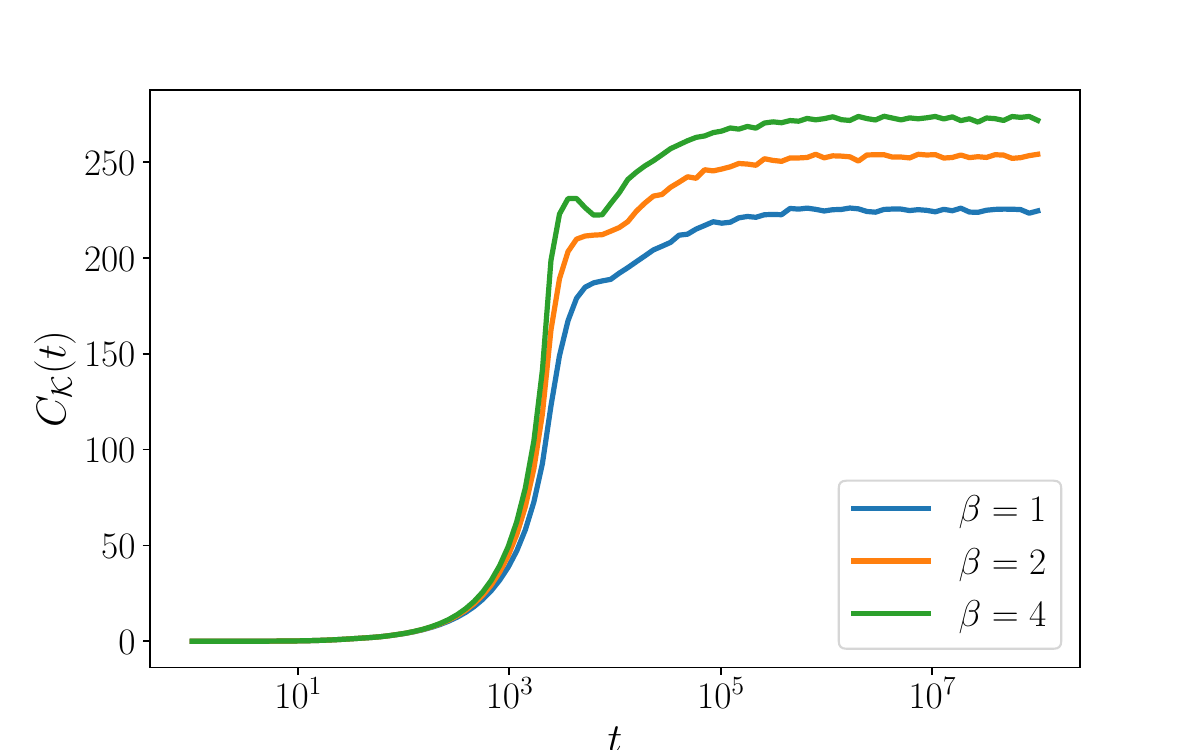}
        \caption{$|n=500\rangle$}
        \label{subfig:krylov500}
    \end{subfigure}
    \begin{subfigure}{0.49\textwidth}
        \centering
        \includegraphics[width=\linewidth]{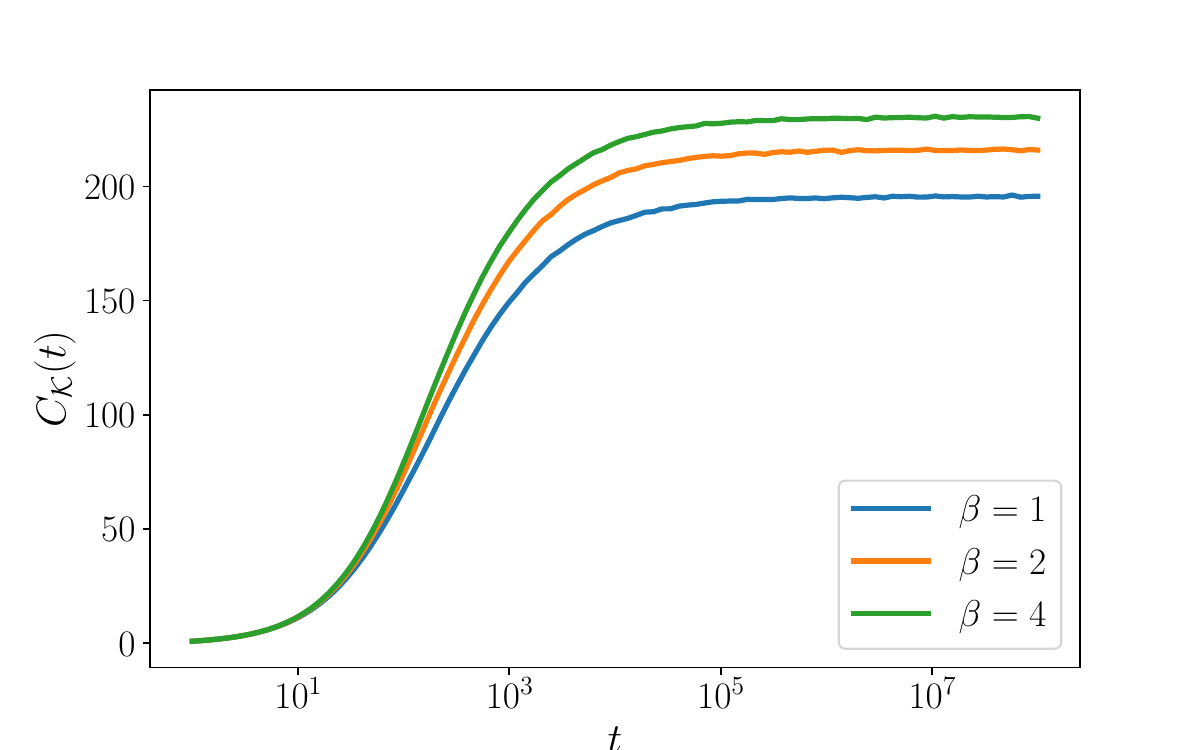}
        \caption{$|n=1000\rangle$}
    \end{subfigure}
    \\
    \begin{subfigure}{0.49\textwidth}
        \centering
        \includegraphics[width=\linewidth]{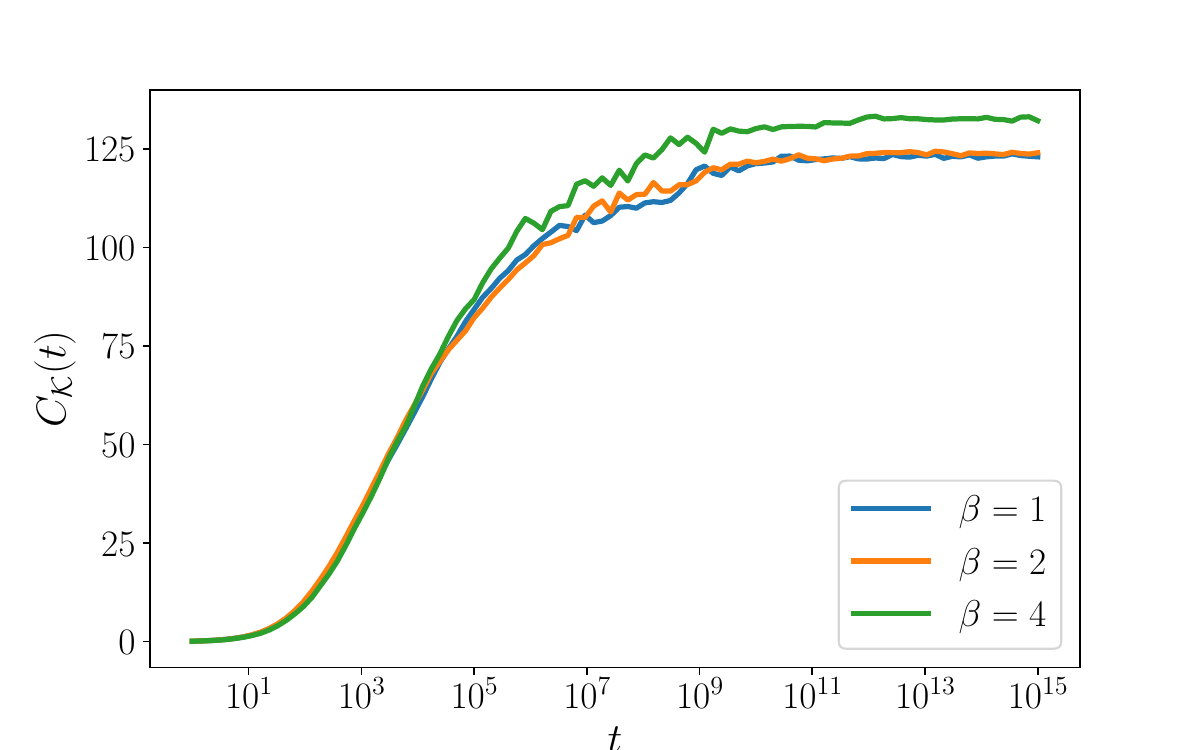}
        \caption{Random seed}
    \end{subfigure}

	\caption{Krylov complexity $\mathcal{C}_{\mathcal{K}}$ for the distribution $g_i = (\chi_{\alpha \beta\,i})^p$, with $\alpha = 55/8$ and $p=10$. A total of $100$ draws are performed for various seed states in a length $L=1000$ spin chain. The (unnormalized) random state is taken of the form $\sum c_\ell|\ell\rangle$ for uniform $c_l\in [0, 10]$, picked anew each draw. We see a rapid growth followed by a plateau, showing that the seed state quickly spreads to a large portion of the one-impurity sector.}
    \label{fig:krylov}
\end{figure}

\begin{figure}[!t]
	\centering
	\includegraphics[width=.8\linewidth]{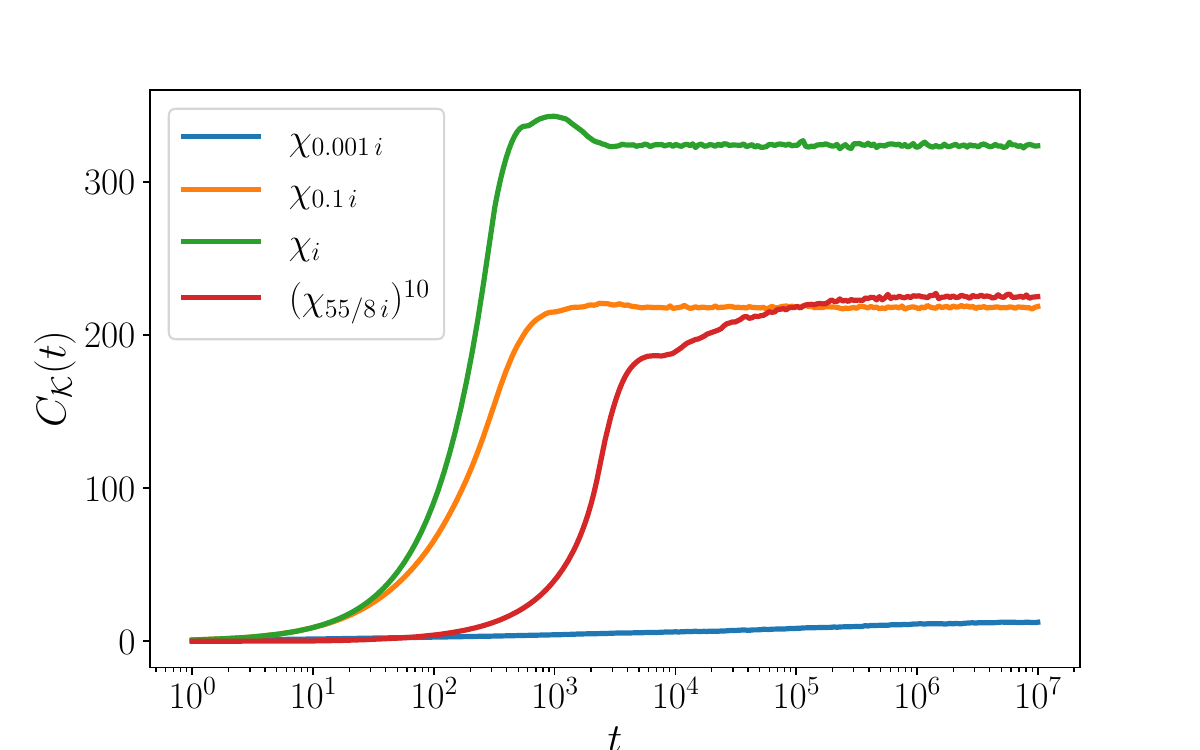}
	\caption{Krylov complexity for the XXX spin chain of length $L=1000$ where coupling constants $b_i$ are chosen in a $\chi$ distribution. A total of $200$ draws for the Hamiltonian are performed for the seed state $|\varphi(t=0)\rangle = | 500\rangle$, and the average over these draw is plotted. We see that even though the distributions $\chi_{0.1i}$ and $(\chi_{55/8 i})^{10}$ both have the same qualitative shape, their spectral statistics are significantly different (compare figure \ref{fig:p-1-figure} and figure \ref{fig:r-without-edge-modes}).}
	\label{fig:krylov_other_chi}
\end{figure}

As a complementary diagnostic of chaotic dynamics, it is helpful to also investigate Krylov complexity
in this class of spin chains. As we now demonstrate, Krylov complexity does not appear to distinguish
between choices of the couplings which exhibit the desired eigenvalue repulsion, and those which instead exhibit different level spacing statistics.

To probe the extent to which Krylov complexity provides an accurate diagnostic for chaos / complexity spread, we seek to understand some of the characteristic features expected, namely a local peak, followed by a dip, followed by a plateau.\footnote{This is to be contrasted with the spectral form factor, which is expected to exhibit a dip, followed by a ramp, followed by a plateau, see e.g., \cite{Cotler:2016fpe}.} To some extent, we have implicitly addressed this issue directly by analyzing the spectral properties of our tuned Hamiltonian system. Our aim here is different: we seek to understand the extent to which Krylov complexity alone can inform us of this behavior.

Along these lines, we consider two situations. First, we randomly draw ``generic'' states as obtained from linear combinations $\sum_{\ell} c_\ell \ket{\ell}$ in the Krylov basis, and with respect to such states, we extract the Krylov complexity as a function of time. The second situation consists of a ``localized'' excitation, as obtained by working with a state in the middle or end of the chain, i.e., $\ket{n = L/2}$ and $\ket{n = L}$. We note that while these states are localized in position space, they involve linear combinations with order one coefficients of the energy eigenstates. In our analysis, we focus on several different choices of spin chain Hamiltonian, some of which exhibit chaotic behavior, and some which do not. The main conclusion we draw from this is that relying on Krylov complexity alone can result in two sorts of issues: on the one hand, it can sometimes fail to detect the onset of chaos, i.e., a ``false negative''. On the other hand, it can sometimes incorrectly signal the presence of chaos even when not present, i.e., a ``false positive''.

Our procedure for constructing the Krylov complexity plots is as follows. First, we randomly draw a Hamiltonian. Given a seed state, we then evolve it with respect to this fixed Hamiltonian, extracting the corresponding Krylov complexity. We repeat this, drawing many independent draws for a Hamiltonian and then average over these choices to obtain our Krylov complexity plots. When we have a randomly drawn seed state, we also average over this.

Let us begin with examples of a false negative. 
In figure \ref{fig:krylov} we consider the tuned case where we take the $b_i \sim (\chi_{\alpha \beta i})^{p}$ with $\alpha = 55/8$ and $p = 10$. We have already seen that this distribution exhibits eigenvalue repulsion, spectral rigidity, and a chaotic spectral form factor.\footnote{We have chosen to focus on $p = 10$ rather than $p = 30$ since it provides us access to examples which are not quite in the Gaussian ensemble, but approach it (as $p$ is increased further).} Using both randomly drawn and localized states, we see the expected ``plateau.''  Indeed, as one can see from figure \ref{fig:krylov}, the distribution given in equation \eqref{eq:gamma-dist} mimics the behavior expected from Gaussian ensembles as shown in the previous section (figure \ref{fig:krylov-theoric}): a rapid growth of $\mathcal{C}_{\mathcal{K}}(t)$ followed by an almost-constant plateau, interpreted as the seed state quickly spreading over the Hilbert space associated with the one-impurity sector. Furthermore, as has been observed in various cases like the XXZ spin chain \cite{Rabinovici:2022beu,Balasubramanian:2022tpr}, the late-time behavior of
$\mathcal{C}_{\mathcal{K}}(t)$ is controlled by the choice of the seed state
$|\psi(t=0)\rangle$. We find that for a random choice, the Krylov complexity
will usually grow up to about a quarter of $L/2$, the theoretical maximum. The plateau value of Krylov complexity is higher if the seed state is chosen to be $\ket{500}$ or $\ket{1000}$.

Missing, however, are some other qualitative features of complexity, in particular the peak and then the dip followed by a plateau in Krylov complexity usually associated with quantum chaos \cite{Balasubramanian:2022tpr}. By inspection, this does not appear for many of the seed states shown in figure \ref{fig:krylov}, and it is debatable whether it is present, e.g., in the $\ket{n = 500}$ state. Since the spectral statistics establish that these Hamiltonians exhibit chaotic behavior irrespective of the choice of seed state, we find that the presence or absence of the peak does not always track the chaotic nature of the spectrum for this class of theories.

Let us now turn to examples with a false positive. 
To illustrate this, we consider a broader class of tuned Hamiltonians in which we draw the off-diagonal entries $b_j$ from other 
ensembles. We illustrate some examples of this sort in figure \ref{fig:krylov_other_chi}. 

As a first example, note that the plateau value of the Krylov complexity for the Anderson localized system ($b_i \sim \chi_{0.001 i}$) is significantly lower due to localization in the Krylov basis \cite{Rabinovici:2021qqt}, in agreement with emergent integrability observed in such systems. Furthermore, both in figure \ref{fig:krylov} and figure \ref{fig:krylov_other_chi}, we observe that the plateau value of Krylov complexity is correlated with the degree of level repulsion in the system, or in other words, with the average level spacing ratio $\langle \tilde{r} \rangle$.

That being said, we stress that Krylov complexity alone is not sufficient to diagnose a chaotic regime. Indeed, other $\chi$ distributions can have qualitatively the same behavior of $\mathcal{C}_{\mathcal{K}}(t)$ as Gaussian ensembles, as shown in figure \ref{fig:krylov_other_chi}. As mentioned above, these distributions do not match the level spacing statistics $P(\tilde{r})$ of Gaussian ensembles and are not associated with chaos. In particular, comparing the growth of Krylov complexity for the distributions $\chi_{0.1i}$ and $(\chi_{55/8i})^{10}$ shows that two different distributions with different $r$-statistics can have very similar behavior and plateau values for Krylov complexity. 

Furthermore, for the case where the $b_i$ are drawn from the $\chi_i$ distribution, we find that the Krylov complexity exhibits the ramp, peak, and plateau that is expected from a chaotic theory. However, the level statistics of this theory do not match a Gaussian ensemble, as seen from the green histogram in figure \ref{fig:p-1-figure}. Thus, this is an example of Krylov complexity providing a false positive for the onset of chaos.

Overall, then, we take this to mean that while Krylov complexity is clearly a valuable tool 
in the study of quantum chaos, it is helpful to supplement this with additional diagnostics. It would be interesting to study these issues further.

\section{Interlude: Target Space Billiards} \label{sec:BILLIARDS}

The analysis of the previous section shows that tuning the marginal couplings of our 4D SCFT results in a
controlled chaotic subsector. In this section we return to the stringy origin of these 4D SCFTs and interpret this chaotic dynamics in terms of fluctuations moving as a chaotic billiard in the target space geometry (see e.g. \cite{Nastase:2026lhz, Fatemiabhari:2025cyy, Fatemiabhari:2025usn, Fatemiabhari:2025poq, Fatemiabhari:2026rob, Fatemiabhari:2026goj, Shaghoulian:2016umj, DeClerck:2023fax, Berenstein:2023vtd}). In section \ref{sec:HOLO} we take the near horizon limit to study analogous questions in the gravity dual.

In the generic case considered in this paper where we have $N$ D3-branes probing  a Calabi--Yau singularity $X$, the marginal parameters correspond to specific closed string twisted sector moduli, i.e., their values are frozen as boundary conditions on $\partial X$. The stringy background is specified on the conical geometry $\mathbb{R}^{3,1} \times \mathbb{R}_{\geq 0} \times Y$, with $Y = \partial X$ the boundary geometry of the Calabi--Yau cone $X$. The presence of the D3-branes leads to a non-trivial backreaction, which in the large $N$ limit leads to a geometry of the form $\mathrm{AdS}_5 \times Y$, but more generally, we simply have a warped throat of finite length.

Consider now the spin chain as well as its spectrum of excitations. To be more concrete, we 
consider the orbifold $\mathbb{C}^{3} / \mathbb{Z}_K$ in which the action on the holomorphic coordinate $x$ of $\mathbb{C}^3$ leaves $x^{K}$ invariant. The locus $x^{K} = 0$ defines a torsional cycle in the boundary $S^{5} / \mathbb{Z}_K$, but close to the tip of the cone it is better visualized as a particular twisted sector state. Fluctuations of this state are dictated by the quantum numbers of the $\mathfrak{su}(2,2\vert 1)$, and these directly translate to geometric actions in the background geometry $\mathbb{R}^{3,1} \times \mathbb{R}_{\geq 0} \times Y$ equipped with a warped metric. Since we can match all of the isometries of this geometry to quantum numbers, we observe that we are just tracking a closed string state as it chaotically bounces around in this ambient geometry. Similar considerations hold for more general D-brane probes of Calabi--Yau cone geometries.

\subsection{Special Case: 4D \texorpdfstring{$\mathcal{N} = 2$}{N=2} SCFTs}

In the special case of 4D $\mathcal{N} = 2$ SCFTs, there are many ways to arrive at essentially the same 
IR SCFT. As such, there are different geometric interpretations available. In this subsection we focus on the 4D $\mathcal{N} = 2$ SCFTs given by the quiver gauge theories:
\begin{equation}
[G_0] \,\text{---}\, G_1 \,\text{---}\, G_2 \,\text{---}\, \cdots \,\text{---}\, G_{L-1} \,\text{---}\, [G_{L}],
\end{equation}
where each link is a hypermultiplet, and we take $G_{j} = SU(N)$ for all $j$. Similar considerations hold for the closed loop as well.

We have already given a geometric interpretation for D3-branes probing this singularity. There are a few other ways to arrive at the same system, each with a corresponding ambient geometry.

To begin, observe that this same quiver gauge theory naturally arises from a compactification of the 6D conformal matter theory \cite{DelZotto:2014hpa, Heckman:2014qba} obtained from $L$ M5-branes 
probing the singularity $\mathbb{C}^{2} / \mathbb{Z}_N$, namely the background M-theory geometry is $\mathbb{R}^{5,1} \times \mathbb{R}_{\bot} \times \mathbb{C}^2 / \mathbb{Z}_N$, with the coincident M5-branes filling the $\mathbb{R}^{5,1}$ factor and sitting at the singularity.  One moves onto the partial tensor branch by keeping the M5-branes on the singularity, but separating them in the transverse direction $\mathbb{R}_{\bot}$. These separation moduli are controlled by tensor multiplet scalar vevs $t_{i}$ for $i = 1,...,L-1$. Compactifying further on a $T^2$ takes us to the background $\mathbb{R}^{3,1} \times T^2 \times \mathbb{R}_{\bot} \times \mathbb{C}^{2} / \mathbb{Z}_N$. 
The gauge couplings of the 4D $\mathcal{N} = 2$ SCFT are given by:
\begin{equation}\label{eq:gsquared}
\frac{1}{g^2_{i}} = t_{i} \times \mathrm{Vol}(T^2).
\end{equation}
The complexification to the $\tau_i$ of the 4D gauge theory follows from including periods of the 2-form potentials of the tensor multiplet over the $T^2$.

In this realization, the ground state of the spin chain corresponds to an M2-brane which is stretched from the very left to the very right of the quiver. Excitations amount to allowing motion of the M2-brane in the geometry, again as dictated by the quantum numbers of the $\mathfrak{su}(2,2\vert2)$ spin chain. In the infinitesimal neighborhood of the M5-branes, observe that the M2 again exhibits chaotic motion as it ``bounces'' off the different spacings between adjacent M5-branes.

There is another well-known generalized class $\mathcal{S}$ \cite{Gaiotto:2009hg,Gaiotto:2009we} construction of the same 4D $\mathcal{N} = 2$ theory where we instead wrap $N$ M5-branes on a sphere with $n$ punctures \cite{Ohmori:2015pua, Ohmori:2015pia}. More precisely, we start from the 6D $\mathcal{N} = (2,0)$ theory of type $A_{N-1}$ and take two full punctures on opposite ends, i.e., each retaining a full $\mathfrak{su}(N)$ and $n - 2$ simple punctures so that there are no additional non-abelian flavor symmetries. The corresponding theory is labelled as $\mathsf{T}_{\mathfrak{g}}\{\mathfrak{g},Y_{\mathrm{simple}}...,Y_{\mathrm{simple}},\mathfrak{g} \}$. Observe that there are $n - 3$ complex structure moduli for this punctured Riemann surface; we can interpret these as the $L - 1 = n-3$ marginal couplings of our quiver gauge theory:
\begin{equation}\label{eq:taupunc}
\tau_i = \frac{4\pi i}{g_i^2} + \frac{\theta_i}{2\pi}, \qquad i=1,\dots,L -1.
\end{equation}
In the degeneration limit, the surface decomposes into $L$ thrice-punctured spheres, i.e., trinions in which tubes attach to neighboring trinions along punctures.\footnote{In other words, this leaves one ``unattached'' puncture for each trinion and the gauge couplings arise geometrically as the complex structure moduli of the degeneration.}
\begin{equation}
q_i = e^{2\pi i \tau_i}.
\end{equation}
In this case, the ambient geometry is of the form $\mathbb{R}^{3,1} \times T^{\ast}C \times \mathbb{R}^3$. Just as in our discussion of 6D conformal matter compactified on a $T^2$, the ground state of the spin chain corresponds to an M2-brane which wraps all of the punctured sphere, and fluctuations of the spin chain correspond to the M2-brane exploring its ambient geometry.

While it is tempting to take the large $L$ limit to reach a holographic dual description, there are some subtleties with simultaneously retaining the single spin chain sector we have focused on up to this point. In the construction with $L$ M5-branes discussed near equation (\ref{eq:gsquared}), we can retain a fixed value of the gauge coupling, but at the expense of decompactifying the volume of the $T^2$. In the construction with $L$ punctures discussed near equation (\ref{eq:taupunc}), the large $L$ limit involves a very large number of punctures, which again obscures the original geometric construction.

We now turn to a more systematic discussion of holographic limits.

\section{Towards a Holographic Dual} \label{sec:HOLO}

We now study possible gravity dual descriptions of our chaotic spin chains. Consider a stack of $N$ D3-branes probing the local Calabi--Yau singularity $X$. Taking the near horizon limit, this yields a candidate gravity dual of the form $\mathrm{AdS}_5 \times Y$ with $X$ given by the cone over $Y = \partial X$. A celebrated feature of the spin chain sectors found in $\mathcal{N} = 4$ SYM theory \cite{Berenstein:2002jq} is that the spectrum of excitations carries over to the gravity dual, even at large 't Hooft coupling $\lambda = g^2_{YM} N$. This is because the low lying excitations of the spin chain of length $L$ are controlled by perturbation theory in $\lambda / L^{2}$, i.e., the length of the spin chain suppresses the energy spectrum.

Now, in the spin chain sectors investigated in the present paper, this extrapolation is unavailable. To see why, we focus in this section on the illustrative example given by $N$ D3-branes probing $X = \mathbb{C}^{3} / \mathbb{Z}_K$, with gravity dual $\mathrm{AdS}_5 \times S^{5} / \mathbb{Z}_K$. Similar considerations hold for other SCFTs realized from D3-branes probing a Calabi--Yau cone.

Consider, then, the backgrounds $\mathrm{AdS}_5 \times S^5 / \mathbb{Z}_K$. The supergravity approximation is valid provided the radius curvature of $S^{5} / \mathbb{Z}_K$ remains sufficiently large. More precisely, since the 't Hooft coupling of the original $\mathrm{AdS}_{5} \times S^{5}$ theory controls the radius of curvature via:
$\lambda_{\mathrm{IIB}}^{1/4} \sim R / \ell_{s} \gg 1$ where $R$ is the AdS$_5$ radius, in the orbifold theory the IIB 't Hooft coupling needs to satisfy: $\lambda_{\mathrm{IIB}}^{1/4} / K \gg 1$ (see e.g., \cite{Horowitz:2007pr}). On the other hand, the controlled single spin chain sector approximation is available provided we have $\lambda_{\mathrm{IIB}} / L^{2} \ll 1$. Since $L$ is at most $K$, the two regimes are clearly incompatible. Said differently, while we can certainly extrapolate to large 't Hooft coupling of the quiver gauge theory, i.e. $\lambda_{j} \gg 1$, with $\lambda_{j} / L^{2} \ll 1$, this is incompatible with having a semi-classical gravity dual.

To some extent, this is to be expected, because the single spin chain sector
descends from a single trace operator of the parent $\mathcal{N}=4$
SYM\ theory. Indeed, one generically expects that in the
AdS/CFT\ correspondence, chaotic behavior should appear for operators above
the threshold $\Delta>N$, i.e., once we cross over to the large black hole
regime. Provided we work with operators such as Tr$(X^{J})$ with $J<N$ in the
$\mathcal{N}=4$ SYM theory, the mixing with multi-particle operators is all
suppressed in the large $N$ limit.\footnote{For $J$ near $N$, the single
particle states are better expressed as $\det X$, and subdeterminants, i.e., as
giant graviton operators \cite{Balasubramanian:2001nh}.}

To make contact with a semi-classical gravity dual, we proceed by considering a broader class of spin chain sectors where the length of the spin chain is far larger than the order of the orbifold group.

\subsection{Multi-Trace Dynamics} \label{ssec:MULTI}

To access chaotic dynamics and have a semi-classical dual, we
consider a broader class of \textquotedblleft multi-trace\textquotedblright%
\ operators. In comparing the parent and
orbifold theory, the operator Tr$(X^{J})$ invariant under the orbifold group
action will have $J=WL$, where $W\in\mathbb{N}$ denotes the \textquotedblleft winding number\textquotedblright\ around the
quiver. In this setting, there is automatically some operator mixing with
multi-trace operators of the form:%
\begin{equation}
\text{Tr}(X^{J_{1}})...\text{Tr}(X^{J_{M}})\sim\left\vert J_{1},...,J_{M}%
\right\rangle
\end{equation}
where $J_{m}=w_{m}L$ and $w_{1}+...+w_{M}=W$, i.e., a partition of $W$. In the
absence of any impurities, these operators are all $1/2$ BPS, and have
exactly the same scaling dimension. Observe also that the multi-trace
operators are not completely orthogonal; their overlap is dictated by $1/N$
counting, i.e., in the overlap between $\left\vert J_{1},...,J_{M}%
\right\rangle $ and $\left\vert J_{1}^{\prime},...,J_{M^{\prime}}^{\prime
}\right\rangle $ (in the obvious notation)\ we have:%
\begin{equation}
\left\langle J_{1}^{\prime},...,J_{M^{\prime}}^{\prime}|J_{1},...,J_{M}%
\right\rangle =\frac{1}{N^{\vert M-M^{\prime} \vert}},
\end{equation}
which is an exact statement due to each operator being BPS. Observe that we
can always consider a suitable orthonormal basis for these $1/2$ BPS states.
This procedure was carried out for $\mathcal{N}=4$ SYM\ in
\cite{Corley:2001zk}, and similar considerations apply here (see also
\cite{McLoughlin:2020zew}).

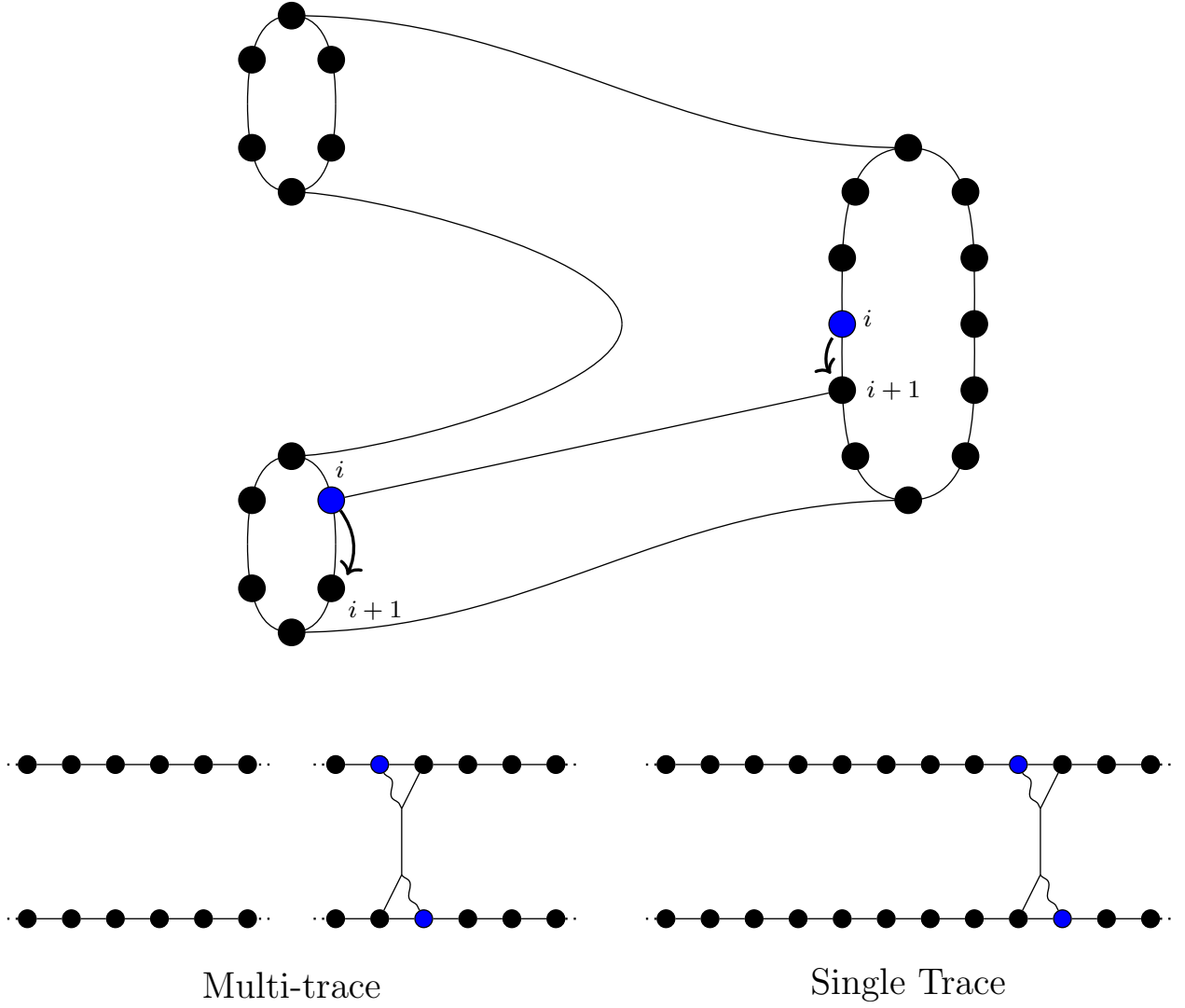
\begin{figure}[httbp]
\centering
\scalebox{1.25}{
\begin{tikzpicture}
	\begin{pgfonlayer}{nodelayer}
		\node [style=none] (0) at (-3.5, 3.5) {};
		\node [style=none] (1) at (-3.5, 1.5) {};
		\node [style=none] (6) at (-3, 2.5) {};
		\node [style=none] (7) at (-4, 2.5) {};
		\node [style=none] (8) at (-3.5, -1.5) {};
		\node [style=none] (9) at (-3.5, -3.5) {};
		\node [style=none] (10) at (-3, -2.5) {};
		\node [style=none] (11) at (-4, -2.5) {};
		\node [style=none] (12) at (3.5, 2) {};
		\node [style=none] (13) at (3.5, -2) {};
		\node [style=none] (14) at (4.25, 0) {};
		\node [style=none] (15) at (2.75, 0) {};
		\node [style=none] (16) at (0.25, 0) {};
		\node [draw, circle, fill=blue, minimum size=3mm] (17) at (-3.05, -2) {};
		\node [draw, circle, fill=black, minimum size=3mm] (18) at (2.75, -0.75) {};
		\node [draw, circle, fill=black, minimum size=3mm] (19) at (-3.5, -1.5) {};
		\node [draw, circle, fill=black, minimum size=3mm] (20) at (-3.95, -2) {};
		\node [draw, circle, fill=black, minimum size=3mm] (21) at (-3.5, -3.5) {};
		\node [draw, circle, fill=black, minimum size=3mm] (22) at (-3.05, -3) {};
		\node [draw, circle, fill=black, minimum size=3mm] (23) at (-3.95, -3) {};
		\node [draw, circle, fill=black, minimum size=3mm] (24) at (-3.5, 3.5) {};
		\node [draw, circle, fill=black, minimum size=3mm] (25) at (-3.5, 1.5) {};
		\node [draw, circle, fill=black, minimum size=3mm] (26) at (-3.05, 2) {};
		\node [draw, circle, fill=black, minimum size=3mm] (27) at (-3.95, 2) {};
		\node [draw, circle, fill=black, minimum size=3mm] (28) at (-3.05, 3) {};
		\node [draw, circle, fill=black, minimum size=3mm] (29) at (-3.95, 3) {};
		\node [draw, circle, fill=blue, minimum size=3mm] (30) at (2.75, 0) {};
		\node [draw, circle, fill=black, minimum size=3mm] (31) at (3.5, 2) {};
		\node [draw, circle, fill=black, minimum size=3mm] (32) at (4.25, 0) {};
		\node [draw, circle, fill=black, minimum size=3mm] (33) at (3.5, -2) {};
		\node [draw, circle, fill=black, minimum size=3mm] (34) at (4.25, -0.75) {};
		\node [draw, circle, fill=black, minimum size=3mm] (35) at (2.75, 0.75) {};
		\node [draw, circle, fill=black, minimum size=3mm] (36) at (4.25, 0.75) {};
		\node [draw, circle, fill=black, minimum size=3mm] (37) at (2.9, 1.5) {};
		\node [draw, circle, fill=black, minimum size=3mm] (38) at (4.15, 1.5) {};
		\node [draw, circle, fill=black, minimum size=3mm] (39) at (4.15, -1.5) {};
		\node [draw, circle, fill=black, minimum size=3mm] (40) at (2.9, -1.5) {};
		\node [style=none, font={\scriptsize}] (41) at (-2.95, -1.65) {$i$};
		\node [style=none, font={\scriptsize}] (42) at (-2.55, -3.25) {$i+1$};
		\node [style=none] (43) at (-2.87, -2.85) {};
		\node [style=none] (44) at (2.75, -0.75) {};
		\node [style=none, font={\scriptsize}, anchor=west] (45) at (2.98, 0.07) {$i$};
		\node [style=none, font={\scriptsize}, anchor=west] (46) at (3.02, -0.75) {$i+1$};
		\node [draw, circle, fill=black, minimum size=2mm] (47) at (-6, -5) {};
		\node [draw, circle, fill=black, minimum size=2mm] (48) at (-6.5, -5) {};
		\node [draw, circle, fill=black, minimum size=2mm] (49) at (-5.5, -5) {};
		\node [draw, circle, fill=black, minimum size=2mm] (50) at (-5, -5) {};
		\node [draw, circle, fill=black, minimum size=2mm] (51) at (-4.5, -5) {};
		\node [draw, circle, fill=black, minimum size=2mm] (52) at (-4, -5) {};
		\node [draw, circle, fill=black, minimum size=2mm] (53) at (-3, -5) {};
		\node [draw, circle, fill=blue, minimum size=2mm] (54) at (-2.5, -5) {};
		\node [draw, circle, fill=black, minimum size=2mm] (55) at (-2, -5) {};
		\node [draw, circle, fill=black, minimum size=2mm] (56) at (-1.5, -5) {};
		\node [draw, circle, fill=black, minimum size=2mm] (57) at (-1, -5) {};
		\node [draw, circle, fill=black, minimum size=2mm] (58) at (-0.5, -5) {};
		\node [style=none] (59) at (-3.75, -5) {};
		\node [style=none] (60) at (-3.25, -5) {};
		\node [style=none] (61) at (-6.75, -5) {};
		\node [style=none] (62) at (-0.25, -5) {};
		\node [draw, circle, fill=black, minimum size=2mm] (63) at (-6, -6.75) {};
		\node [draw, circle, fill=black, minimum size=2mm] (64) at (-6.5, -6.75) {};
		\node [draw, circle, fill=black, minimum size=2mm] (65) at (-5.5, -6.75) {};
		\node [draw, circle, fill=black, minimum size=2mm] (66) at (-5, -6.75) {};
		\node [draw, circle, fill=black, minimum size=2mm] (67) at (-4.5, -6.75) {};
		\node [draw, circle, fill=black, minimum size=2mm] (68) at (-4, -6.75) {};
		\node [draw, circle, fill=black, minimum size=2mm] (69) at (-3, -6.75) {};
		\node [draw, circle, fill=black, minimum size=2mm] (70) at (-2.5, -6.75) {};
		\node [draw, circle, fill=blue, minimum size=2mm] (71) at (-2, -6.75) {};
		\node [draw, circle, fill=black, minimum size=2mm] (72) at (-1.5, -6.75) {};
		\node [draw, circle, fill=black, minimum size=2mm] (73) at (-1, -6.75) {};
		\node [draw, circle, fill=black, minimum size=2mm] (74) at (-0.5, -6.75) {};
		\node [style=none] (75) at (-3.75, -6.75) {};
		\node [style=none] (76) at (-3.25, -6.75) {};
		\node [style=none] (77) at (-6.75, -6.75) {};
		\node [style=none] (78) at (-0.25, -6.75) {};
		\node [draw, circle, fill=black, minimum size=2mm] (79) at (3.25, -5) {};
		\node [draw, circle, fill=black, minimum size=2mm] (80) at (3.75, -5) {};
		\node [draw, circle, fill=black, minimum size=2mm] (81) at (4.25, -5) {};
		\node [draw, circle, fill=blue, minimum size=2mm] (82) at (4.75, -5) {};
		\node [draw, circle, fill=black, minimum size=2mm] (83) at (2.75, -5) {};
		\node [draw, circle, fill=black, minimum size=2mm] (84) at (2.25, -5) {};
		\node [draw, circle, fill=black, minimum size=2mm] (85) at (1.75, -5) {};
		\node [draw, circle, fill=black, minimum size=2mm] (86) at (1.25, -5) {};
		\node [draw, circle, fill=black, minimum size=2mm] (87) at (5.25, -5) {};
		\node [draw, circle, fill=black, minimum size=2mm] (88) at (5.75, -5) {};
		\node [draw, circle, fill=black, minimum size=2mm] (89) at (0.75, -5) {};
		\node [draw, circle, fill=black, minimum size=2mm] (90) at (6.25, -5) {};
		\node [style=none] (91) at (0.5, -5) {};
		\node [style=none] (92) at (6.5, -5) {};
		\node [draw, circle, fill=black, minimum size=2mm] (93) at (3.25, -6.75) {};
		\node [draw, circle, fill=black, minimum size=2mm] (94) at (3.75, -6.75) {};
		\node [draw, circle, fill=black, minimum size=2mm] (95) at (4.25, -6.75) {};
		\node [draw, circle, fill=black, minimum size=2mm] (96) at (4.75, -6.75) {};
		\node [draw, circle, fill=black, minimum size=2mm] (97) at (2.75, -6.75) {};
		\node [draw, circle, fill=black, minimum size=2mm] (98) at (2.25, -6.75) {};
		\node [draw, circle, fill=black, minimum size=2mm] (99) at (1.75, -6.75) {};
		\node [draw, circle, fill=black, minimum size=2mm] (100) at (1.25, -6.75) {};
		\node [draw, circle, fill=blue, minimum size=2mm] (101) at (5.25, -6.75) {};
		\node [draw, circle, fill=black, minimum size=2mm] (102) at (5.75, -6.75) {};
		\node [draw, circle, fill=black, minimum size=2mm] (103) at (0.75, -6.75) {};
		\node [draw, circle, fill=black, minimum size=2mm] (104) at (6.25, -6.75) {};
		\node [style=none] (105) at (0.5, -6.75) {};
		\node [style=none] (106) at (6.5, -6.75) {};
		\node [style=none] (107) at (-2.25, -5.5) {};
		\node [style=none] (108) at (-2.25, -6.25) {};
		\node [style=none] (109) at (5, -5.5) {};
		\node [style=none] (110) at (5, -6.25) {};
		\node [style=none] (111) at (-3.5, -7.5) {Multi-trace};
		\node [style=none] (112) at (3.5, -7.5) {Single Trace};
	\end{pgfonlayer}
	\begin{pgfonlayer}{edgelayer}
		\draw [in=90, out=0] (0.center) to (6.center);
		\draw [in=0, out=-90] (6.center) to (1.center);
		\draw [in=-90, out=180] (1.center) to (7.center);
		\draw [in=180, out=90] (7.center) to (0.center);
		\draw [in=90, out=0] (8.center) to (10.center);
		\draw [in=0, out=-90] (10.center) to (9.center);
		\draw [in=-90, out=180] (9.center) to (11.center);
		\draw [in=180, out=90] (11.center) to (8.center);
		\draw [in=90, out=0] (12.center) to (14.center);
		\draw [in=0, out=-90] (14.center) to (13.center);
		\draw [in=-90, out=180] (13.center) to (15.center);
		\draw [in=180, out=90] (15.center) to (12.center);
		\draw [in=180, out=0] (0.center) to (12.center);
		\draw [in=180, out=0] (9.center) to (13.center);
		\draw [in=-90, out=0, looseness=0.50] (8.center) to (16.center);
		\draw [in=90, out=0, looseness=0.50] (1.center) to (16.center);
		\draw (17) to (18);
		\draw [->, line width=1pt, bend left] (17) to (43.center);
		\draw [->, line width=1pt, bend right=35, shorten <=1pt, shorten >=2pt] (30) to (18);
		\draw (48) to (47);
		\draw (47) to (49);
		\draw (49) to (50);
		\draw (50) to (51);
		\draw (51) to (52);
		\draw (53) to (54);
		\draw (54) to (55);
		\draw (55) to (56);
		\draw (56) to (57);
		\draw (57) to (58);
		\draw [style=DottedLine] (59.center) to (52);
		\draw [style=DottedLine] (60.center) to (53);
		\draw [style=DottedLine] (48) to (61.center);
		\draw [style=DottedLine] (58) to (62.center);
		\draw (64) to (63);
		\draw (63) to (65);
		\draw (65) to (66);
		\draw (66) to (67);
		\draw (67) to (68);
		\draw (69) to (70);
		\draw (70) to (71);
		\draw (71) to (72);
		\draw (72) to (73);
		\draw (73) to (74);
		\draw [style=DottedLine] (75.center) to (68);
		\draw [style=DottedLine] (76.center) to (69);
		\draw [style=DottedLine] (64) to (77.center);
		\draw [style=DottedLine] (74) to (78.center);
		\draw (89) to (86);
		\draw (86) to (85);
		\draw (85) to (84);
		\draw (84) to (83);
		\draw (83) to (79);
		\draw (79) to (80);
		\draw (80) to (81);
		\draw (81) to (82);
		\draw (82) to (87);
		\draw (87) to (88);
		\draw (88) to (90);
		\draw [style=DottedLine] (89) to (91.center);
		\draw [style=DottedLine] (90) to (92.center);
		\draw (103) to (100);
		\draw (100) to (99);
		\draw (99) to (98);
		\draw (98) to (97);
		\draw (97) to (93);
		\draw (93) to (94);
		\draw (94) to (95);
		\draw (95) to (96);
		\draw (96) to (101);
		\draw (101) to (102);
		\draw (102) to (104);
		\draw [style=DottedLine] (103) to (105.center);
		\draw [style=DottedLine] (104) to (106.center);
		\draw [decorate, decoration={snake, amplitude=0.45mm}] (54) to (107.center);
		\draw (107.center) to (55);
		\draw (107.center) to (108.center);
		\draw (108.center) to (70);
		\draw [decorate, decoration={snake, amplitude=0.45mm}] (108.center) to (71);
		\draw [decorate, decoration={snake, amplitude=0.45mm}] (82) to (109.center);
		\draw (109.center) to (87);
		\draw (109.center) to (110.center);
		\draw (110.center) to (96);
		\draw [decorate, decoration={snake, amplitude=0.45mm}] (110.center) to (101);
	\end{pgfonlayer}
\end{tikzpicture}
}
\caption{Depiction of operator mixing in multi-trace sectors. There can be hopping within a single
spin chain, as well as splitting and joining with hopping to nearest neighbors on another spin chain. Provided all spin chains are sufficiently large, the one-loop approximation still holds, even in the holographic limit of large 't Hooft coupling. However, there are a large number of smaller spin chains where a perturbative analysis is inadequate. In this figure the impurity (blue) is a derivative insertion that can mix via nearest neighbor interactions governed by the leading order Feynman diagram. See Appendix \ref{app:SpinChainComputation} for further details.
}
\label{fig:Multitrace}
\end{figure}

What happens when we include impurities? Provided we are still working in the
limit of a small number of impurities, we still have a large charge expansion
available, and so can in principle proceed much as we did before. The
complication is that in the semi-classical holographic limit with multi-trace
sectors, we will also need to entertain values of $J_{i}\sim L$. In these
cases, the 't Hooft coupling is large enough that any one-loop approximation
will break down. See figure \ref{fig:Multitrace} for a depiction of these multi-trace effects.

Our aim in the remainder of this section
will be to first consider a simple illustrative example where the value of $J$
remains close to $L$, i.e., where the one-loop approximation is only valid
provided we have sufficiently small 't Hooft coupling, i.e., away from a
semi-classical gravity dual. We then turn to the case where the 't Hooft
coupling is large enough to have a semi-classical gravity dual. In this case,
the spin chain approximation still characterizes some, though not all of the
operator mixing.

\subsubsection{Small 't Hooft Coupling}

As a warmup, let us first consider the case of multi-trace dynamics with small
't Hooft coupling. To illustrate the main idea, it already suffices to
consider the case of two traces in the parent gauge theory, i.e., we focus on
impurities inserted into the operators Tr$(X^{2L})$ and Tr$(X^{L})$Tr$(X^{L}%
)$. The first operator can be viewed as the ground state for a single long
spin chain of length $2L$, while the second can be viewed as a pair of spin
chains, each of length $L$. Suppose we now insert impurities of the sort
already considered in the previous sections. We assume we have inserted a
sufficient number to make \textquotedblleft generic\textquotedblright%
\ statements, i.e., that an impurity inserted at locations separated by $L$
spin chain sites in Tr$(X^{2L})$ are distinguishable.\footnote{This would fail
in the strictly one impurity case, but can be addressed straightforwardly. For
ease of exposition we opt to focus on the generic situation.}

We now turn to operator mixing in this basis. In the one-loop approximation,
we still observe nearest neighbor hopping within a given spin chain. This
yields block diagonal contributions to the dilatation operator, i.e.,
$\Delta^{(2L)}$, the spin chain Hamiltonian for impurities inserted into
Tr$(X^{2L})$, and $\Delta^{(L)}$, the spin chain Hamiltonian for impurities
inserted into Tr$(X^{L})$. There are also off-diagonal blocks in which
impurities inserted in Tr$(X^{2L})$ transition to impurities inserted into
Tr$(X^{L})$Tr$(X^{L})$. To characterize these transitions, observe that we can
reference the location of an impurity inserted inside the quiver, i.e., it has
a designated position $i$, where this position coordinate ranges from
$i=1,...,L$. In Tr$(X^{L})$, this is simply the location in the spin chain,
whereas in Tr$(X^{2L})$, there is in principle an ambiguity since the spin
chain has $2L$ sites, i.e., there is also a \textquotedblleft winding
number\textquotedblright\ around the quiver. Nevertheless, by inspection of
the operator mixing, we observe that the single impurity hopping term
contribution to the Hamiltonian now splits up as:%
\begin{equation}
\Delta_{\text{single}}=\left[
\begin{array}
[c]{ccc}%
\Delta^{(L)} & \varepsilon & \frac{1}{N}\Delta^{(L)}\\
\varepsilon & \Delta^{(L)} & \frac{1}{N}\Delta^{(L)}\\
\frac{1}{N}\Delta^{(L)} & \frac{1}{N}\Delta^{(L)} & \Delta^{(L)}%
\end{array}
\right]  .
\end{equation}
where the upper left $2L\times2L$ block is strictly not quite block diagonal;
there is instead a term $\varepsilon$ which connects site $L$ to $L+1$; this
is a subleading effect at large $L$, however. The approximate tensor product
structure is manifestly:%
\begin{equation}
\Delta_{\text{single}}\simeq\Delta^{(L)}\otimes\left[
\begin{array}
[c]{ccc}%
1 & 0 & 1/N\\
0 & 1 & 1/N\\
1/N & 1/N & 1
\end{array}
\right]  .
\end{equation}
In particular, the second factor of the tensor product is independent of the
't Hooft coupling, and diagonalizing it amounts to simply picking an
orthonormal basis for the BPS\ operators Tr$(X^{2L})$ and Tr$(X^{L})$%
Tr$(X^{L})$ anyway. As such, we conclude that the spin chain Hamiltonian
approach naturally extends to the case $J=2L$ as well. Provided we can work in
a regime where $\lambda/L^{2}$ remains small, we then have a controlled
chaotic system, even including multi-trace dynamics. As already remarked,
however, the semi-classical gravity dual requires the opposite regime where
$\lambda/L^{4}$ is large. We therefore turn to this situation next.

\subsubsection{Large 't Hooft Coupling}

So long as we remain at comparatively small 't Hooft coupling, our analysis of
chaotic dynamics provides a controlled subsector for operator mixing in a
broad class of 4D\ SCFTs. On the other hand, it is also natural to ask whether
we can maintain control over these dynamics at larger 't Hooft coupling, to
make contact with holography. More precisely, we need to take the 't Hooft
coupling of the parent theory sufficiently large such that $\lambda/L^{4}\gg1$
so that we have a valid semi-classical gravity dual.

Our nearest neighbor approximation for hopping on a chaotic spin chain will
remain valid provided we work with a sufficiently large spin chain, i.e.,
operators of the form Tr$(X^{J})$ for $J=WL$ with $W$ sufficiently large. At
the very least, we require $\lambda/J^{2}\ll1$, but as already remarked, the
high winding number now requires us to inevitably deal with multi-trace
states. Observe that there is always a ground state sector $\left(
\text{Tr}(X^{L})\right)  ^{W}$, so there will always be a part of the spectrum
which has a high degree of operator mixing; $L$ can be \textquotedblleft
large\textquotedblright\ but still needs to be sufficiently small so that we
retain a semi-classical gravity dual.

To gain further insight into this case, we study the \textquotedblleft
typical\textquotedblright\ state at large $W$. Since we have a large number of
possible partitions available to us, we begin by recalling some basic features
of random partitions with the uniform distribution. First of all, recall that
in a partition into $M$ parts, the average size of $M$, and the standard
deviation are \cite{10.1215/S0012-7094-41-00826-8}:\footnote{Following \cite{10.1215/S0012-7094-41-00826-8}, the number of parts $M$ of a uniformly random partition of $W$ is
\begin{equation}
M = \frac{\sqrt{W}}{C}\log W + x\sqrt{W} \equiv \mu_M + x\sqrt{W},
\end{equation}
with $C=\pi\sqrt{2/3}$, and where the fluctuation variable $x$ has limiting CDF
\begin{equation}
    F(x)=\exp\left\{-\tfrac{2}{C}e^{-Cx/2}\right\}.
\end{equation}
I.e., $x$ is a Gumbel variable of scale $\beta = 2/C = \sqrt{6}/\pi$. We have that
\begin{equation}
\mu_M = \frac{\sqrt{W}}{C}\log W  = \frac{\sqrt{6W}}{2\pi}\log W,
\end{equation}
to leading order. The standard deviation of $x$ is $\sigma_x = \left(\pi/\sqrt{6}\right)\,\beta = 1$. Since $W$ is fixed, $\sigma_M = \sqrt{W}\,\sigma_x = \sqrt{W}$, so that $\mu_M \propto \sqrt{W}\log W$ and $\sigma_M \propto \sqrt{W}$ to leading order.}
\begin{equation}
\mu_{M}\propto\sqrt{W}\log W+...\text{ \ \ \ and \ \ }\sigma_{M}\propto
\sqrt{W}+...,
\end{equation}
where the \textquotedblleft...\textquotedblright\ denote subleading
contributions. In other words, the distribution experiences large
fluctuations, though we do have $\sigma_{M}/\mu_{M}\ll1$. The typical size of
the length of a spin chain ground state can now be extracted via the average:%
\begin{equation}
\overline{w}\pm\delta w=\frac{1}{M}\underset{m=1}{\overset{M}{\sum}}%
w_{m}=\frac{W}{M}\simeq\frac{W}{\mu_{M}\pm\sigma_{M}}\sim\frac{W}{\mu_{M}%
}\left(  1\pm\frac{\sigma_{M}}{\mu_{M}}\right)  .
\end{equation}
To use the spin chain approximation for the typical value $\overline{w}$, and
to also have a semi-classical gravity dual, we thus require:%
\begin{equation}
\frac{\lambda}{(\overline{w}L)^{2}}\ll1\ll\frac{\lambda}{L^{4}}%
\end{equation}
or:%
\begin{equation}
L\ll\frac{\sqrt{W}}{\log W}.
\end{equation}
At sufficiently large $W$, then, we have a chaotic spin chain sector where we
can still control the chaos. Inevitably, however, there are multi-trace
sectors where the effects of strong coupling lead to non-local interactions
and large transitions in the spin chain. While we also expect chaotic dynamics
to be in play for such multi-trace operators, it is hard to quantify the
operator mixing in this regime. Note also that mixing between this single
particle spin chain sector with Tr$(X^{LW})$ is naively suppressed by large
powers of $1/N$. Since, however, we must now contemplate all orders of the 't
Hooft coupling (including non-perturbative contributions) it is unclear to us
whether planarity affords much computational control in this regime.

In any case, we have now shown that even at very large values of $W$, some of
our chaotic spin chain analysis carries through, and that moreover, this
embeds in a semi-classical gravity dual. As one might expect, full control
over the operator mixing is more challenging here, a task we leave for future work.

\section{Conclusions} \label{sec:CONC}

In this work, we have developed a framework for studying quantum chaos directly in 4D SCFTs. We have identified a
closed subsector in which one-loop operator mixing is controlled by an effective spin chain, with the strength
of nearest neighbor interactions controlled by the marginal gauge couplings of the SCFT. By tuning the strength of these
marginal couplings, we have shown that this subsector exhibits chaotic dynamics, but more broadly realizes Anderson localization.
We identified this behavior by tracking the spectral statistics of the effective Hamiltonian in the dilute gas (i.e., low impurity) regime
for the spin chain. We also saw that complementary diagnostics for chaos such as Krylov complexity do not always distinguish
between chaotic behavior and more generic (non-chaotic) phenomena. The excitations of the 
spin chain dictate the motion of a chaotic billiard in the extra-dimensional geometry. We have also established a holographic interpretation; this requires extending our analysis to include the splitting and joining of spin chain sectors. In the remainder of this section we discuss some avenues of future investigation.

We have presented a class of tuned Hamiltonians with off-diagonal entries controlled by powers of the chi-distribution, i.e. $b_i \sim \chi_{i \alpha_{0}(p) \beta}^p$. At large $p$ we saw that this behavior approaches the same universality class as random matrices drawn from a Gaussian ensemble. It would be interesting to study further whether as $p$ scales with $L$, the length of the spin chain, we do indeed approach the same universality class.

It would be interesting to further study the compactification of the $\mathcal N=2$ / AdS$_5$ defect down to supersymmetric quantum mechanics / AdS$_2$. In doing so, we anticipate a connection to the vast literature on diagnosing chaos from out-of-time-order correlators in $\mathrm{AdS}_2$ / CFT$_1$ (see \cite{Altland:2026tog} for a recent review).

It would be interesting to extend our analysis to include marginal irrelevant deformations of the SCFT. Provided the RG flow back to the conformal fixed point is sufficiently slow, this would produce an even larger arena to study chaotic behavior with an effective spin chain description, closely paralleling the conformal case. However, there is no well-defined dilatation operator generating scaling dimensions away from the conformal fixed point. As a result, the spin chain Hamiltonian can no longer be identified directly with the anomalous dimension matrix and is instead a more general mixing matrix. It would be informative to understand the details of this construction further.

Finally, the results of this paper suggest that there are certain paths through the parameter space of gauge couplings that lead to chaotic dynamics and other paths that lead to Anderson localization. This suggests that geometric phases accumulated along such paths may serve as an additional diagnostic of quantum chaos.\footnote{See for example \cite{Chen:2026vml}.} It would be interesting to study this possibility further and compare the results to predictions from spectral statistics and Krylov complexity. This may also help in finding other distributions that lead to chaotic dynamics. One reason to suspect this is possible is that the $\chi$ and power-of-$\chi$ distributions only explore a particular slicing of the generalized gamma distribution (see Appendix \ref{app:Distributions}), and we suspect that relaxing this restriction should lead to chaos more generally.

\section*{Acknowledgments}

We thank W. Chan,  M. Claassen, C. Cummings, C. Lawrie, F. Mantegazza, S.N. Meynet, M. Montero, and J. Sonner for helpful discussions. We thank V. Balasubramanian, T. McLoughlin, and J. Sonner for helpful comments on an earlier draft. AC, VC, and JJH thank the 2025 Simons Summer workshop for hospitality during part of this work. FB is partially supported by the Deutsche Forschungsgemeinschaft under Germany’s Excellence Strategy -- EXC 2121 ``Quantum Universe'' -- 390833306, the Collaborative Research Center --- SFB 1624 ``Higher Structures, Moduli Spaces, and Integrability'' --- 506632645, and by the German Research Foundation through a German-Israeli Project Cooperation (DIP) grant ``Holography and the Swampland''. The work of AC is supported by the Turkish Fulbright Commission's PhD grant. The work of VC is supported by an NSF Graduate Research Fellowship. The work of VC and JJH is supported by DOE (HEP) Award DE-SC0013528. The work of JJH is also supported by BSF grant 2022100 and a University Research Foundation grant at the University of Pennsylvania.


\appendix

\section{Operator Mixing / Spin Chains} \label{app:SpinChainComputation}

In this Appendix we review the one-loop anomalous-dimension matrices governing the one-impurity sector of the 4D $\mathcal N=2$ quiver gauge theory studied in the main text.

\subsection{\texorpdfstring{$SU(2)_R$}{SU(2)R} Sector} \label{app:SU(2)RImpurity}

We begin with the $SU(2)_R$ impurity sector. The starting point is the 4D $\mathcal{N} = 2$ quiver gauge theory with all gauge groups $SU(N)$, which we index as $G_{i} = SU(N_i)$ for $i = 1,...,L$. I.e., we have a linear quiver with gauge groups $SU(N)$ at each interior node and bifundamental hypermultiplets connecting adjacent nodes:
\begin{equation}
[G_0]-G_1-\cdots-G_{L-1}-[G_{L}] .
\end{equation}
This is the open linear quiver where $G_1,\dots,G_{L-1}$ are dynamical gauge groups and $G_0,G_L$ are flavor groups. Each link carries bifundamental scalars $X_i$ and $Y_i^\dagger$ which transform under the group $G_{i-1} \times G_i$. In $\mathcal N=1$ language the field content consists of a pair of chiral multiplets $X_i\oplus Y_i^\dagger$, and each gauge node carries an adjoint chiral multiplet $\Phi_i$ inside the $\mathcal N=2$ vector multiplet.

A key simplification comes from the $SU(2)_R$ symmetry of the $\mathcal N=2$ theory. The fields $(X_i,Y_i^\dagger)$ form a doublet under $SU(2)_R$, and one may therefore construct a highest-weight state given by the ``vacuum'' operator
\begin{equation}\label{vacuum}
O_{\rm pure}=\sqrt{Z_L}\,X_1X_2\cdots X_L .
\end{equation}
This operator lies in a protected short multiplet and has vanishing anomalous dimension.

The one-impurity sector is given by flipping one $X_j$ to a $Y_j^\dag$:
\begin{equation}
O_i=\sqrt{Z_L}X_1 X_2 \cdots X_{i-1}\,Y_i^\dagger X_{i+1}\cdots X_L .
\label{eq:SingleImpurityOperator}
\end{equation}
Observe that there is one protected linear combination as it comes from applying the $SU(2)_R$ lowering operator $J_- = J^{(1)}_{-} + ... + J^{(L)}_{-}$ to $O_{\rm pure}$:
\begin{equation}
J_- O_{\rm pure}\propto \sum_{i=1}^L O_i.
\end{equation}

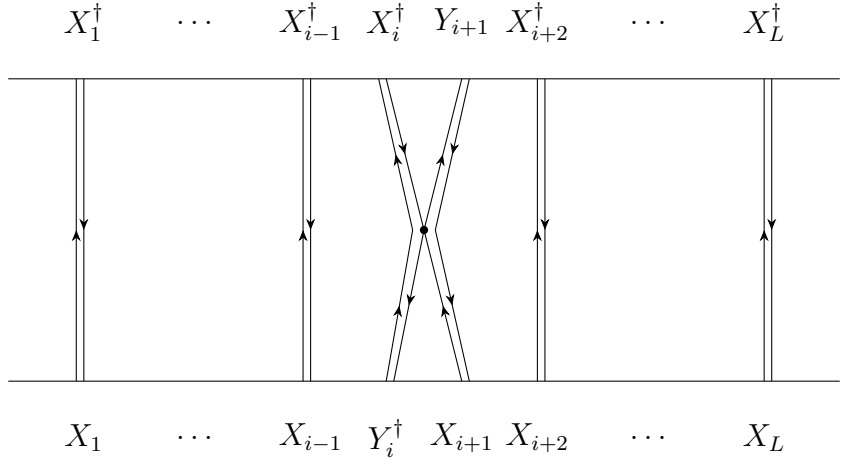
\begin{figure}[!t]
\centering
\begin{tikzpicture}
	\begin{pgfonlayer}{nodelayer}
		\node [style=none] (1) at (-6, 9) {};
		\node [style=none] (2) at (5, 9) {};
		\node [style=none] (3) at (-6, 5) {};
		\node [style=none] (4) at (5, 5) {};
		\node [style=none] (5) at (-5, 9.75) {$X_1^\dagger$};
		\node [style=none] (6) at (-5, 4.25) {$X_1$};
		\node [style=none] (7) at (-5, 9) {};
		\node [style=none] (7b) at (-5.1, 9) {};
		\node [style=none] (8) at (-5.1, 5) {};
		\node [style=none] (8b) at (-5, 5) {};
		\node [style=none] (9) at (5, 9.75) {};
		\node [style=none] (10) at (4, 9.75) {$X_L^\dagger$};
		\node [style=none] (11) at (4, 4.25) {$X_L$};
		\node [style=none] (12) at (4, 9) {};
		\node [style=none] (12b) at (4.1, 9) {};
		\node [style=none] (13) at (4.1, 5) {};
		\node [style=none] (13b) at (4, 5) {};
		\node [style=none] (14) at (-1, 9.75) {$X_{i}^\dagger$};
		\node [style=none] (15) at (0, 9.75) {$Y_{i+1}$};
		\node [style=none] (16) at (1, 9.75) {$X_{i+2}^\dagger$};

		\node [style=none] (Ab) at (-0.65, 7) {};
		\node [shape=circle, fill, inner sep=1.1pt] (A) at (-0.5, 7) {};
		\node [style=none] (Ac) at (-0.35, 7) {};

		\node [style=none] (17) at (-1, 4.25) {$Y_{i}^\dagger$};
		\node [style=none] (18) at (0, 4.25) {$X_{i+1}$};
		\node [style=none] (19) at (1, 4.25) {$X_{i+2}$};
		\node [style=none] (20b) at (1.1, 9) {};
		\node [style=none] (20) at (1, 9) {};
		\node [style=none] (21) at (1, 5) {};
		\node [style=none] (21b) at (1.1, 5) {};
		\node [style=none] (22) at (0, 5) {};
		\node [style=none] (22b) at (0.1, 5) {};
		\node [style=none] (23) at (0, 9) {};
		\node [style=none] (23b) at (0.1, 9) {};
		\node [style=none] (24) at (-1, 9) {};
		\node [style=none] (24b) at (-1.1, 9) {};
		\node [style=none] (25) at (-1, 5) {};
		\node [style=none] (25b) at (-0.9, 5) {};
		\node [style=none] (26) at (2.5, 9.75) {$\dots$};
		\node [style=none] (27) at (2.5, 4.25) {$\dots$};
		\node [style=none] (28) at (-3.5, 9.75) {$\dots$};
		\node [style=none] (29) at (-3.5, 4.25) {$\dots$};
		\node [style=none] (30) at (-2.1, 9) {};
		\node [style=none] (30b) at (-2, 9) {};
		\node [style=none] (31) at (-2, 5) {};
		\node [style=none] (31b) at (-2.1, 5) {};
		\node [style=none] (32) at (-2, 9.75) {$X_{i-1}^\dagger$};
		\node [style=none] (33) at (-2, 4.25) {$X_{i-1}$};
	\end{pgfonlayer}
	\begin{pgfonlayer}{edgelayer}
		\draw (1.center) to (2.center);
		\draw (3.center) to (4.center);

		\draw[mid arrow] (7.center) to (8b.center);
		\draw[mid arrow] (8.center) to (7b.center);

		\draw[mid arrow] (13b.center) to (12.center);
		\draw[mid arrow] (12b.center) to (13.center);

		\draw[mid arrow] (21.center) to (20.center);
		\draw[mid arrow] (20b.center) to (21b.center);

		\draw[mid arrow] (31b.center) to (30.center);
		\draw[mid arrow] (30b.center) to (31.center);

		\draw[mid arrow] (Ab.center) to (24b.center);
		\draw[mid arrow] (24.center) to (A.center);
		\draw[mid arrow] (22.center) to (A.center);
		\draw[mid arrow] (Ac.center) to (22b.center);
		\draw[mid arrow] (25.center) to (Ab.center);
		\draw[mid arrow] (A.center) to (25b.center);
		\draw[mid arrow] (A.center) to (23.center);
		\draw[mid arrow] (23b.center) to (Ac.center);
	\end{pgfonlayer}
\end{tikzpicture}
\caption{Feynman diagram associated to the leading order contribution to the correlator $\left<  \mathcal{O}^\dagger_{i+1}\mathcal{O}_{i}\right>$ generated by the F-term potential. Arrows flow from fundamental-representation indices towards anti-fundamental indices.}
\label{fig:HoppingDiagram}
\end{figure}

With this in hand, we now review the relevant computations of reference \cite{Baume:2020ure}, which show that the dynamics of the one-impurity sector of 4D $\mathcal{N}=2$ theories have a natural interpretation in terms of an XXX spin chain. In our conventions, 
the free scalar propagator is
\begin{equation}
\langle (X_i^\dagger)^{A_{i}}{}_{B_{i-1}}(x) (X_i)^{A_{i-1}}{}_{B_{i}}(0)\rangle= \frac{1}{4\pi^2}\frac{\delta^{A_{i}}{}_{B_{i}}   \delta^{A_{i-1}}{}_{B_{i-1}}}{|x|^{2\Delta_X}},\qquad\Delta_X=1 .
\end{equation}
Throughout we factor out the tree–level normalization
\begin{equation}\label{2-pt-function-anomalous-dimension}
\langle O_i^\dagger(x)O_j(0)\rangle=\frac{1}{|x|^{2\Delta_0}}\Big( \delta_{ij}-\gamma_{ij}\log(|x|^2\Lambda^2)+\cdots\Big),
\end{equation}
where $\Delta_0=L$ is the classical dimension. The relevant scalar interactions arise from the F–term potential
\begin{equation}
V_F=\frac{1}{(4\pi^2)^2}\sum_{i=1}^{L-1}\sum_{a\in\text{adj}(G_i)} |F_{i,a}|^2 ,
\end{equation}
with
\begin{equation}
F_{i,a}=\sqrt2\,g_i\Big(\Tr_{i+1}(Y_{i+1} T_{i,a} X_{i+1})-\Tr_{i-1}(X_{i}T_{i,a}Y_{i})\Big) .
\end{equation}
Define
\begin{equation}
A_i=\Tr_{i+1}(Y_{i+1} T_{i,a} X_{i+1}) \, ,\qquad B_i=\Tr_{i-1}(X_{i}T_{i,a}Y_{i}) \, ,
\end{equation}
so that
\begin{equation}
|F_{i,a}|^2=2g_i^2(A_i-B_i)(A_i^\dagger-B_i^\dagger).
\end{equation}
Expanding,
\begin{equation}
|F_{i,a}|^2=2g_i^2\Big(A_iA_i^\dagger+B_iB_i^\dagger-A_iB_i^\dagger-B_iA_i^\dagger
\Big).
\end{equation}
The cross–terms $A_iB_i^\dagger$ and $B_iA_i^\dagger$ generate mixing between operators with impurities on neighboring sites.\footnote{These terms have a direct physical interpretation: they correspond to the impurity hopping between neighboring sites in the operator chain. This is the microscopic origin of the tridiagonal structure of the anomalous-dimension matrix.} The leading contribution comes from the interaction vertex
\begin{equation}
2g_i^2\,A_i(z)B_i^\dagger(z).
\end{equation}
See figure \ref{fig:HoppingDiagram}. The operators contain the local structures
\begin{align}
O_{i+1}^\dagger(x)&\supset(X_{i}^\dagger)^{\alpha}{}_{\beta}(x)(Y_{i+1})^{u}{}_{v}(x),\\
O_{i}(0)&\supset(Y_{i}^\dagger)^{\gamma}{}_{\delta}(0)(X_{i+1})^{r}{}_{s}(0).
\end{align}
The relevant four Wick contractions are
\begin{align}
\langle Y_{i+1}(x)Y_{i+1}^\dagger(z)\rangle&\propto\frac{1}{|x-z|^{2\Delta_X}},\\
\langle X_{i}^\dagger(x)X_{i}(z)\rangle&\propto\frac{1}{|x-z|^{2\Delta_X}},\\
\langle Y_{i}(z)Y_{i}^\dagger(0)\rangle&\propto\frac{1}{|z|^{2\Delta_X}},\\
\langle X_{i+1}(z)X_{i+1}^\dagger(0)\rangle&\propto\frac{1}{|z|^{2\Delta_X}} .
\end{align}
Thus the scalar propagators give
\begin{equation}
\frac{1}{|x-z|^{4\Delta_X}|z|^{4\Delta_X}} .
\end{equation}

The propagators also produce Kronecker deltas that identify the fundamental indices of the two generators $T_a$. Schematically the color structure becomes
\begin{equation}
\sum_a(T_a)^m{}_n(T_a)^p{}_q\delta^n{}_p\delta^q{}_m .
\end{equation}
The delta functions collapse the indices,
\begin{equation}
\sum_a(T_a)^m{}_n(T_a)^n{}_m=\sum_a\Tr_{\rm fund}(T_aT_a).
\end{equation}
Dividing by the fundamental dimension $d_F$ yields
\begin{equation}
\widetilde C_i=\frac{1}{d_F}\sum_{a\in\text{adj}(G_i)}\Tr_{\rm fund}(T_aT_a).
\end{equation}
For $G_i=SU(N)$,\footnote{Comparing with the notation and conventions of references \cite{Baume:2020ure, Baume:2022cot}, we note that the present expression is larger by a factor of $N$. This is simply because the ``gauge coupling'' of those references is simply the 't Hooft coupling $\lambda = g_{i}^2 N$, whereas here we have opted to work in terms of $g_i^2$. A further comment is that in the gravity dual estimate for the ``gauge coupling'' the backreaction from the gauge theory branes is already taken into account in the conventions of \cite{Baume:2020ure, Baume:2022cot}.}
\begin{equation}
\widetilde C_i=\frac{N^2-1}{2N}.
\end{equation}

The full correlator is written relative to the tree–level normalization
\begin{equation}
\frac{1}{|x|^{2\Delta_0}}.
\end{equation}
At tree level two propagators would connect the impurity fields directly from $x$ to $0$,
\begin{equation}
\frac{1}{|x|^{2\Delta_X}}\frac{1}{|x|^{2\Delta_X}}=\frac{1}{|x|^{4\Delta_X}} .
\end{equation}
In the interacting diagram these are replaced by
\begin{equation}
\frac{1}{|x-z|^{2\Delta_X}}\frac{1}{|z|^{2\Delta_X}}\quad\text{and}\quad\frac{1}{|x-z|^{2\Delta_X}}\frac{1}{|z|^{2\Delta_X}} .
\end{equation}
Relative to the tree–level normalization the correction therefore
contains the factor
\begin{equation}
\frac{|x|^{4\Delta_X}}{|x-z|^{4\Delta_X}|z|^{4\Delta_X}} .
\end{equation}
The resulting correction to the correlator is
\begin{equation}
\delta\langle O_{i+1}^\dagger(x)O_{i}(0)\rangle=\frac{1}{|x|^{2\Delta_0}}\frac{2g_i^2\widetilde C_i}{(4\pi^2)^2}\int d^4z\,\frac{|x|^{4\Delta_X}}{|x-z|^{4\Delta_X}|z|^{4\Delta_X}} .
\end{equation}
Following \cite{Berenstein:2002jq, Minahan:2002ve}, define:
\begin{equation}
I(x)=\int d^4z\frac{|x|^{4\Delta_X}}{|x-z|^{4\Delta_X}|z|^{4\Delta_X}} .
\end{equation}
For $D=4$ and $\Delta_X=1$ this integral is logarithmically divergent,
\begin{equation}
I(x)=\pi^2\log(|x|^2\Lambda^2)+\cdots .
\end{equation}
Thus
\begin{equation}
\langle O_{i+1}^\dagger(x)O_{i}(0)\rangle=\frac{1}{|x|^{2\Delta_0}}\left(1+\frac{g_i^2\widetilde C_i}{8\pi^2}\log(|x|^2\Lambda^2)+\cdots\right).
\end{equation}
Comparing with the definition of $\gamma_{ij}$ gives
\begin{equation}
\gamma_{i+1,i}=-\frac{g_i^2\widetilde C_i}{8\pi^2}.
\end{equation}

With this in hand, we turn our attention to the diagonal components of the anomalous dimension matrix. The diagonal correlator receives many contributions, including scalar interactions and vector boson exchange diagrams. Instead of computing all diagrams explicitly, one can determine the result using a symmetry argument. Consider again the operator
\begin{equation}
O_{\rm pure}=\sqrt{Z_L}\,X_1X_2\cdots X_L .
\end{equation}
This operator is a highest–weight state under the global $SU(2)_R$ symmetry and therefore belongs to a protected short multiplet. 
Since there is a protected descendant:
\begin{equation}
J_- O_{\rm pure}\propto\sum_{i=1}^L O_i 
\end{equation}
which lies in the same short multiplet as $O_{\rm pure}$, it must also have protected dimension. Therefore the anomalous–dimension matrix in the one–impurity sector must annihilate the vector $(1,1,\ldots,1)$.

The off–diagonal matrix elements are
\begin{equation}
\gamma_{i+1,i}=-\frac{g_i^2\widetilde C_i}{8\pi^2},\qquad\gamma_{i,i+1}=-\frac{g_{i}^2\widetilde C_{i}}{8\pi^2}.
\end{equation}
Requiring that
\begin{equation}
\sum_j \gamma_{ij}=0
\end{equation}
fixes the diagonal element,
\begin{equation}
\gamma_{ii}=\frac{g_{i-1}^2\widetilde C_{i-1}+g_{i}^2\widetilde C_{i}}{8\pi^2}.
\end{equation}
Thus
\begin{equation}
\langle O_i^\dagger(x)O_i(0)\rangle=\frac{1}{|x|^{2\Delta_0}}\left(1-\frac{g_{i-1}^2\widetilde C_{i-1}+g_{i}^2\widetilde C_{i}}{8\pi^2}\log(|x|^2\Lambda^2)+\cdots\right).
\end{equation}

The anomalous–dimension matrix therefore has the structure
\begin{equation}
\gamma_{ij}=\frac{\widetilde C}{8\pi^2}\left(- g_{i-1}^2 \delta_{i,j+1}+(g_{i-1}^2+g_{i}^2)\delta_{ij}-g_{i}^2 \delta_{i,j-1}\right),
\end{equation}
which is the lattice Laplacian for an open XXX spin chain. Written explicitly, we have
\begin{equation}\label{spin chain-hamiltonian-su2}
\gamma^{(\mathcal N=2)}=\frac{\widetilde{C}}{8\pi^2}
\begin{pmatrix}
g_1^2 & -g_1^2 & 0 & \cdots & 0\\
-g_1^2 & g_1^2+g_2^2 & -g_2^2 & \cdots & 0\\
0 & -g_2^2 & g_2^2+g_3^2 & \ddots & \vdots\\
\vdots & \vdots & \ddots & \ddots & -g_{L-1}^2\\
0 & 0 & \cdots & -g_{L-1}^2 & g_{L-1}^2
\end{pmatrix}.
\end{equation}
If all bulk couplings are equal, $g_i=g$, then
\begin{equation}
\gamma^{(\mathcal N=2)}=\frac{\widetilde{C}\,g^2}{8\pi^2}
\begin{pmatrix}
1 & -1 & 0 & \cdots & 0\\
-1 & 2 & -1 & \cdots & 0\\
0 & -1 & 2 & \ddots & \vdots\\
\vdots & \vdots & \ddots & \ddots & -1\\
0 & 0 & \cdots & -1 & 1
\end{pmatrix}\,,
\end{equation}
which is equivalent to the Hamiltonian of the one-impurity sector of the integrable Heisenberg spin chain. Finally, note that this is the case of the open linear quiver; the case of the closed periodic quiver follows exactly as above, and the anomalous dimension matrix is given by
\begin{equation}
\gamma^{(\mathcal N=2)}_\textrm{closed}=\frac{\widetilde{C}}{8\pi^2}
\begin{pmatrix}
g_{L}^2 + g_1^2 & -g_1^2 & 0 & \cdots & -g_L^2 \\
-g_1^2 & g_1^2 + g_2^2 & -g_2^2 & \cdots & 0\\
0 & -g_2^2 & g_2^2 + g_3^2 & \ddots & \vdots\\
\vdots & \vdots & \ddots & \ddots & -g_{L-1}^2 \\
-g_L^2 & 0 & \cdots & -g_{L-1}^2 & g_{L-1}^2 + g_{L}^2
\end{pmatrix}.
\end{equation}

\subsection{\texorpdfstring{$SL(2)$}{SL(2)} Sector} \label{app:SL(2)Impurity}

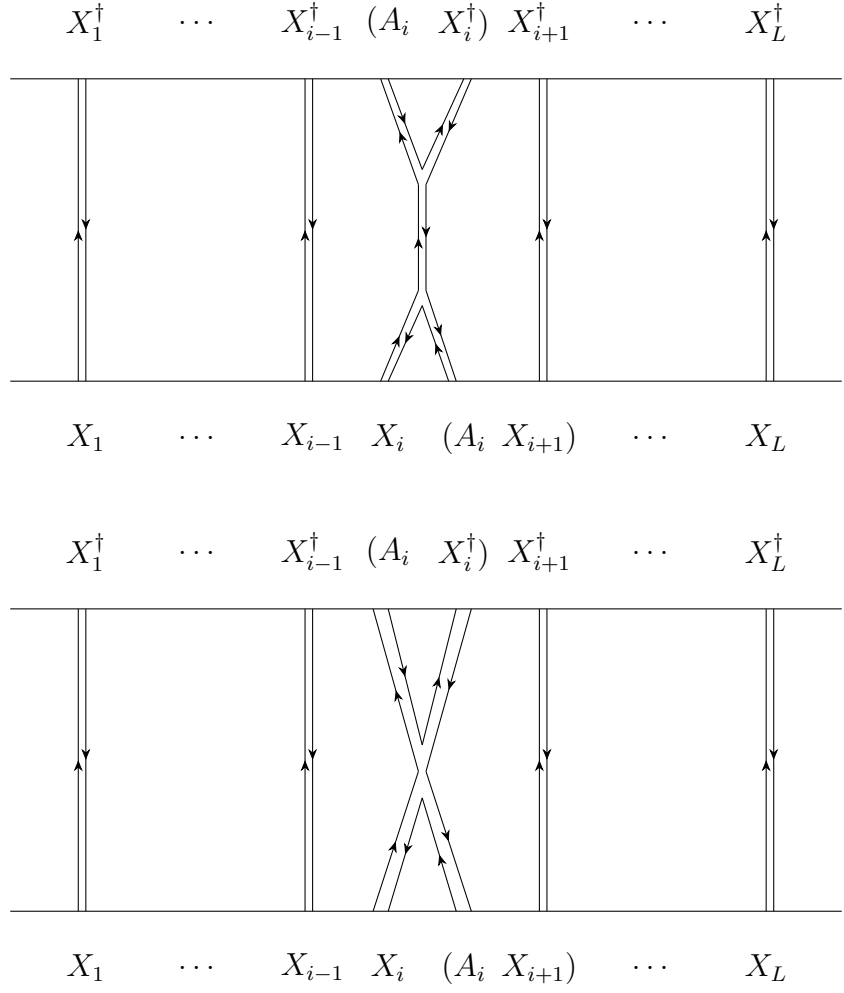
\begin{figure}[!t]
\centering
\begin{tikzpicture}
	\begin{pgfonlayer}{nodelayer}
		\node [style=none] (1) at (-6, 9) {};
		\node [style=none] (2) at (5, 9) {};
		\node [style=none] (3) at (-6, 5) {};
		\node [style=none] (4) at (5, 5) {};
		\node [style=none] (5) at (-5, 9.75) {$X_1^\dagger$};
		\node [style=none] (6) at (-5, 4.25) {$X_1$};
		\node [style=none] (7) at (-5, 9) {};
		\node [style=none] (7b) at (-5.1, 9) {};
		\node [style=none] (8) at (-5.1, 5) {};
		\node [style=none] (8b) at (-5, 5) {};
		\node [style=none] (9) at (5, 9.75) {};
		\node [style=none] (10) at (4, 9.75) {$X_L^\dagger$};
		\node [style=none] (11) at (4, 4.25) {$X_L$};
		\node [style=none] (12) at (4, 9) {};
		\node [style=none] (12b) at (4.1, 9) {};
		\node [style=none] (13) at (4.1, 5) {};
		\node [style=none] (13b) at (4, 5) {};
		\node [style=none] (14) at (-1, 9.75) {$(A_{i}$};
		\node [style=none] (15) at (0, 9.75) {$X_i^\dagger)$};
		\node [style=none] (16) at (1, 9.75) {$X_{i+1}^\dagger$};

		\node [style=none] (A) at (-0.5, 7.6) {};
		\node [style=none] (Ab) at (-0.6, 7.6) {};
		\node [style=none] (Ac) at (-0.55, 7.8) {};
		\node [style=none] (B) at (-0.5, 6.5) {};
		\node [style=none] (Bb) at (-0.6, 6.2) {};
		\node [style=none] (Bc) at (-0.55, 6.0) {};
		
		\node [style=none] (C) at (-0.50, 6.2) {};

		\node [style=none] (17) at (-1, 4.25) {$X_{i}$};
		\node [style=none] (18) at (0, 4.25) {$(A_{i}$};
		\node [style=none] (19) at (1, 4.25) {$X_{i+1})$};
		\node [style=none] (20) at (1, 9) {};
		\node [style=none] (20b) at (1.1, 9) {};
		\node [style=none] (20) at (1, 9) {};
		\node [style=none] (21) at (1, 5) {};
		\node [style=none] (21b) at (1.1, 5) {};
		\node [style=none] (22) at (-0.1, 5) {};
		\node [style=none] (22b) at (-0.2, 5) {};
		\node [style=none] (23) at (0, 9) {};
		\node [style=none] (23b) at (0.1, 9) {};
		\node [style=none] (24) at (-1, 9) {};
		\node [style=none] (24b) at (-1.1, 9) {};
		\node [style=none] (25) at (-1, 5) {};
		\node [style=none] (25b) at (-1.1, 5) {};
		\node [style=none] (26) at (2.5, 9.75) {$\dots$};
		\node [style=none] (27) at (2.5, 4.25) {$\dots$};
		\node [style=none] (28) at (-3.5, 9.75) {$\dots$};
		\node [style=none] (29) at (-3.5, 4.25) {$\dots$};
		\node [style=none] (30) at (-2.1, 9) {};
		\node [style=none] (30b) at (-2, 9) {};
		\node [style=none] (31) at (-2, 5) {};
		\node [style=none] (31b) at (-2.1, 5) {};
		\node [style=none] (32) at (-2, 9.75) {$X_{i-1}^\dagger$};
		\node [style=none] (33) at (-2, 4.25) {$X_{i-1}$};
	\end{pgfonlayer}
	\begin{pgfonlayer}{edgelayer}
		\draw (1.center) to (2.center);
		\draw (3.center) to (4.center);
		\draw[mid arrow] (7.center) to (8b.center);
		\draw[mid arrow] (8.center) to (7b.center);
		\draw[mid arrow] (13b.center) to (12.center);
		\draw[mid arrow] (12b.center) to (13.center);
		\draw[mid arrow] (21.center) to (20.center);
		\draw[mid arrow] (20b.center) to (21b.center);
		\draw[mid arrow] (31b.center) to (30.center);
		\draw[mid arrow] (30b.center) to (31.center);
		%
		\draw[mid arrow] (Ac.center) to (23.center);
		\draw[mid arrow] (24.center) to (Ac.center);
		\draw[mid arrow] (Ab.center) to (24b.center);
		\draw[mid arrow] (23b.center) to (A.center);
		%
		\draw[mid arrow] (22b.center) to (Bc.center);
		\draw[mid arrow] (Bc.center) to (25.center);
		\draw[mid arrow] (25b.center) to (Bb.center);
		\draw[mid arrow] (C.center) to (22.center);
		\draw[mid arrow] (A.center) to (C.center);
		\draw[mid arrow] (Bb.center) to (Ab.center);
	\end{pgfonlayer}
\end{tikzpicture} \\
\vspace{1cm}
\begin{tikzpicture}
	\begin{pgfonlayer}{nodelayer}
		\node [style=none] (1) at (-6, 9) {};
		\node [style=none] (2) at (5, 9) {};
		\node [style=none] (3) at (-6, 5) {};
		\node [style=none] (4) at (5, 5) {};
		\node [style=none] (5) at (-5, 9.75) {$X_1^\dagger$};
		\node [style=none] (6) at (-5, 4.25) {$X_1$};
		\node [style=none] (7) at (-5, 9) {};
		\node [style=none] (7b) at (-5.1, 9) {};
		\node [style=none] (8) at (-5.1, 5) {};
		\node [style=none] (8b) at (-5, 5) {};
		\node [style=none] (9) at (5, 9.75) {};
		\node [style=none] (10) at (4, 9.75) {$X_L^\dagger$};
		\node [style=none] (11) at (4, 4.25) {$X_L$};
		\node [style=none] (12) at (4, 9) {};
		\node [style=none] (12b) at (4.1, 9) {};
		\node [style=none] (13) at (4.1, 5) {};
		\node [style=none] (13b) at (4, 5) {};
		\node [style=none] (14) at (-1, 9.75) {$(A_{i}$};
		\node [style=none] (15) at (0, 9.75) {$X_i^\dagger)$};
		\node [style=none] (16) at (1, 9.75) {$X_{i+1}^\dagger$};

		\node [style=none] (A) at (-0.55, 6.5) {};

		\node [style=none] (Ab) at (-0.50, 6.85) {}; 
		\node [style=none] (Ad) at (-0.6, 6.85) {}; 

		\node [style=none] (Ac) at (-0.55, 7.20) {};

		\node [style=none] (17) at (-1, 4.25) {$X_{i}$};
		\node [style=none] (18) at (0, 4.25) {$(A_{i}$};
		\node [style=none] (19) at (1, 4.25) {$X_{i+1})$};
		\node [style=none] (20) at (1, 9) {};
		\node [style=none] (20b) at (1.1, 9) {};
		\node [style=none] (20) at (1, 9) {};
		\node [style=none] (21) at (1, 5) {};
		\node [style=none] (21b) at (1.1, 5) {};
		\node [style=none] (22) at (0.1, 5) {}; 
		\node [style=none] (22b) at (-0.1, 5) {};
		\node [style=none] (23) at (-0.1, 9) {};
		\node [style=none] (23b) at (0.1, 9) {};
		\node [style=none] (24) at (-1, 9) {};
		\node [style=none] (24b) at (-1.2, 9) {};
		\node [style=none] (25) at (-1, 5) {};
		\node [style=none] (25b) at (-1.2, 5) {};
		\node [style=none] (26) at (2.5, 9.75) {$\dots$};
		\node [style=none] (27) at (2.5, 4.25) {$\dots$};
		\node [style=none] (28) at (-3.5, 9.75) {$\dots$};
		\node [style=none] (29) at (-3.5, 4.25) {$\dots$};
		\node [style=none] (30) at (-2.1, 9) {};
		\node [style=none] (30b) at (-2, 9) {};
		\node [style=none] (31) at (-2, 5) {};
		\node [style=none] (31b) at (-2.1, 5) {};
		\node [style=none] (32) at (-2, 9.75) {$X_{i-1}^\dagger$};
		\node [style=none] (33) at (-2, 4.25) {$X_{i-1}$};
	\end{pgfonlayer}
	\begin{pgfonlayer}{edgelayer}
		\draw (1.center) to (2.center);
		\draw (3.center) to (4.center);

		\draw[mid arrow] (7.center) to (8b.center);
		\draw[mid arrow] (8.center) to (7b.center);

		\draw[mid arrow] (13b.center) to (12.center);
		\draw[mid arrow] (12b.center) to (13.center);

		\draw[mid arrow] (21.center) to (20.center);
		\draw[mid arrow] (20b.center) to (21b.center);

		\draw[mid arrow] (31b.center) to (30.center);
		\draw[mid arrow] (30b.center) to (31.center);
		
		\draw[mid arrow] (24.center) to (Ac.center);
		\draw[mid arrow] (Ac.center) to (23.center);

		\draw[mid arrow] (A.center) to (25.center);
		\draw[mid arrow] (22b.center) to (A.center);

		\draw[mid arrow] (Ad.center) to (24b.center);
		\draw[mid arrow] (25b.center) to (Ad.center);
	
		\draw[mid arrow] (Ab.center) to (22.center);
		\draw[mid arrow] (23b.center) to (Ab.center);
	\end{pgfonlayer}
\end{tikzpicture}
\caption{Feynman diagrams contributing to the off-diagonal correlator $\langle\mathcal{O}^\dagger_i\mathcal{O}_{i+1}\rangle$ of the $SL(2)$ sector via a gauge boson involved in covariant derivatives. Arrows flow from fundamental-representation indices towards anti-fundamental indices.}
\label{fig:HoppingDiagram-sl2}
\end{figure}

We now repeat the one-impurity analysis in the derivative, or $SL(2)$, sector. The operators of interest are obtained by inserting a single covariant derivative into the long bifundamental word,
\begin{equation}
    \mathcal{O}^{D}_{i}=\sqrt{Z_L}X_1 X_2 \cdots X_{i-1} (D X_i) X_{i+1}\cdots X_L ,
\end{equation}
where $D X$ is a covariant derivative with (suppressed) vector index pointing along a null (i.e., lightcone direction) of the 4D spacetime. 
There are two diagrams to consider, shown in figure \ref{fig:HoppingDiagram-sl2}. 
The important point is that its coupling dependence can be fixed without explicitly evaluating the loop integral.

To see this, let us write the color indices carefully. The bifundamental field $X_i$ transforms under the adjacent gauge groups $G_{i-1} \times G_{i}$, so we write
\begin{equation}
    (X_i)^{a_{i-1}}_{a_{i}} ,
\end{equation}
where $a_{i-1}$ is a fundamental index of $G_{i-1}$ and $a_{i}$ is an antifundamental index of $G_{i}$. With this convention,
\begin{equation}
    (D_\mu X_i)^{a_{i-1}}_{a_{i}} = \partial_\mu (X_i)^{a_{i-1}}_{a_{i}}- i g_{i-1} (A_{i-1,\mu})^{a_{i-1}}_{b_{i-1}}(X_i)^{b_{i-1}}_{a_{i}}
    + i g_{i}(X_i)^{a_{i-1}}_{b_{i}}(A_{i,\mu})^{b_{i}}_{a_{i}} \, .
\label{eq:CovariantDerivativeXi}
\end{equation}
Similarly,
\begin{equation}
    (D_\mu X_{i+1})^{a_{i}}_{a_{i+1}}=\partial_\mu (X_{i+1})^{a_{i}}_{a_{i+1}}- i g_{i}(A_{i,\mu})^{a_{i}}_{b_{i}}(X_{i+1})^{b_{i}}_{a_{i+1}}  + i g_{i+1}(X_{i+1})^{a_{i}}_{b_{i+1}} (A_{i+1,\mu})^{b_{i+1}}_{a_{i+1}} \, .
\label{eq:CovariantDerivativeXi+1}
\end{equation}
The off-diagonal matrix element $\gamma_{i,i+1}$ describes the hopping of the derivative impurity from $X_i$ to $X_{i+1}$. Therefore the relevant contraction comes from the gauge field appearing in both $D_\mu X_i$ and $D_\mu X_{i+1}$. From \eqref{eq:CovariantDerivativeXi} and \eqref{eq:CovariantDerivativeXi+1}, the only common gauge field is $A_{i,\mu}$. More explicitly, the relevant pair of terms is
\begin{equation}
    i g_{i}(X_i)^{a_{i-1}}_{b_{i}}(A_{i,\mu})^{b_{i}}_{a_{i}} \qquad \text{and} \qquad - i g_{i}(A_{i,\mu})^{a_{i}}_{c_{i}} (X_{i+1})^{c_{i}}_{a_{i+1}} .
\end{equation}
These two insertions have the correct $G_{i}$ color indices to be contracted by the gauge-field propagator. In the large-$N$ limit, the propagator is given by:
\begin{equation}
    \big\langle(A_{i,\mu})^{b}_{a}(x)(A_{i,\nu})^{c}_{d}(y) \big\rangle \propto \delta^{b}_{d}\,\delta^{c}_{a} \, .
\end{equation}
By contrast, the $A_{i-1}$ term in $D_\mu X_i$ carries $G_{i-1}$ gauge indices and has no field with matching color indices in $D_\mu X_{i+1}$. Likewise, the $A_{i+1}$ term in $D_\mu X_{i+1}$ carries $G_{i+1}$ indices and has no matching partner in $D_\mu X_i$. Therefore, in the planar nearest-neighbor hopping matrix element, the only allowed contraction is the one involving two $A_{i}$ insertions.

It follows immediately that the right-moving hopping matrix element is
proportional to $g_{i}^2$. Thus
\begin{equation}
    \gamma_{i,i+1}=- \frac{\widetilde{C}}{8\pi^2} g_{i}^2 ,
\label{eq:SL2RightHop}
\end{equation}
where $\widetilde{C}$ is a numerical coefficient fixed by the remaining one-loop integral and group-theory normalization. Similarly, the left-moving hopping matrix element comes from the common gauge field $A_{i-1}$ appearing in $D X_{i-1}$ and $D X_i$, giving
\begin{equation}
    \gamma_{i,i-1}= - \frac{\widetilde{C}}{8\pi^2} g_{i-1}^2 .
\label{eq:SL2LeftHop}
\end{equation}

The diagonal entries can then be fixed without an independent diagrammatic calculation. The sum over all possible derivative insertions is the covariant derivative of the vacuum operator,
\begin{equation}
    D_\mu \big( X_1 X_2 \cdots X_L \big) = \sum_i X_1 \cdots X_{i-1} (D_\mu X_i) X_{i+1}\cdots X_L=\frac{1}{\sqrt{Z_L}} \sum_i \mathcal{O}^{D}_{i} .
\label{eq:DerivativeDescendant}
\end{equation}
This operator is a descendant of the protected vacuum operator and therefore has vanishing anomalous dimension. Hence the anomalous dimension matrix must annihilate the uniform vector $(1,1,\cdots, 1)$,
\begin{equation}
        \sum_j \gamma_{ij} = 0 .
\end{equation}
Using the nearest-neighbor form of the mixing matrix together with
\eqref{eq:SL2RightHop} and \eqref{eq:SL2LeftHop}, this fixes
\begin{equation}
    \gamma_{ii}=- \gamma_{i,i-1} - \gamma_{i,i+1}=\frac{\widetilde{C}}{8\pi^2}\left( g_{i-1}^2 + g_{i}^2 \right) .
\label{eq:SL2Diagonal}
\end{equation}

Thus the one-impurity anomalous dimension matrix in the derivative sector has the tridiagonal form
\begin{equation}
    \gamma^{(D)}_{ij}= \frac{\widetilde{C}}{8\pi^2}\left[ \left(g_{i-1}^2 + g_{i}^2\right)\delta_{ij} - g_i^2 \delta_{i,j-1} - g_{i-1}^2 \delta_{i,j+1} \right] .
\label{eq:sl2-spin chain-hamiltonian}
\end{equation}
Equivalently, the derivative impurity sector is governed by the same nearest-neighbor spin chain Hamiltonian structure as the $SU(2)$ impurity sector, with the hopping across the link between sites $i$ and $i+1$ weighted by the gauge coupling $g_{i}^2$. The overall coefficient $\widetilde{C}$ is fixed by the explicit evaluation of the remaining one-loop diagram in figure \ref{fig:HoppingDiagram-sl2}.

\section{Chi and Generalized Gamma Distributions} \label{app:Distributions}

In this Appendix, we review some properties of the chi distribution and the generalized gamma distribution, which are the probability distributions used in section \ref{sec:SpinChaos}.

The chi distribution $\chi_k$ with $k$ degrees of freedom is defined as the distribution according to which the magnitude of a vector of length $k$ is distributed if each element of the vector is drawn independently from the standard normal distribution, meaning that the quantity
\begin{equation}
    Y = \left( \sum_{n=1}^k X_n^2 \right)^{1/2}
\end{equation}
is chi-distributed if each $X_n$ is independently drawn from the standard normal distribution. The probability density function of $\chi_k$ is given by
\begin{equation}
    \rho_Y(y;k) = \begin{cases}
        \frac{1}{2^{k/2 - 1} \Gamma(k/2)} y^{k-1}e^{-y^2/2} & y \geq 0 \, , \\
        0 & y<0 \, ,
    \end{cases}
\end{equation}
and its moments are
\begin{equation}
    \mu_n = \int_0^\infty \rho_Y(y;k) y^n \, \mathrm{d}y = 2^{\frac{n}{2}} \frac{\Gamma\left(\frac12 (k+n)\right)}{\Gamma\left(\frac12 k\right)} \, .
\end{equation}

We define the $p$'th power of the chi distribution, $\chi_k^p$, as the probability distribution for the $p$'th power of a random variable drawn from the chi distribution $\chi_k$. Defining $Z = Y^p$ (for $p > 0$), with $Y$ being a random variable drawn from $\chi_k$, the probability distribution function for $Z$ is given by
\begin{equation}
    \rho_Z(z;k) = \frac{1}{p \, 2^{k/2 -1} \Gamma(k/2)} z^{\frac{k}{p} - 1} e^{-\frac{z^{2/p}}{2}} \, , \quad z>0 \, .
\end{equation}
The mean of the $p$'th power of the chi distribution is equal to the $p$'th moment of the chi distribution:
\begin{equation}
    \langle Z \rangle = \mu_p = 2^{\frac{p}{2}} \frac{\Gamma\left(\frac12 (k+p)\right)}{\Gamma\left(\frac12 k\right)} \, .
\end{equation}
On the other hand, the variance is given by
\begin{equation}
    \textrm{Var}(Z) = \langle Z^2 \rangle - \langle Z \rangle^2 =2^p\left( \frac{\Gamma\left(\frac{1}{2}(k+2p)\right)}{\Gamma\left( \frac{1}{2}k \right)} - \left[\frac{\Gamma\left(\frac{1}{2}(k+p)\right)}{\Gamma\left( \frac{1}{2}k \right)}\right]^2 \right).
\end{equation}
In the large $k$ limits discussed in the main text, the mean and variance respectively simplify to
\begin{align}
    \mu&=\langle Z \rangle = k^{p/2} + \mathcal{O}(k^{p/2 -1})\,,\\[1em]
    \sigma^2&=\textrm{Var}(Z) = \frac{1}{2}p^2 k^{p-1} + \mathcal{O}(k^{p-2})\,.
\end{align}
Additionally, we have that\footnote{We note that this expression is only valid if $k>p$, which is violated for $i=1$ when $k= \alpha \beta i$ as in the main text, and $\alpha < p$. However, the contribution of the first few sites is negligible in the large spin chain limit, and the argument that the sum blows up in the $L\rightarrow \infty$ limit still holds.}
\begin{equation}
    \left\langle \frac{1}{Z}\right\rangle = 2^{-\frac{p}{2}} \frac{\Gamma\left(\frac12 (k-p)\right)}{\Gamma\left(\frac12 k\right)} \, ,
\end{equation}
which in the large $k$ limit simplifies to
\begin{equation}
    \left\langle \frac{1}{Z}\right\rangle = k^{-p/2} + \mathcal{O}(k^{-p/2 -1})\,.
\end{equation}

Up to a scale factor, powers of the chi distribution, $\chi_k^p$, are in one-to-one correspondence with a particular slice of the generalized gamma distribution $\text{GeneralizedGamma}(a,d,c)$, the probability density function for which is given by
\begin{equation}
    \rho_{\text{GeneralizedGamma}}(x ; a, d, c) = \frac{c}{a^d \, \Gamma(d/c)} x^{d-1} e^{-(x/a)^c} \, , \quad x>0 \, ,
\end{equation}
where $a$ is a scale parameter and $d$ and $c$ are shape parameters. Comparing the two expressions, one can write the relation as
\begin{equation}
    \chi_k^p = \text{GeneralizedGamma}\left(2^{\frac{p}{2}}, \frac{k}{p}, \frac{2}{p}\right) \, .
\end{equation}

In section \ref{sec:SpinChaos}, we showed that if the couplings $b_i$ of the XXX spin chain are drawn from the distribution $(\chi_{\alpha \beta i})^p$, the dynamics in the 1-impurity sector is chaotic. One can make the same statement by expressing the distribution as a generalized gamma distribution:
\begin{equation}
    b_i \sim \text{GeneralizedGamma}\left( 2^{\frac{p}{2}}, \frac{\alpha  \beta \, i}{p}, \frac{2}{p} \right) \, .
\end{equation}

Equivalently, one can make the statement that drawing the couplings $b_i$ from the distribution
\begin{equation}
    b_i \sim \text{GeneralizedGamma}\left( 2^{1/c}, \alpha_c \, \beta \, i, c \right)
\end{equation}
leads to chaotic dynamics for a sufficiently small $c$ and an appropriately chosen $\alpha_c$.

\section{Spectral Rigidity} \label{app:SpectralRigidity}

In the main text we checked that the local level statistics of our chaotic spin chain models agree with those expected from random matrix theory. In particular, the $r$-statistics probe correlations between nearby energy levels. However, to claim that the corresponding tridiagonal matrices lie in the same universality class as a given ensemble of random matrices, it is also useful to test long-range spectral
correlations. A standard diagnostic for this purpose is the Dyson--Mehta
spectral rigidity, denoted $\Delta_3$ \cite{DysonMehta1963IV, Berry85}.

Let $\{E_i\}$ be the spectrum and define the raw spectral staircase function
\begin{equation}
  N(E) = \#\{i : E_i \leq E\} = \sum_{i} \theta(E - E_i) \, ,
\end{equation}
where $\theta(E)$ is the Heaviside step function. This staircase contains both the smooth variation of the density of states and the fluctuations of the individual levels. We denote the smooth part by $\overline{N}(E)$.

Before computing spectral rigidity, one should perform a so-called unfolding of the spectrum, corresponding to a reparametrization of the energy axis. To each level $E_i$ we assign an unfolded coordinate
\begin{equation}
  x_i = \overline{N}(E_i).
\end{equation}
Since $\overline{N}(E)$ counts the average number of levels below $E$, the unfolded coordinate measures energy in units of the local mean level spacing. Equivalently, the unfolded levels have unit mean density,
\begin{equation}
  \langle d(x) \rangle = 1.
\end{equation}

After this reparametrization, we define the unfolded staircase function by
\begin{equation}
  N_{\mathrm{unf}}(x) = \#\{i : x_i \leq x\} = \sum_i \theta(x-x_i).
\end{equation}

The Dyson--Mehta spectral rigidity is then defined by fitting the unfolded staircase function to the best straight line over a long interval, and measuring the mean-square deviation from this fit. Using a symmetric interval of the form
$[\alpha-\mathcal{E}/2,\alpha+\mathcal{E}/2]$, we define
\begin{equation}
  \Delta_3(\mathcal{E}) = \left\langle \frac{1}{\mathcal{E}}\min_{A,B} \int_{-\mathcal{E}/2}^{\mathcal{E}/2} \Big(N_{\mathrm{unf}}(\alpha + x)-A-Bx \Big)^2 \mathrm{d}x \right\rangle_{\alpha}. \label{eq:Delta3_def}
\end{equation}
Here the minimization over $A$ and $B$ subtracts the best local linear approximation to the staircase function, while the outer average $\langle \cdots \rangle_{\alpha}$ denotes a local average over the center $\alpha$ of the interval. Minimizing with respect to $A$ and $B$ yields:
\begin{equation}
    \Delta_3(\mathcal{E}) = \left\langle \frac{K}{\mathcal{E}} - \frac{J_0^2}{\mathcal{E}^2} - \frac{12 J_1^2}{\mathcal{E}^4} \right\rangle_\alpha
\end{equation}
where
\begin{align}
    J_0 &= \int_{-\mathcal{E}/2}^{\mathcal{E}/2} N_{\text{unf}}(\alpha + x) \, \mathrm{d}x \, , \\
    J_1 &= \int_{-\mathcal{E}/2}^{\mathcal{E}/2} x \, N_{\text{unf}}(\alpha + x) \, \mathrm{d}x \, , \\
    K &= \int_{-\mathcal{E}/2}^{\mathcal{E}/2} N_{\text{unf}}(\alpha + x)^2 \, \mathrm{d}x \, .
\end{align}
Once the unfolded spectrum $\{x_i\}$ is obtained, these integrals are straightforward to compute as they reduce to discrete sums.

The mean $\langle \cdots \rangle_\alpha$ denotes a local averaging over a central value $\alpha$, which can be interpreted as the mean in a small interval $[\alpha - \epsilon/2, \alpha + \epsilon/2]$: 
\begin{equation}
\langle \cdots \rangle_\alpha = \frac{1}{\epsilon} \int_{\alpha - \epsilon/2}^{\alpha + \epsilon/2} (\cdots)_{\alpha_0} \mathrm{d}\alpha_0 \, ,
\end{equation}
such that the width $\epsilon$ of the interval is small compared to the size of the Hilbert space but large compared to the mean spacing. This means that one slides the window $[\alpha-\mathcal{E}/2,\alpha+\mathcal{E}/2]$ across this interval and averages the resulting values. In practice, the local averaging $\langle \cdots \rangle_\alpha$ is done by a discrete approximation of the integral.

$\Delta_3(\mathcal{E})$ measures the stiffness of the spectrum against long-wavelength fluctuations. For an uncorrelated, Poisson spectrum one finds
\begin{equation}
  \Delta_3(\mathcal{E}) \sim \frac{\mathcal{E}}{15}.
\end{equation}
By contrast, for Wigner--Dyson random matrix ensembles, $\Delta_3(\mathcal{E})$ grows only logarithmically with $\mathcal{E}$, and its large-$\mathcal{E}$ asymptotics are given by
\cite{MehtaBook}:
\begin{align}
  \Delta_3^{\mathrm{GOE}}(\mathcal{E}) &=
    \frac{1}{\pi^2}\left[\ln(\mathcal{E}) +\ln(2\pi)+ \gamma - \frac{5}{4}
    - \frac{\pi^2}{8}\right] + \mathcal{O}(\mathcal{E}^{-1}) \simeq \frac{1}{\pi^2}\left[ \ln(\mathcal{E})-0.0687 \right],
    \label{eq:Delta3_GOE}\\
  \Delta_3^{\mathrm{GUE}}(\mathcal{E}) &=
    \frac{1}{2\pi^2}\left[\ln(\mathcal{E})+\ln(2\pi) + \gamma - \frac{5}{4}\right]
    + \mathcal{O}(\mathcal{E}^{-1})\simeq \frac{1}{2\pi^2}\left[ \ln(\mathcal{E})+1.1651 \right],
    \label{eq:Delta3_GUE}\\
  \Delta_3^{\mathrm{GSE}}(\mathcal{E}) &=
    \frac{1}{4\pi^2}\left[\ln(\mathcal{E})+\ln(4\pi) + \gamma - \frac{5}{4}
    + \frac{\pi^2}{8}\right] + \mathcal{O}(\mathcal{E}^{-1})\simeq \frac{1}{4\pi^2}\left[ \ln(\mathcal{E})+3.0919 \right],
    \label{eq:Delta3_GSE}
\end{align}
where $\gamma$ is the Euler--Mascheroni constant. The coefficient of the logarithm is universal, $1/\beta\pi^2$ with $\beta = 1, 2, 4$ for the GOE, GUE, and GSE respectively, while the additive $\mathcal{O}(1)$ constants are ensemble-dependent. Note that chaotic spectra are more rigid than integrable or Poisson spectra, as the long-range fluctuations are parametrically suppressed.

Therefore, while the $r$-statistics test short-range level repulsion, $\Delta_3(\mathcal{E})$ probes long-range spectral correlations. In the main body, we have exhibited clear indications of chaotic dynamics. Moreover, in the large $p$ limit, this behavior tends to the same universality class as random matrices drawn from a Gaussian ensemble.

\subsection{Numerical Procedure}

We now summarize our procedure for numerically extracting the spectral rigidity shown in figure \ref{fig:SpectralRigidities}. 

The first step is to obtain the smoothed-out staircase $\overline{N}(E)$ and compute the unfolded spectrum. We compute the 1-impurity eigenvalues of a spin chain of length $L = 10^4$ with 1000 draws. We then obtain a discrete approximation for $\overline{N}(E)$ by collecting the eigenvalues from all runs together and sorting them into $2 \times 10^5$ bins of equal width.\footnote{Picking a large number of bins gives us higher precision for the estimate of the density of states at low $E$, as the density of states for this system is highly concentrated towards low $E$, and gets even more concentrated as $p$ increases. We picked the number of data points (i.e., the number of bins for the histogram) via trial and error to get a more precise unfolding.

We also note that for the computation of the spectral form factor, we choose the edges of the bins of the histogram to be logarithmically spaced instead of uniformly spaced, with the first bin ranging from 0 to 10 times the smallest spacing between two consecutive eigenvalues in the spectrum. For this computation, we use $10^4$ bins. Using logarithmically spaced edges for the bins results in a more accurate unfolding across the whole spectrum, which is necessary for the computation of the spectral form factor.} We then normalize the count by dividing it by the number of runs, and then fit an interpolating spline curve to the data. Running the simulations with many draws allows us to obtain a more precise approximation for the density of states of the single-impurity Hamiltonian with random couplings.

Then, we compute the spectral rigidity separately for each run (for each draw for a random Hamiltonian). We obtain the unfolded spectrum by the procedure explained above, and compute the spectral rigidity $\Delta_3$. We do the local averaging in $\alpha$ by doing the computations for all integer values of $\alpha$ in the interval $\alpha \in [7900, 8100]$,\footnote{Due to numerical challenges with the estimation of the density of states and unfolding of the spectrum, we picked the midpoint $\alpha$ to be 8000, as the unfolding of the spectrum is more accurate around larger eigenvalues because the density of states before unfolding is highly concentrated towards lower values of $E$. In principle, the spectral form factor does not depend on the choice of the midpoint, as long as we are far away from the edges.} and averaging the results, meaning that
\begin{equation}
    \langle \cdots \rangle_\alpha = \frac{1}{201} \sum_{\alpha = 7900}^{8100} (\cdots)_\alpha \, .
\end{equation}
As the final step, we average the spectral rigidity $\Delta_3$ over all 1000 runs, and show the results in figure \ref{fig:SpectralRigidities}.

\newpage

\bibliographystyle{utphys}
\bibliography{SpinChaos}

\end{document}